\documentclass[aps,prl,twocolumn,superscriptaddress,showpacs]{revtex4}
\usepackage{graphicx}
\usepackage{latexsym}
\usepackage{amssymb}
\usepackage{amsmath}
\usepackage{amsfonts}
\usepackage{bbm}
\usepackage{bm}
\usepackage{multirow}
\usepackage{color}
\usepackage{xcolor}

\usepackage[colorlinks=true, citecolor={blue!80!black}, urlcolor={blue!50!black}, linkcolor = {blue!80!black}]{hyperref}

\usepackage[percent]{overpic}

\begin{document}

\title{Quantum simulation of generic spin exchange models in Floquet-engineered Rydberg atom arrays}

\author{Naveen Nishad}
\affiliation{Physics Department, Technion - Israel Institute of Technology, Haifa 32000, Israel}
\author{Anna Keselman}
\thanks{The Lawrence S. Jackier Fellow}
\affiliation{Physics Department, Technion - Israel Institute of Technology, Haifa 32000, Israel}
\author{Thierry Lahaye}
\affiliation{Université Paris-Saclay, Institut d’Optique Graduate School, CNRS, Laboratoire Charles Fabry, Palaiseau Cedex 91127, France}
\author{Antoine Browaeys}
\affiliation{Université Paris-Saclay, Institut d’Optique Graduate School, CNRS, Laboratoire Charles Fabry, Palaiseau Cedex 91127, France}
\author{Shai Tsesses}
\affiliation{Andrew and Erna Viterbi Department of Electrical \& Computer Engineering, Technion - Israel Institute of Technology, Haifa 32000, Israel}
\affiliation{Department of Physics and Research Laboratory of Electronics, Massachusetts Institute of Technology, Cambridge, MA 02139, USA}

\begin{abstract}

Although quantum simulation can give insight into elusive or intractable physical phenomena, many quantum simulators are unavoidably limited in the models they mimic. Such is also the case for atom arrays interacting via Rydberg states – a platform potentially capable of simulating any kind of spin exchange model, albeit with currently unattainable experimental capabilities. Here, we propose a new route towards simulating generic spin exchange Hamiltonians in atom arrays, using Floquet engineering with both global and local control. To demonstrate the versatility and applicability of our approach, we numerically investigate the generation of several spin exchange models which have yet to be realized in atom arrays, using only previously-demonstrated experimental capabilities. Our proposed scheme can be readily explored in many existing setups, providing a path to investigate a large class of exotic quantum spin models.

\end{abstract}

\maketitle

\section{I. Introduction}

Quantum simulation \cite{Feynman1982,Lloyd1996} is a promising technology for solving complex problems \cite{Abrams1999} and performing optimization \cite{farhi2014} beyond the capabilities of classical computers. For the past 20 years, quantum simulation has managed to expose the behavior of matter during phase transitions \cite{greiner2002} and in extreme out-of-equilibrium scenarios \cite{schreiber2015,Kyprianidis2021,Randall2021}, emulate relativistic \cite{gerritsma2010} or gravitational \cite{steinhauer2016observation}  effects and accurately calculate molecular properties \cite{lanyon2010towards,peruzzo2014}. Though many quantum systems can, in principle, be employed for quantum simulation \cite{Georgescu2014}, neutral atoms have emerged as a leading platform for this purpose \cite{bloch2012}, given their long coherence times and the ability to accurately control their quantum state optically, electrically and magnetically.

Traditionally, quantum simulation with neutral atoms was performed utilizing their controlled collisions in an ensemble \cite{regal2004,bartenstein2004,Zwierlein2004,Bourdel2004}, or the interplay between their tunneling and on-site energies in an optical lattice \cite{jordens2008,Schneider2008,bakr2009,sherson2010,Struck2011,Aidelsburger2013,Miyake2013}. In recent years, however, more focus has been given to atom arrays \cite{Dumke2002} – assortments of single atoms \cite{Nogrette2014} or atom clouds \cite{periwal2021} with a controllable 1D \cite{Endres2016}, 2D \cite{barredo2016} or 3D \cite{barredo2018synthetic} geometry. In atom arrays, the simulated Hamiltonian is usually based on exciting the atoms to Rydberg states \cite{Saffman2010} and inducing an effective spin exchange via van der Waals \cite{Jaksch2001,urban2009observation,gaetan2009observation} or resonant dipole-dipole \cite{walker2005zeros,Barredo2015,ravets2014coherent} interactions. This rapidly developing architecture has already been used to design and build quantum computers \cite{bluvstein2022quantum,graham2022multi}, simulate intractable quantum phases \cite{bernien2017probing,ebadi2021quantum,scholl2021quantum,chen2023continuous} and explore topological properties of matter \cite{Deleseluc2019,semeghini2021}.

Atom arrays have also been proposed as a candidate system to simulate any type of spin model \cite{weimer2010rydberg}. However, even with recent advances towards this goal \cite{Sheng2022,Singh2022,singh2022mid}, no realization of this proposal has thus far been presented. Indeed, atom arrays – and any contemporary quantum simulator, in this regard – can only simulate specific spin exchange Hamiltonians \cite{Altman2021}. Therefore, certain important physical phenomena, such as chiral topological solitons \cite{roessler2006,Lohani2019} or Majorana edge modes \cite{kitaev2006}, are still impossible to controllably generate in experiment.  

Here, we propose a scheme using Floquet engineering \cite{Goldman2014,bukov2015} of atom arrays for analog quantum simulation of generic spin exchange Hamiltonians \cite{tsesses2022}. Our scheme is based on the periodic application of global \cite{Scholl2022} and local \cite{deLeseluc2017} temporal modulation to atoms interacting through resonant dipole-dipole interaction, effectively generating any desired spin exchange. Using exact diagonalization (ED) and Matrix Product State (MPS) simulations, we showcase the versatility of our scheme by producing controlled Dzyaloshinskii-Moriya (DM) \cite{dzyaloshinsky1958,moriya1960} and Kitaev \cite{kitaev2006} interactions, all with currently available experimental capabilities. Our proposal is readily applicable in many experimental setups using atom arrays, laying the ground for a new route for quantum simulation of exotic quantum spin models.

\section{II. Theoretical Formalism}

\begin{figure*}
  \includegraphics[width=\textwidth]{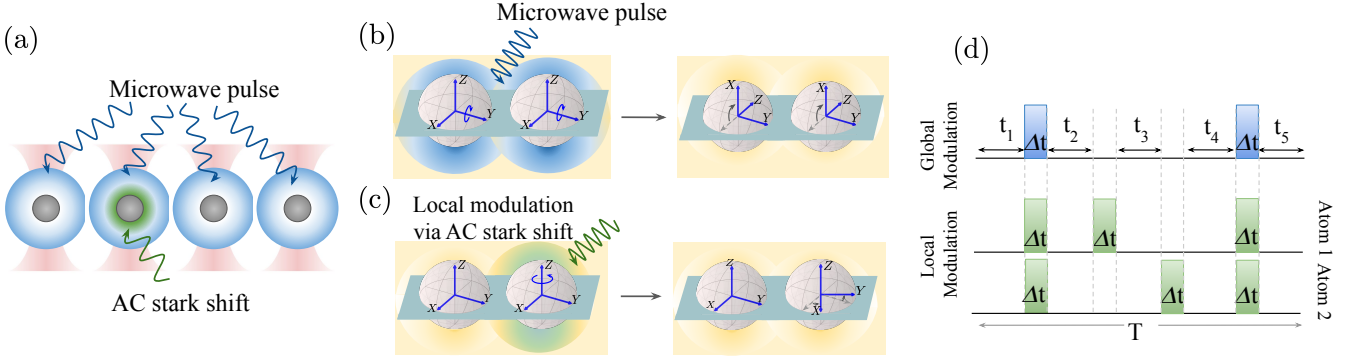}
  \caption{Generating arbitrary spin exchange interactions between Rydberg atoms with Floquet engineering: concept illustration. (a) An array of optically-trapped atoms is excited to a manifold of two Rydberg states with different parities, which interact under a spin exchange Hamiltonian given in Eq.~\eqref{eq:1}. Microwave pulses and AC Stark shift pulses are then periodically applied to the atom array, in order to create the desired effective interaction. (b) When a microwave pulse is applied to all Rydberg atoms simultaneously, it rotates the reference frame of their interaction around the X or Y axis (an example for a rotation of the reference frame about the Y axis is shown in the figure). 
  Such rotations allow for an interaction along all three axes, with controllable magnitudes. (c) When an AC Stark shift is applied to a single atom via an optical pulse, it creates a relative phase between the Rydberg ladder operators of that atom and its neighbours. This phase translates into a rotation about the Z axis in the interaction reference frame, enabling control over the magnitude, sign or symmetry of the exchange interaction. (d) An example for a pulse sequence during one modulation period $T$. Global microwave modulation pulses (blue) are applied on all atoms. Local optical modulation pulses (green) are applied on each atom separately, and can occur simultaneously with global modulation or irrespective of it. In between pulses, in the times marked $t_1,\dots,t_5$, the system evolves freely.}
  \label{fig:1}
\end{figure*}

Figure \ref{fig:1} illustrates the general concept of our method, which is based on the interaction of atoms in an ordered array. The atoms are excited to a manifold of two Rydberg states with different parities, wherein they interact through resonant dipole-dipole interactions (Fig. \ref{fig:1}(a)). Before any modulation to the atoms, when considering the rotating wave approximation and with a sufficiently large interatomic distance, their interaction Hamiltonian is accurately captured by an XY Heisnberg spin exchange of the form \cite{browaeys2020}
\begin{equation} \label{eq:1}
    H_{\rm XY}=\sum_{i \neq j}J_{ij}(\sigma_i^x\sigma_{j}^x+\sigma_i^y\sigma_{j}^y)=\sum_{i \neq j}J_{0}^{ij}(\sigma_{i}^+\sigma_{j}^-+\sigma_{i}^-\sigma_{j}^+),
\end{equation}
where $\sigma^x$, $\sigma^y$ are the Pauli-X and Pauli-Y matrices; $\sigma^+=(\sigma^x+i\sigma^y)/2$ and $\sigma^-=(\sigma^x-i\sigma^y)/2$ are ladder operators in the Rydberg manifold; $i$,$j$ are indices representing atoms at different positions in the array; and $J_{0}^{ij}=2J_{ij}=C_{3}/R_{ij}^3$ is the bare resonant-dipole-dipole interaction strength, which depends on the interatomic distance $R_{ij}$ and the Rydberg state-dependent coefficient $C_3$. Throughout the manuscript, we assume that $C_3$ is isotropic, implying that an out-of-plane magnetic field is applied to the atoms, though controlling the interaction anisotropy can add further degrees of freedom for Hamiltonian engineering \cite{Deleseluc2019}. For simplicity, we consider only the nearest-neighbor interactions in Eq.~\eqref{eq:1} throughout the manuscript 
(i.e., $J_{0}^{\langle ij\rangle }=J_0$ and $J_{0}^{ij}=0$ for $ij\neq \langle ij\rangle$), 
which is a fairly good approximation for the cases considered below (see Supplemental Material). That said, this is not inherently necessary, and generalizations of our method to include interactions between farther neighbors are possible.

To sculpt the interaction in Eq.~\eqref{eq:1} into a generic spin exchange, we employ the concept of Floquet engineering \cite{Goldman2014, bukov2015}, which was successfully implemented in the past for quantum simulation purposes, particularly in order to produce artificial gauge fields \cite{Aidelsburger2013,Miyake2013, jotzu2014,Struck2011}. Floquet engineering involves periodically modulating a physical system in time, and in the latter case its Hamiltonian becomes time-dependent, such that $H(t)=H(t+T)$, where $T$ is the modulation period. By performing the modulation in the high-frequency regime \cite{bukov2015}, which implies $J_0 T \ll 2 \pi$ in our case, one obtains the effective Hamiltonian
\begin{equation} \label{eq:2}
    H_{\rm{eff}}=\frac{1}{T}\int_0^TH(t)dt +\mathcal{O}(T).
\end{equation} 
The leading-order correction to $H_{\rm eff}$, stated explicitly above, arises due to non-commuting terms of $H(t)$ at different times within the modulation period \cite{Goldman2014,bukov2015}. We note that higher-order corrections to $H_{\rm eff}$ decrease polynomially with $T$ (see Supplemental Material) and that Eq.~\eqref{eq:2} converts the problem of reaching an effective Hamiltonian, $H_{{\rm{eff}}}$, into the problem of engineering an instantaneous Hamiltonian, $H(t)$. To this end, we combine two forms of modulation: one applied \emph{globally} on all of the atoms, while another is applied \emph{locally} on each atom separately.

Global modulation can be used to 
rotate the interaction frame of reference
\cite{geier2021}, affecting the $\sigma^x$, $\sigma^y$ operators in Eq.~\eqref{eq:1} via the relation $\tilde{\sigma}^{x,y}=e^{i\Theta\hat{n}\cdot\vec{\sigma}}\sigma^{x,y}e^{-i\Theta\hat{n}\cdot\vec{\sigma}}$, where $\hat{n}$ is the rotation axis, $\Theta$ is the rotation angle, $\vec{\sigma}$ is the Pauli matrix-vector and $\tilde{\sigma}^{x,y}$ are the rotated Pauli operators. Such a modulation has been extensively explored in the context of magnetic resonances in atoms \cite{haeberlen1968coherent,vandersypen2005nmr} or solid-state spins \cite{de2010universal,ryan2010robust}, and was recently applied for Floquet engineering of atoms in Rydberg states via microwave driving \cite{geier2021,Scholl2022}, as illustrated in Fig. \ref{fig:1}(b). 

On the other hand, local modulation can be used to generate a relative phase $\Delta\phi_{ij}$ between the ladder operators $\sigma^+$, $\sigma^-$ of neighboring atoms $i$ and $j$.
This may be achieved by directly modulating the interaction energy \cite{roushan2017chiral} or by locally detuning the energy level structure in a given site \cite{wang2019synthesis}. The latter was previously implemented via AC Stark shifts of Rydberg atoms \cite{deLeseluc2017}, as is illustrated in Fig. \ref{fig:1}(c). It is worth noting that this relative phase, termed the Peierls phase in certain scenarios, is associated with an effective magnetic flux \cite{Lienhard2020Peierls}, enabling the simulation of unique physical phenomena in and of itself, such as quantum gauge fields \cite{ohler2022self} and fractional Chern insulators \cite{Weber2022}. 

When combined, both modulations transform Eq.~\eqref{eq:1} to the instantaneous interaction Hamiltonian
\begin{equation} \label{eq:3} \begin{split}
H(t)=\sum_{i \neq j}J_{ij}[&\cos(\Delta\phi_{ij}(t))(\tilde{\sigma}_i^x(t)\tilde{\sigma}_{j}^x(t)+\tilde{\sigma}_i^y(t)\tilde{\sigma}_{j}^y(t))
\\&+\sin(\Delta\phi_{ij}(t))(\tilde{\sigma}_i^x(t)\tilde{\sigma}_{j}^y(t)-\tilde{\sigma}_i^y(t)\tilde{\sigma}_{j}^x(t))]
\end{split} 
\end{equation}
The effect each modulation has on Eq.~\eqref{eq:1} is apparent from Eq.~\eqref{eq:3}: a global modulation can introduce coupling along the z axis and controls the anisotropy between interaction energies along different axes; while the local modulation can change the sign of the coupling and turn the exchange from symmetric to anti-symmetric, as well as locally determine its magnitude. Eqs.~\eqref{eq:2} and \eqref{eq:3} therefore imply that any desired two-body spin-exchange Hamiltonian may be generated via our method, given a suitable series of optical and microwave pulses is applied to the atoms.

\section{III. Numerical Analysis}

The total modulation period of the system, as seen in Fig. \ref{fig:1}(d), is thus $T=\sum_i t_i+n\Delta t$, where $\Delta t$ denotes the duration of an applied pulse, $n$ is the number of applied pulses, and $t_i$ ,  $i=1,2,\dots, n+1$ are the free evolution times of the system in between pulse applications. We numerically simulate the modulated atomic evolution within a single period via the unitary $U_F$, defined as  
\begin{equation} \label{eq:4}
    U_{F}(t_1,\dots,t_{n+1},\Delta t)=\mathcal{T}\exp(-i\int_0^TH(t)dt),
\end{equation}
where $H(t)$ is the time-dependent Hamiltonian of Eq.~\eqref{eq:3} and $\mathcal{T}$ is the time ordering operator. We implement the time evolution of states using ED for small system sizes (with number of atoms $L\leq10$), and using time-dependent variational principle (TDVP)~\cite{tdvp1,tdvp2} in an MPS framework~\cite{Schollwock2011} for larger system sizes ($L>10$), employing the ITensor library~\cite{itensor}. In all of our MPS calculations, a bond dimension of 50 is being used, which has been found to be sufficient for the time scales at which the results are presented.

In the ideal Floquet engineering scenario, i.e. $\Delta t=0$, reaching a target Hamiltonian $H_S$ is only a matter of determining the free evolution times $t_i$, controlling in turn the coupling strength of different interaction terms. This procedure inevitably depends on both the geometry and boundary conditions of the system, while exhibiting inherent errors stemming from higher-order corrections to the effective Hamiltonian picture or experimental errors due to noise sources. Thus, it is important to design the pulse sequence while employing dynamic decoupling schemes, which are capable of mitigating both issues \cite{choi2020robust}. 
In all of our results below, for the timescales shown, the dynamics induced by the ideal Floquet engineering closely follows that of the target Hamiltonian, allowing us to use it as a basis for comparison with more practical scenarios. 

In contrast to the ideal case, any practical Floquet engineering scenario includes pulses with a finite width, potentially hindering the success of correct Hamiltonian engineering, as the system continues to evolve during pulse application. Thus, determining the right free evolution times becomes a more tasking problem, which we solve by optimizing the sequence on a small number of atoms through ED, before proceeding to MPS simulations. In our optimization process, we minimize $\| U_F^{\dagger} U_S - \mathbbm{1}\|$, where $U_{S}=\exp(-iH_{S}T)$ is the target time evolution unitary, thus maximizing the similarity between the target and engineered time evolution within a single modulation period.  

\begin{figure*}
  \includegraphics[width=0.95\textwidth]{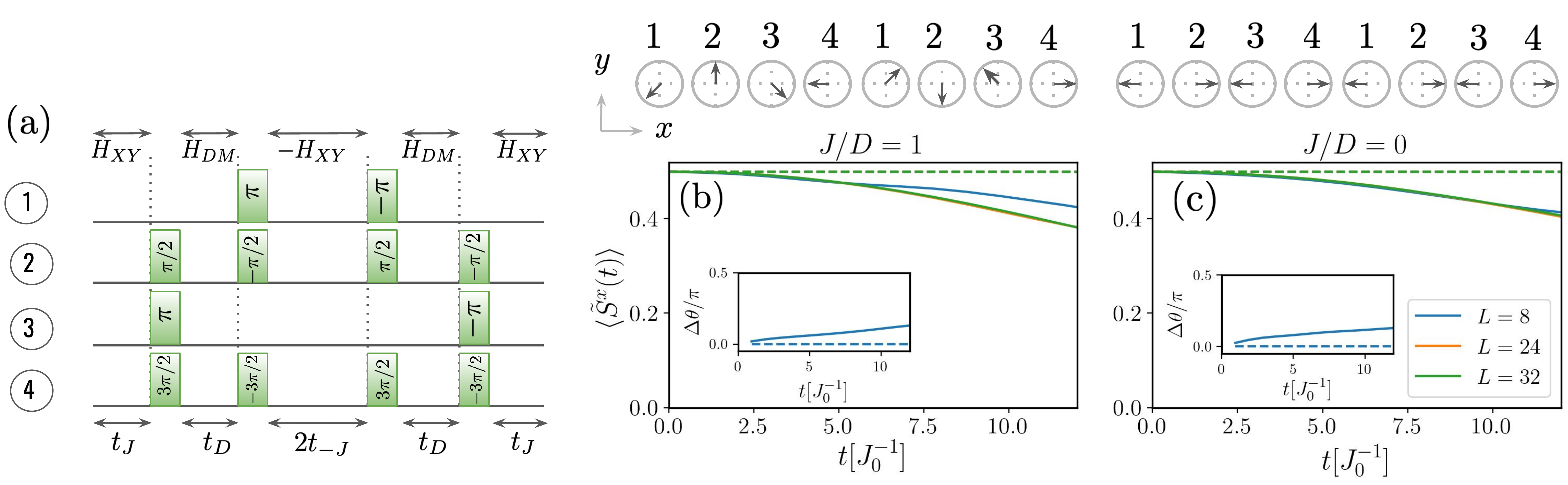}
  \caption{Engineering an effective Dzyaloshinskii-Moriya (DM) interaction between Rydberg atoms with a controllable magnitude. (a) Pulse sequence required to generate the Hamiltonian in Eq.~\eqref{eq:XY+DM} with tunable Heisenberg and DM couplings on a ring of atoms. The pulse sequence is applied simultaneously on each 4-atom segment of the ring and involves only local modulation. Each pulse produces a rotation around the z axis in the interaction reference frame, with a specific rotation angle. During the free evolution times of the sequence, the instantaneous Hamiltonian of the system is either an XY (with positive or negative sign of the coupling) or a purely DM Hamiltonian. (b) Time-dependent local magnetization $\langle \tilde{S}^x(t)\rangle$ for $J/D=1$, when the system is initialized in the zero-energy eigenstate $|\psi_0\rangle$ and is thus expected to stay stationary. Results are plotted for system sizes of 8 (blue), 24 (green) and 32 (orange) atoms, with dashed (solid) lines corresponding to a pulse sequence with $\Delta t = 0$ ($\Delta t = 20$ns). A graphic representation of the initial state in the x-y plane is given above the plot. (c) Same as (b), but for $J/D=0$ (i.e., only DM interaction). Insets in (b) and (c) are the time-dependent coherence figure of merit, $\Delta\theta (t)$, with the dashed (solid) line representing the result for an ideal (finite pulse) modulation scheme. In both (b) and (c), we assume the bare interaction strength between the atoms to be $J_0=2\pi \cdot 250 \rm{kHz}$.}
  \label{fig:2}
\end{figure*}

In practice, our optimization is akin to minimizing the higher-order corrections to the effective Hamiltonian picture (see Supplemental Material). To quantify the influence of the higher-order corrections, we define the unitary $(U_F^{\dagger})^N (U_S)^N$, with $N=t/T$ being the number of modulation periods and denote its time-dependent eigenvalues by $\{\exp(i\theta_l(N))\}_{l=1..2^L}$. Deviations of $\theta_l$ from zero correspond to deviations of the engineered unitary $U_F$ from the desired unitary $U_S$. We thus define the average, time-dependent deviation 
\begin{equation} \label{eq:5}
    \Delta \theta (t) =\sqrt{\frac{1}{2^L} \sum_{l=1}^{2^L}\theta_l^2(t/T)},
\end{equation}
where $\{\theta_l\}\in (-\pi,\pi]$. The figure of merit in Eq.~\eqref{eq:5} gives a quantitative measure for the time-dependent error in the engineered Hamiltonian. Only in the small angle regime ($\Delta \theta \ll \pi$) can one say that the Hamiltonian was sufficiently well engineered, yet this definition is not quantitative. We define here the coherence time of the engineered Hamiltonian as the time it takes $\Delta \theta$ to reach the value $\pi/7$, which corresponds to a ten percent deviation of $\cos(\Delta\theta)$ from unity.

Importantly, the main source of error in a practical scenario stems from the way time-dependent Hamiltonian parameters in Eq.~\eqref{eq:3}  evolve during pulse application times $\Delta t$. For simplicity, and to show the robustness of our method, we chose a square pulse leading to a linear change with time, although such a pulse shape is known to be sub-optimal. 
Further information about the optimization process appears in the Supplemental Material.

In all simulations below, we consider every pulse width in a practical case to be $\Delta t=20$ns, while the bare atom-atom interaction strength and modulation period are $J_0=2\pi \cdot 250 \rm{kHz}$ and $T \approx 0.6 \rm{\mu s}$, respectively. Notably, all chosen parameters are experimentally feasible~\cite{deLeseluc2017,Scholl2022} and ensure that the high-frequency regime is satisfied ($J_0 T \ll 2 \pi$).

\section{IV. Dzyaloshinskii-Moriya and XYZ interactions via either local or global modulation}

We begin exploring the capabilities of Hamiltonian engineering using our method by generating Dzyaloshinskii-Moriya (DM) interaction between Rydberg atoms. DM interaction was initially discovered as the source of weak ferromagnetism in certain antiferromagnets \cite{dzyaloshinsky1958,moriya1960}, and is an antisymmetric interaction term of the form $\sum_{i \neq j}\vec{D_{ij}}\cdot(\vec{\sigma_i} \times \vec{\sigma_j})$, with $\vec{D_{ij}}$ being the interaction strength vector. It is a direct manifestation of spin-orbit coupling \cite{moriya1960}, giving rise to chiral magnetic solitons \cite{bogdanov1994thermodynamically} such as magnetic skyrmions \cite{roessler2006, muhlbauer2009skyrmion, yu2010real}, which hold great promise for applications in magnetic information processing and storage \cite{nagaosa2013topological,fert2013skyrmions,back20202020}. 

Typically, DM interaction is a weak effect in magnetic materials compared to other effects, limiting the interaction regimes one can naturally achieve. Hence, to explore the full range of physical phenomena it can manifest, as well as the consequent quantum magnetic phases \cite{Lohani2019,Sotnikov2021,Psaroudaki2021,Siegl2022,Haller2022}, it is imperative to not only generate DM interaction, but also to control its strength relative to other interaction terms.

Fig. \ref{fig:2} presents the engineering of an effective Hamiltonian with a controlled ratio of an in-plane (XY) Heisenberg interaction with strength $J$ and an out-of-plane (Z) DM interaction with strength $D$, taking the form 
\begin{equation} \label{eq:XY+DM}
H_{\rm XY+DM} = \sum_{\langle i,j \rangle}J(\sigma_i^x\sigma_{j}^x+\sigma_i^y\sigma_{j}^y)+D(\sigma_i^x\sigma_{j}^y-\sigma_i^y\sigma_{j}^x).
\end{equation}
We consider a 1D chain geometry with periodic boundary conditions (i.e., a ring of atoms), and reach the target Hamiltonian of Eq. \eqref{eq:XY+DM} using only local modulation. Fig. \ref{fig:2}(a) presents the modulation sequence, which is applied simultaneously on every 4-atom segment in the chain (see Supplemental Material for the considerations in its construction). Other than the applied modulation, the sequence consists of three free evolution times, denoted as $t_J,t_D$ and $t_{-J}$, which are used to determine the effective interaction strengths. In the ideal case ($\Delta t=0$ ), the couplings in the effective Hamiltonian are given by $J=J_0(t_J-t_{-J})/2(t_J+t_{-J}+t_D)$ and $D=J_0t_D/2(t_J+t_{-J}+t_D)$, where $J_0$ is the bare interaction strength between the atoms. The ratio $J/D$ in this case can be tuned to any value between 0 and $\infty$. 

To validate the effective Hamiltonian for a given ratio $J/D$, we initialize the system in the state $|\psi_0\rangle=V(J/D)|\psi_x\rangle$, where $|\psi_x\rangle$ denotes a ferromagnetic state along the x axis and $V(J/D) = \otimes_{l=1}^L e^{il(\phi/2)\sigma_l^z}$, with $\phi = \tan^{-1}(J/D)+\pi$. This state is a zero-energy eigenstate of the Hamiltonian in Eq.~\eqref{eq:XY+DM}, and should thus remain stationary in time if the Hamiltonian was engineered correctly. The dynamics of this zero-energy state for two values of $J/D$ is shown in Figs. \ref{fig:2}(b,c), where we plot the local magnetization of the time evolved state $|\psi (t)\rangle$ with respect to the initial state $|\psi_0\rangle$. Namely, $\langle \tilde{S}^x(t)\rangle = \frac{1}{L} \sum_{l=1}^L\langle \tilde{\psi} (t)|S_i^x|\tilde{\psi} (t)\rangle$, where $|\tilde{\psi} (t)\rangle =V^{-1}(J/D)|\psi (t)\rangle$.

In the ideal case we see no visible dynamics, as expected from a perfectly engineered Hamiltonian. However, for a practical scenario with finite pulse widths, we see deviations of the local magnetization with respect to the initial state, albeit the time scale for this dynamics is fairly long, with $\langle \tilde{S}^x(t)\rangle$ reducing to half of its initial value when $t \approx 24 J_0^{-1}$ for both cases considered. The engineered Hamiltonian coherence time, as defined above using $\Delta \theta$, was about $14 J_0^{-1}$ for both cases considered. Thus, the Hamiltonian is correctly engineered for times well above the experimentally measured decoherence time of untrapped Rydberg atoms \cite{Barredo2020,Xu2021}, which is $15 \rm{\mu s}$ at the most, or less than $4J_0^{-1}$ for our chosen parameters. We note that no significant dependence on system size is observed in the dynamics (explored here for $L=8,24$ and $32$, see Supplemental Material for more system sizes). Thus, Fig. \ref{fig:2} illustrates that our approach for Hamiltonian engineering is both practically possible and scalable, while providing engineered Hamiltonian values beyond what is naturally possible in magnetic materials (where $D \ll J$).

\begin{figure}
  \includegraphics[width=0.38\textwidth]{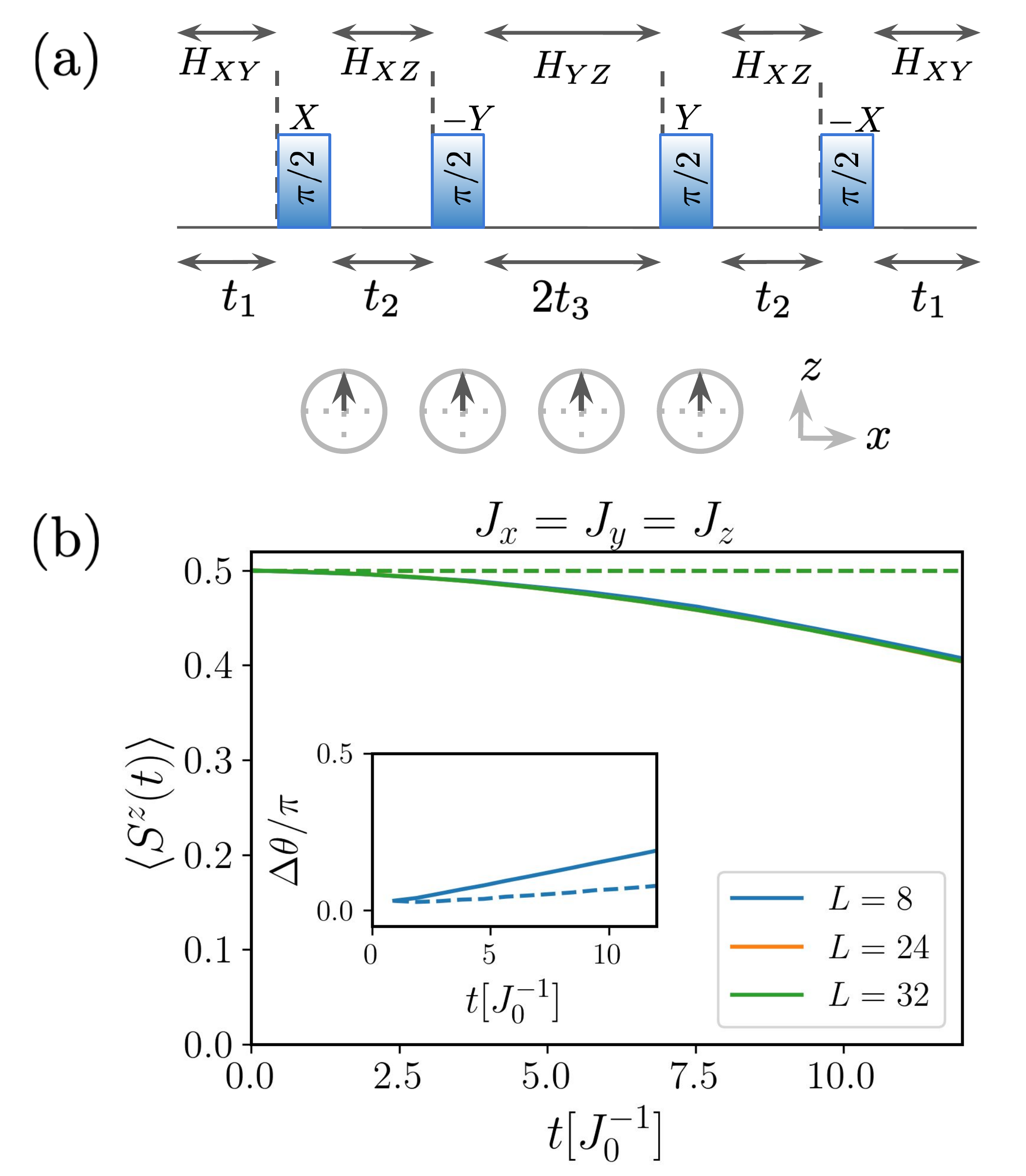}
  \caption{Engineering an effective XYZ Heisenberg Hamiltonian in a Rydberg atoms array. (a) Pulse sequence required to generate the XYZ interaction \cite{geier2021,Scholl2022}, termed the WAHUHA sequence in dynamic decoupling theory \cite{choi2020robust}. Each pulse rotates the interaction frame of reference for all atoms around either the x or y axes, such that during each free evolution time period the instantaneous interaction is along a different set of two axes. (b) Average z magnetization $\langle S^{z}(t)\rangle$ for $J_x=J_y=J_z$, when the system is initialized in a ferromagnetic state along the z axis (illustrated above the plot). The initial state is expected to be stationary  \cite{Scholl2022}. Results are plotted for system sizes of 8 (blue), 24 (green) and 32 (orange) atoms, with dashed (solid) lines corresponding to a pulse sequence with $\Delta t = 0$ ($\Delta t = 20$ns). A graphic representation of the initial state in the x-z plane is given above the plot. Inset is the time-dependent coherence figure of merit, $\Delta\theta (t)$, with the dashed (solid) line representing the result for an ideal (finite pulse) modulation scheme. In both (b) and (c), we assume the bare interaction strength between the atoms to be $J_0=2\pi \cdot 250 \rm{kHz}$.}
\label{fig:3}
\end{figure}

For completeness we investigate the operation of global modulations within our scheme, 
engineering an effective XYZ interaction in a similar ring of Rydberg atoms, as has been recently demonstrated experimentally \cite{Scholl2022}. The effective Hamiltonian takes the form
\begin{equation} \label{eq:XYZ}
H_{\rm XYZ} = \sum_{\langle i,j \rangle}J_x\sigma_i^x\sigma_{j}^x+J_y\sigma_i^y\sigma_{j}^y+J_z\sigma_i^z\sigma_{j}^z,
\end{equation}
where $\sigma_i^z$ is the Pauli-Z operator of the atom $i$, and $J_x,J_y,J_z$ denote the coupling strength along each axis. The results of Hamiltonian engineering appear in Fig. \ref{fig:3}, and although the pulse sequence is essentially the same as in \cite{geier2021,Scholl2022} (Fig. \ref{fig:3}(a)), here we numerically optimize the free evolution times to compensate for the finite duration of applied pulses in a practical scenario. In the ideal case, $J_x=J_0\frac{t_1+t_2}{2(t_1+t_2+t_3)},J_y=J_0\frac{t_1+t_3}{2(t_1+t_2+t_3)}$ and $J_z=J_0\frac{t_2+t_3}{2(t_1+t_2+t_3)}$ where $t_1,t_2$ and $t_3$ are the free evolution times defined in Fig. \ref{fig:3}(a) and $J_0$ is the bare interaction strength between the Rydberg atoms. Note that this pulse sequence dictates the relations $J_x+J_y+J_z=J_0$ and $J_x,J_y,J_z \leq J_0/2$, which constrain the achievable anisotropy using global modulation alone.

Fig. \ref{fig:3}(b) shows the magnetization dynamics of a system engineered to have isotropic interaction (i.e., $J_x=J_y=J_z$). We initialize the system in a ferromagnetic state along the z axis (as shown above Fig. \ref{fig:3}(b)), which is an eigenstate of the target Hamiltonian, and should therefore remain stationary. Results with different initial states, as well as with anisotropic interaction, are given in the Supplemental Material.  

Contrary to the case of local modulation, even in the case of ideal (infinitely short) pulses there is a visible increase in $\Delta \theta$ over time, even though no change in the local magnetization can be observed. This is due to the existence of non-commuting terms between instantaneous Hamiltonians in different times. The engineered Hamiltonian coherence time in a practical scenario is shorter than in the case of local modulation, and stands at about $~9J_0^{-1}$. Notably, it is still much larger than currently achievable decoherence times of Rydberg atoms, as mentioned above. Interestingly, the magnetization reduces to half of its initial value at a far larger timescale ($t \approx 30J_0^{-1}$). 

The results of both local and global modulation, therefore, highlight that measuring magnetization dynamics of a single state is not always a good method to verify Hamiltonian engineering. Additionally, the accuracy of Hamiltonian engineering may be increased even when using sub-optimal pulses, by optimizing free evolution times (see Supplemental Material for a comparison to the result with parameters in \cite{Scholl2022}).

\begin{figure*}
  \includegraphics[width=\textwidth]{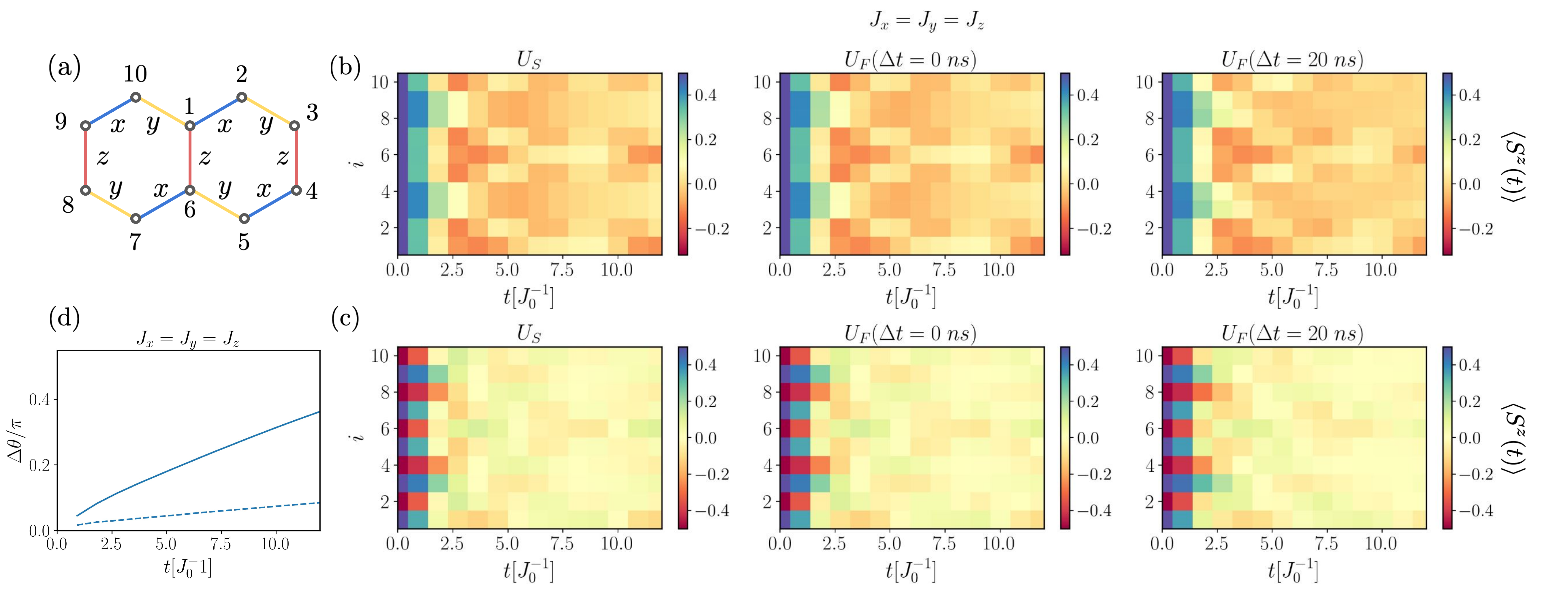}
  \caption{Engineering an effective Kitaev interaction between Rydberg atoms. (a) Pictorial representation of the Kitaev couplings. Different $\alpha$-bonds are color-coded in the following way: blue (x axis interaction); yellow (y axis interaction); and red (z axis interaction). The illustration shows two plaquettes of the lattice, corresponding exactly to the simulated atomic system. (b) A heatmap of the z magnetization dynamics $\langle S^{z}(t)\rangle$ for all 10 simulated atoms in the two plaquette system. In the target Hamiltonian, all interaction strengths in each direction are equal ($J_x=J_y=J_z=J_0/3$), and the initial state is ferromagnetic along the z axis. Three subplots are shown: dynamics with the target Hamiltonian ($U_S$), dynamics with an ideal engineered Hamiltonian ($U_F(\Delta t =0)$) and dynamics with an engineered Hamiltonian with finite pulse widths ($U_F(\Delta t =20ns)$). A yellow color represents positive magnetization and a blue color represents negative magnetization. (c) the same as (b), but for an antiferromagnetic initial state. (d) Time-dependent coherence figure of merit, $\Delta\theta(t)$, with the dashed (solid) line representing the result for an ideal (finite pulse) modulation scheme.  Similarly to fig. \ref{fig:2}, we assumed here $J_0=2\pi \cdot 250 \rm{kHz}$.}
  \label{fig:Kitaev_figure}
\end{figure*}

\section{V. Kitaev interaction via combined global and local modulation}
Finally, we combine both local and global modulation to produce a generic spin exchange, specifically choosing the Kitaev interaction \cite{kitaev2006} . As can be seen in Fig. \ref{fig:Kitaev_figure}(a), the Kitaev interaction is a direction-dependent spin exchange on a honeycomb lattice, which can be written compactly as 
\begin{equation} \label{eq:kitaev}
H_{\rm Kitaev} = \sum_{\alpha=x,y,z} J_\alpha \sum_{\langle ij\rangle_\alpha} \sigma_{i}^\alpha\sigma_{j}^\alpha,
\end{equation}
where $\langle ij\rangle_\alpha$ denote nearest-neighbors on the $\alpha$-bond. The Kitaev honeycomb model is one of the hallmarks of modern condensed matter physics and is analytically exactly solvable. It is known to host a quantum spin liquid in its ground state \cite{balents2010spin}, and exhibits a quantum phase transition when the absolute value of the interaction strength along one direction exceeds the sum of the absolute value of the others. Furthermore, applying a magnetic field in this model is known to drive the system into a gapped quantum spin liquid with non-abelian anyonic excitations \cite{kitaev2006}.

While the search for magnetic materials that exhibit a dominant Kitaev exchange is still ongoing \cite{Jackeli2009,kim2009phase,Singh2012,Choi2012,Plumb2014,hwan2015direct,banerjee2016proximate,kasahara2018majorana,sears2020ferromagnetic,yokoi2021half}, it is natural to ask whether this type of interaction could be engineered, to allow a more detailed investigation. In fact, two recent proposals to do just that also rely on cold neutral atoms and a different form of Floquet engineering \cite{kalinowski2022non,sun2022engineering}. Our proposal to engineer Kitaev interaction requires a complex pulse sequence with a large number of parameters, and is therefore appended in the Supplemental Material. It relies on the global modulation introduced in Fig. \ref{fig:3}a, along with local modulation during and in-between the globally-applied pulses. The results for simulations of two plaquettes of the honeycomb lattice (10 atoms) are given in Fig. \ref{fig:Kitaev_figure}b,c. The figures show the dynamics of the magnetization along the z axis for two different initial states, one ferromagnetic and the other antiferromagnetic, where the target Hamiltonian was a pure Kitaev interaction with $J_x=J_y=J_z$. These initial states were chosen as simple, experimentally feasible examples, since no eigenstate of this Hamiltonian is a product state. Similar dynamics, yet in the other possible phase of the Kitaev model, are plotted in the Supplemental Material. 

We compare the target Hamiltonian dynamics to those of the engineered Hamiltonian, finding a good agreement between the results for all presented times (i.e., $t \leq 12J_0^{-1}$), for the application of either ideal or practical pulses. However, when $\Delta \theta$ is analyzed (Fig. \ref{fig:Kitaev_figure}(d)), the coherence time for the engineered Hamiltonian is found to be about $4J_0^{-1}$ in the practical case and about $23J_0^{-1}$ in the ideal case. This relatively low coherence, stemming from the combination of local and global modulation, as well as the system geometry, is nevertheless sufficient for meaningful quantum simulation, and can be extended by further optimizing the applied pulse shape or the pulse sequence. The results in Fig. \ref{fig:Kitaev_figure} serve to illustrate that our method can indeed generate a generic spin exchange between Rydberg atoms.    

\section{VI. Discussion}

In summary, we proposed a method to simulate generic spin exchange Hamiltonians using Floquet engineering of atom arrays, when combining both global and local temporal modulation. Through numerical simulation, we demonstrated the generation of DM and XYZ interactions for over 30 atoms in a ring, and Kitaev interaction for 10 atoms in two plaquettes of a honeycomb lattice. Rather than just consider an idealized case, our simulations included currently achievable experimental parameters \cite{deLeseluc2017,Scholl2022}, showing that our proposal can be employed already in existing experimental setups.  

In principle, our scheme is entirely scalable, since the length of the pulse sequence required for Hamiltonian engineering scales only with the number of atoms interacting in a single unit cell (e.g., a chain, a triangular lattice, a square lattice etc.), and not with the number of atoms (as can be seen in Figs. 2,3). This is also true for the decrease in engineered Hamiltonian coherence, as adding more interactions for each atom increases the effect of higher-order corrections to the effective Hamiltonian. It is further worth noting that different engineered interaction terms experience decoherence differently, as evident from Figs. 2-4, implying that some spin models will be easier to simulate then others. For local modulation purposes, the number of addressing beams does scale linearly with the number of atoms, but this is not out of the ordinary in the field of atom arrays \cite{young2020half}. 

Our proposal then serves as an alternative route for universal quantum simulation of spin models \cite{weimer2010rydberg}, with the main advantage being simplicity in design and implementation, making it more accessible to a wider range of experimental systems. Its main disadvantages stem from the need to modulate the system both locally and globally, limiting the possible simulation time while exacerbating the risk of scattering atoms out of the required Rydberg manifold. That said, both issues can be mitigated by carefully choosing the parameters of the atomic system (Rydberg levels, energy detunings etc.), while requiring that it decoheres faster than the engineered Hamiltonian (which is quite possible, as we have shown). Ultimately, with improvements to the coherence of Rydberg atoms \cite{Barredo2020} and an increase in modulation speed, one can expect quantum simulation with our scheme to span timescales wherein atoms undergo hundreds, or even thousands, of interactions.  

Finally, our method can readily be used for several other Hamiltonian engineering functionalities, even though we do not directly demonstrate them in this work. Using only global modulation, for example, one can compensate for residual van der Waals interaction between Rydberg atoms, potentially enabling an XY model with faster interaction times \cite{browaeys2020}. On the other hand, employing only local modulation can engineer the ratio between nearest- and next-nearest-neighbor interactions, a highly important parameter for models of frustrated quantum magnets \cite{balents2010spin}. Furthermore, local modulation can artificially produce a different functional dependence for the interaction between the atoms, allowing a transition between long-range and short-range interactions. Finally, using both local and global modulation allows for the engineering of three-dimensional Heisenberg and DM interactions, giving rise to quantum magnetic topological solitons \cite{Lohani2019,Sotnikov2021,Psaroudaki2021,Siegl2022,Haller2022}.

\section{Acknowledgements}
A.K. acknowledges funding by the Israeli Council for Higher Education support program for hiring outstanding faculty members in quantum science and technology in research universities and by the Israel Science Foundation (Grant No. 2443/22). S.T. is grateful for the support of the Yad Hanadiv Foundation through the Rothschild Fellowship, the Israeli Council for Higher Education through the Quantum Science and Technology Post-Doctoral Fellowship and the Adams Fellowship Program of the Israel Academy of Science and Humanities.

\bibliography{refs}

\begin{thebibliography}{105}
\expandafter\ifx\csname natexlab\endcsname\relax\def\natexlab#1{#1}\fi
\expandafter\ifx\csname bibnamefont\endcsname\relax
  \def\bibnamefont#1{#1}\fi
\expandafter\ifx\csname bibfnamefont\endcsname\relax
  \def\bibfnamefont#1{#1}\fi
\expandafter\ifx\csname citenamefont\endcsname\relax
  \def\citenamefont#1{#1}\fi
\expandafter\ifx\csname url\endcsname\relax
  \def\url#1{\texttt{#1}}\fi
\expandafter\ifx\csname urlprefix\endcsname\relax\def\urlprefix{URL }\fi
\providecommand{\bibinfo}[2]{#2}
\providecommand{\eprint}[2][]{\url{#2}}

\bibitem[{\citenamefont{Feynman}(1982)}]{Feynman1982}
\bibinfo{author}{\bibfnamefont{R.~P.} \bibnamefont{Feynman}},
  \bibinfo{journal}{International Journal of Theoretical Physics}
  \textbf{\bibinfo{volume}{21}}, \bibinfo{pages}{467} (\bibinfo{year}{1982}).

\bibitem[{\citenamefont{Lloyd}(1996)}]{Lloyd1996}
\bibinfo{author}{\bibfnamefont{S.}~\bibnamefont{Lloyd}},
  \bibinfo{journal}{Science} \textbf{\bibinfo{volume}{273}},
  \bibinfo{pages}{1073} (\bibinfo{year}{1996}).

\bibitem[{\citenamefont{Abrams and Lloyd}(1999)}]{Abrams1999}
\bibinfo{author}{\bibfnamefont{D.~S.} \bibnamefont{Abrams}} \bibnamefont{and}
  \bibinfo{author}{\bibfnamefont{S.}~\bibnamefont{Lloyd}},
  \bibinfo{journal}{Phys. Rev. Lett.} \textbf{\bibinfo{volume}{83}},
  \bibinfo{pages}{5162} (\bibinfo{year}{1999}).

\bibitem[{\citenamefont{Farhi et~al.}(2014)\citenamefont{Farhi, Goldstone, and
  Gutmann}}]{farhi2014}
\bibinfo{author}{\bibfnamefont{E.}~\bibnamefont{Farhi}},
  \bibinfo{author}{\bibfnamefont{J.}~\bibnamefont{Goldstone}},
  \bibnamefont{and} \bibinfo{author}{\bibfnamefont{S.}~\bibnamefont{Gutmann}},
  \bibinfo{journal}{arXiv preprint arXiv:1411.4028}  (\bibinfo{year}{2014}).

\bibitem[{\citenamefont{Greiner et~al.}(2002)\citenamefont{Greiner, Mandel,
  Esslinger, H{\"a}nsch, and Bloch}}]{greiner2002}
\bibinfo{author}{\bibfnamefont{M.}~\bibnamefont{Greiner}},
  \bibinfo{author}{\bibfnamefont{O.}~\bibnamefont{Mandel}},
  \bibinfo{author}{\bibfnamefont{T.}~\bibnamefont{Esslinger}},
  \bibinfo{author}{\bibfnamefont{T.~W.} \bibnamefont{H{\"a}nsch}},
  \bibnamefont{and} \bibinfo{author}{\bibfnamefont{I.}~\bibnamefont{Bloch}},
  \bibinfo{journal}{nature} \textbf{\bibinfo{volume}{415}}, \bibinfo{pages}{39}
  (\bibinfo{year}{2002}).

\bibitem[{\citenamefont{Schreiber et~al.}(2015)\citenamefont{Schreiber,
  Hodgman, Bordia, Lüschen, Fischer, Vosk, Altman, Schneider, and
  Bloch}}]{schreiber2015}
\bibinfo{author}{\bibfnamefont{M.}~\bibnamefont{Schreiber}},
  \bibinfo{author}{\bibfnamefont{S.~S.} \bibnamefont{Hodgman}},
  \bibinfo{author}{\bibfnamefont{P.}~\bibnamefont{Bordia}},
  \bibinfo{author}{\bibfnamefont{H.~P.} \bibnamefont{Lüschen}},
  \bibinfo{author}{\bibfnamefont{M.~H.} \bibnamefont{Fischer}},
  \bibinfo{author}{\bibfnamefont{R.}~\bibnamefont{Vosk}},
  \bibinfo{author}{\bibfnamefont{E.}~\bibnamefont{Altman}},
  \bibinfo{author}{\bibfnamefont{U.}~\bibnamefont{Schneider}},
  \bibnamefont{and} \bibinfo{author}{\bibfnamefont{I.}~\bibnamefont{Bloch}},
  \bibinfo{journal}{Science} \textbf{\bibinfo{volume}{349}},
  \bibinfo{pages}{842} (\bibinfo{year}{2015}).

\bibitem[{\citenamefont{Kyprianidis et~al.}(2021)\citenamefont{Kyprianidis,
  Machado, Morong, Becker, Collins, Else, Feng, Hess, Nayak, Pagano
  et~al.}}]{Kyprianidis2021}
\bibinfo{author}{\bibfnamefont{A.}~\bibnamefont{Kyprianidis}},
  \bibinfo{author}{\bibfnamefont{F.}~\bibnamefont{Machado}},
  \bibinfo{author}{\bibfnamefont{W.}~\bibnamefont{Morong}},
  \bibinfo{author}{\bibfnamefont{P.}~\bibnamefont{Becker}},
  \bibinfo{author}{\bibfnamefont{K.~S.} \bibnamefont{Collins}},
  \bibinfo{author}{\bibfnamefont{D.~V.} \bibnamefont{Else}},
  \bibinfo{author}{\bibfnamefont{L.}~\bibnamefont{Feng}},
  \bibinfo{author}{\bibfnamefont{P.~W.} \bibnamefont{Hess}},
  \bibinfo{author}{\bibfnamefont{C.}~\bibnamefont{Nayak}},
  \bibinfo{author}{\bibfnamefont{G.}~\bibnamefont{Pagano}},
  \bibnamefont{et~al.}, \bibinfo{journal}{Science}
  \textbf{\bibinfo{volume}{372}}, \bibinfo{pages}{1192} (\bibinfo{year}{2021}).

\bibitem[{\citenamefont{Randall et~al.}(2021)\citenamefont{Randall, Bradley,
  van~der Gronden, Galicia, Abobeih, Markham, Twitchen, Machado, Yao, and
  Taminiau}}]{Randall2021}
\bibinfo{author}{\bibfnamefont{J.}~\bibnamefont{Randall}},
  \bibinfo{author}{\bibfnamefont{C.~E.} \bibnamefont{Bradley}},
  \bibinfo{author}{\bibfnamefont{F.~V.} \bibnamefont{van~der Gronden}},
  \bibinfo{author}{\bibfnamefont{A.}~\bibnamefont{Galicia}},
  \bibinfo{author}{\bibfnamefont{M.~H.} \bibnamefont{Abobeih}},
  \bibinfo{author}{\bibfnamefont{M.}~\bibnamefont{Markham}},
  \bibinfo{author}{\bibfnamefont{D.~J.} \bibnamefont{Twitchen}},
  \bibinfo{author}{\bibfnamefont{F.}~\bibnamefont{Machado}},
  \bibinfo{author}{\bibfnamefont{N.~Y.} \bibnamefont{Yao}}, \bibnamefont{and}
  \bibinfo{author}{\bibfnamefont{T.~H.} \bibnamefont{Taminiau}},
  \bibinfo{journal}{Science} \textbf{\bibinfo{volume}{374}},
  \bibinfo{pages}{1474} (\bibinfo{year}{2021}).

\bibitem[{\citenamefont{Gerritsma et~al.}(2010)\citenamefont{Gerritsma,
  Kirchmair, Z{\"a}hringer, Solano, Blatt, and Roos}}]{gerritsma2010}
\bibinfo{author}{\bibfnamefont{R.}~\bibnamefont{Gerritsma}},
  \bibinfo{author}{\bibfnamefont{G.}~\bibnamefont{Kirchmair}},
  \bibinfo{author}{\bibfnamefont{F.}~\bibnamefont{Z{\"a}hringer}},
  \bibinfo{author}{\bibfnamefont{E.}~\bibnamefont{Solano}},
  \bibinfo{author}{\bibfnamefont{R.}~\bibnamefont{Blatt}}, \bibnamefont{and}
  \bibinfo{author}{\bibfnamefont{C.}~\bibnamefont{Roos}},
  \bibinfo{journal}{Nature} \textbf{\bibinfo{volume}{463}}, \bibinfo{pages}{68}
  (\bibinfo{year}{2010}).

\bibitem[{\citenamefont{Steinhauer}(2016)}]{steinhauer2016observation}
\bibinfo{author}{\bibfnamefont{J.}~\bibnamefont{Steinhauer}},
  \bibinfo{journal}{Nature Physics} \textbf{\bibinfo{volume}{12}},
  \bibinfo{pages}{959} (\bibinfo{year}{2016}).

\bibitem[{\citenamefont{Lanyon et~al.}(2010)\citenamefont{Lanyon, Whitfield,
  Gillett, Goggin, Almeida, Kassal, Biamonte, Mohseni, Powell, Barbieri
  et~al.}}]{lanyon2010towards}
\bibinfo{author}{\bibfnamefont{B.~P.} \bibnamefont{Lanyon}},
  \bibinfo{author}{\bibfnamefont{J.~D.} \bibnamefont{Whitfield}},
  \bibinfo{author}{\bibfnamefont{G.~G.} \bibnamefont{Gillett}},
  \bibinfo{author}{\bibfnamefont{M.~E.} \bibnamefont{Goggin}},
  \bibinfo{author}{\bibfnamefont{M.~P.} \bibnamefont{Almeida}},
  \bibinfo{author}{\bibfnamefont{I.}~\bibnamefont{Kassal}},
  \bibinfo{author}{\bibfnamefont{J.~D.} \bibnamefont{Biamonte}},
  \bibinfo{author}{\bibfnamefont{M.}~\bibnamefont{Mohseni}},
  \bibinfo{author}{\bibfnamefont{B.~J.} \bibnamefont{Powell}},
  \bibinfo{author}{\bibfnamefont{M.}~\bibnamefont{Barbieri}},
  \bibnamefont{et~al.}, \bibinfo{journal}{Nature chemistry}
  \textbf{\bibinfo{volume}{2}}, \bibinfo{pages}{106} (\bibinfo{year}{2010}).

\bibitem[{\citenamefont{Peruzzo et~al.}(2014)\citenamefont{Peruzzo, McClean,
  Shadbolt, Yung, Zhou, Love, Aspuru-Guzik, and O’Brien}}]{peruzzo2014}
\bibinfo{author}{\bibfnamefont{A.}~\bibnamefont{Peruzzo}},
  \bibinfo{author}{\bibfnamefont{J.}~\bibnamefont{McClean}},
  \bibinfo{author}{\bibfnamefont{P.}~\bibnamefont{Shadbolt}},
  \bibinfo{author}{\bibfnamefont{M.-H.} \bibnamefont{Yung}},
  \bibinfo{author}{\bibfnamefont{X.-Q.} \bibnamefont{Zhou}},
  \bibinfo{author}{\bibfnamefont{P.~J.} \bibnamefont{Love}},
  \bibinfo{author}{\bibfnamefont{A.}~\bibnamefont{Aspuru-Guzik}},
  \bibnamefont{and} \bibinfo{author}{\bibfnamefont{J.~L.}
  \bibnamefont{O’Brien}}, \bibinfo{journal}{Nature communications}
  \textbf{\bibinfo{volume}{5}}, \bibinfo{pages}{4213} (\bibinfo{year}{2014}).

\bibitem[{\citenamefont{Georgescu et~al.}(2014)\citenamefont{Georgescu, Ashhab,
  and Nori}}]{Georgescu2014}
\bibinfo{author}{\bibfnamefont{I.~M.} \bibnamefont{Georgescu}},
  \bibinfo{author}{\bibfnamefont{S.}~\bibnamefont{Ashhab}}, \bibnamefont{and}
  \bibinfo{author}{\bibfnamefont{F.}~\bibnamefont{Nori}},
  \bibinfo{journal}{Rev. Mod. Phys.} \textbf{\bibinfo{volume}{86}},
  \bibinfo{pages}{153} (\bibinfo{year}{2014}).

\bibitem[{\citenamefont{Bloch et~al.}(2012)\citenamefont{Bloch, Dalibard, and
  Nascimbene}}]{bloch2012}
\bibinfo{author}{\bibfnamefont{I.}~\bibnamefont{Bloch}},
  \bibinfo{author}{\bibfnamefont{J.}~\bibnamefont{Dalibard}}, \bibnamefont{and}
  \bibinfo{author}{\bibfnamefont{S.}~\bibnamefont{Nascimbene}},
  \bibinfo{journal}{Nature Physics} \textbf{\bibinfo{volume}{8}},
  \bibinfo{pages}{267} (\bibinfo{year}{2012}).

\bibitem[{\citenamefont{Regal et~al.}(2004)\citenamefont{Regal, Greiner, and
  Jin}}]{regal2004}
\bibinfo{author}{\bibfnamefont{C.~A.} \bibnamefont{Regal}},
  \bibinfo{author}{\bibfnamefont{M.}~\bibnamefont{Greiner}}, \bibnamefont{and}
  \bibinfo{author}{\bibfnamefont{D.~S.} \bibnamefont{Jin}},
  \bibinfo{journal}{Phys. Rev. Lett.} \textbf{\bibinfo{volume}{92}},
  \bibinfo{pages}{040403} (\bibinfo{year}{2004}).

\bibitem[{\citenamefont{Bartenstein et~al.}(2004)\citenamefont{Bartenstein,
  Altmeyer, Riedl, Jochim, Chin, Denschlag, and Grimm}}]{bartenstein2004}
\bibinfo{author}{\bibfnamefont{M.}~\bibnamefont{Bartenstein}},
  \bibinfo{author}{\bibfnamefont{A.}~\bibnamefont{Altmeyer}},
  \bibinfo{author}{\bibfnamefont{S.}~\bibnamefont{Riedl}},
  \bibinfo{author}{\bibfnamefont{S.}~\bibnamefont{Jochim}},
  \bibinfo{author}{\bibfnamefont{C.}~\bibnamefont{Chin}},
  \bibinfo{author}{\bibfnamefont{J.~H.} \bibnamefont{Denschlag}},
  \bibnamefont{and} \bibinfo{author}{\bibfnamefont{R.}~\bibnamefont{Grimm}},
  \bibinfo{journal}{Phys. Rev. Lett.} \textbf{\bibinfo{volume}{92}},
  \bibinfo{pages}{120401} (\bibinfo{year}{2004}).

\bibitem[{\citenamefont{Zwierlein et~al.}(2004)\citenamefont{Zwierlein, Stan,
  Schunck, Raupach, Kerman, and Ketterle}}]{Zwierlein2004}
\bibinfo{author}{\bibfnamefont{M.~W.} \bibnamefont{Zwierlein}},
  \bibinfo{author}{\bibfnamefont{C.~A.} \bibnamefont{Stan}},
  \bibinfo{author}{\bibfnamefont{C.~H.} \bibnamefont{Schunck}},
  \bibinfo{author}{\bibfnamefont{S.~M.~F.} \bibnamefont{Raupach}},
  \bibinfo{author}{\bibfnamefont{A.~J.} \bibnamefont{Kerman}},
  \bibnamefont{and} \bibinfo{author}{\bibfnamefont{W.}~\bibnamefont{Ketterle}},
  \bibinfo{journal}{Phys. Rev. Lett.} \textbf{\bibinfo{volume}{92}},
  \bibinfo{pages}{120403} (\bibinfo{year}{2004}).

\bibitem[{\citenamefont{Bourdel et~al.}(2004)\citenamefont{Bourdel, Khaykovich,
  Cubizolles, Zhang, Chevy, Teichmann, Tarruell, Kokkelmans, and
  Salomon}}]{Bourdel2004}
\bibinfo{author}{\bibfnamefont{T.}~\bibnamefont{Bourdel}},
  \bibinfo{author}{\bibfnamefont{L.}~\bibnamefont{Khaykovich}},
  \bibinfo{author}{\bibfnamefont{J.}~\bibnamefont{Cubizolles}},
  \bibinfo{author}{\bibfnamefont{J.}~\bibnamefont{Zhang}},
  \bibinfo{author}{\bibfnamefont{F.}~\bibnamefont{Chevy}},
  \bibinfo{author}{\bibfnamefont{M.}~\bibnamefont{Teichmann}},
  \bibinfo{author}{\bibfnamefont{L.}~\bibnamefont{Tarruell}},
  \bibinfo{author}{\bibfnamefont{S.~J. J. M.~F.} \bibnamefont{Kokkelmans}},
  \bibnamefont{and} \bibinfo{author}{\bibfnamefont{C.}~\bibnamefont{Salomon}},
  \bibinfo{journal}{Phys. Rev. Lett.} \textbf{\bibinfo{volume}{93}},
  \bibinfo{pages}{050401} (\bibinfo{year}{2004}).

\bibitem[{\citenamefont{J{\"o}rdens et~al.}(2008)\citenamefont{J{\"o}rdens,
  Strohmaier, G{\"u}nter, Moritz, and Esslinger}}]{jordens2008}
\bibinfo{author}{\bibfnamefont{R.}~\bibnamefont{J{\"o}rdens}},
  \bibinfo{author}{\bibfnamefont{N.}~\bibnamefont{Strohmaier}},
  \bibinfo{author}{\bibfnamefont{K.}~\bibnamefont{G{\"u}nter}},
  \bibinfo{author}{\bibfnamefont{H.}~\bibnamefont{Moritz}}, \bibnamefont{and}
  \bibinfo{author}{\bibfnamefont{T.}~\bibnamefont{Esslinger}},
  \bibinfo{journal}{Nature} \textbf{\bibinfo{volume}{455}},
  \bibinfo{pages}{204} (\bibinfo{year}{2008}).

\bibitem[{\citenamefont{Schneider et~al.}(2008)\citenamefont{Schneider,
  Hackermüller, Will, Best, Bloch, Costi, Helmes, Rasch, and
  Rosch}}]{Schneider2008}
\bibinfo{author}{\bibfnamefont{U.}~\bibnamefont{Schneider}},
  \bibinfo{author}{\bibfnamefont{L.}~\bibnamefont{Hackermüller}},
  \bibinfo{author}{\bibfnamefont{S.}~\bibnamefont{Will}},
  \bibinfo{author}{\bibfnamefont{T.}~\bibnamefont{Best}},
  \bibinfo{author}{\bibfnamefont{I.}~\bibnamefont{Bloch}},
  \bibinfo{author}{\bibfnamefont{T.~A.} \bibnamefont{Costi}},
  \bibinfo{author}{\bibfnamefont{R.~W.} \bibnamefont{Helmes}},
  \bibinfo{author}{\bibfnamefont{D.}~\bibnamefont{Rasch}}, \bibnamefont{and}
  \bibinfo{author}{\bibfnamefont{A.}~\bibnamefont{Rosch}},
  \bibinfo{journal}{Science} \textbf{\bibinfo{volume}{322}},
  \bibinfo{pages}{1520} (\bibinfo{year}{2008}).

\bibitem[{\citenamefont{Bakr et~al.}(2009)\citenamefont{Bakr, Gillen, Peng,
  F{\"o}lling, and Greiner}}]{bakr2009}
\bibinfo{author}{\bibfnamefont{W.~S.} \bibnamefont{Bakr}},
  \bibinfo{author}{\bibfnamefont{J.~I.} \bibnamefont{Gillen}},
  \bibinfo{author}{\bibfnamefont{A.}~\bibnamefont{Peng}},
  \bibinfo{author}{\bibfnamefont{S.}~\bibnamefont{F{\"o}lling}},
  \bibnamefont{and} \bibinfo{author}{\bibfnamefont{M.}~\bibnamefont{Greiner}},
  \bibinfo{journal}{Nature} \textbf{\bibinfo{volume}{462}}, \bibinfo{pages}{74}
  (\bibinfo{year}{2009}).

\bibitem[{\citenamefont{Sherson et~al.}(2010)\citenamefont{Sherson, Weitenberg,
  Endres, Cheneau, Bloch, and Kuhr}}]{sherson2010}
\bibinfo{author}{\bibfnamefont{J.~F.} \bibnamefont{Sherson}},
  \bibinfo{author}{\bibfnamefont{C.}~\bibnamefont{Weitenberg}},
  \bibinfo{author}{\bibfnamefont{M.}~\bibnamefont{Endres}},
  \bibinfo{author}{\bibfnamefont{M.}~\bibnamefont{Cheneau}},
  \bibinfo{author}{\bibfnamefont{I.}~\bibnamefont{Bloch}}, \bibnamefont{and}
  \bibinfo{author}{\bibfnamefont{S.}~\bibnamefont{Kuhr}},
  \bibinfo{journal}{Nature} \textbf{\bibinfo{volume}{467}}, \bibinfo{pages}{68}
  (\bibinfo{year}{2010}).

\bibitem[{\citenamefont{Struck et~al.}(2011)\citenamefont{Struck, Ölschläger,
  Targat, Soltan-Panahi, Eckardt, Lewenstein, Windpassinger, and
  Sengstock}}]{Struck2011}
\bibinfo{author}{\bibfnamefont{J.}~\bibnamefont{Struck}},
  \bibinfo{author}{\bibfnamefont{C.}~\bibnamefont{Ölschläger}},
  \bibinfo{author}{\bibfnamefont{R.~L.} \bibnamefont{Targat}},
  \bibinfo{author}{\bibfnamefont{P.}~\bibnamefont{Soltan-Panahi}},
  \bibinfo{author}{\bibfnamefont{A.}~\bibnamefont{Eckardt}},
  \bibinfo{author}{\bibfnamefont{M.}~\bibnamefont{Lewenstein}},
  \bibinfo{author}{\bibfnamefont{P.}~\bibnamefont{Windpassinger}},
  \bibnamefont{and}
  \bibinfo{author}{\bibfnamefont{K.}~\bibnamefont{Sengstock}},
  \bibinfo{journal}{Science} \textbf{\bibinfo{volume}{333}},
  \bibinfo{pages}{996} (\bibinfo{year}{2011}).

\bibitem[{\citenamefont{Aidelsburger et~al.}(2013)\citenamefont{Aidelsburger,
  Atala, Lohse, Barreiro, Paredes, and Bloch}}]{Aidelsburger2013}
\bibinfo{author}{\bibfnamefont{M.}~\bibnamefont{Aidelsburger}},
  \bibinfo{author}{\bibfnamefont{M.}~\bibnamefont{Atala}},
  \bibinfo{author}{\bibfnamefont{M.}~\bibnamefont{Lohse}},
  \bibinfo{author}{\bibfnamefont{J.~T.} \bibnamefont{Barreiro}},
  \bibinfo{author}{\bibfnamefont{B.}~\bibnamefont{Paredes}}, \bibnamefont{and}
  \bibinfo{author}{\bibfnamefont{I.}~\bibnamefont{Bloch}},
  \bibinfo{journal}{Phys. Rev. Lett.} \textbf{\bibinfo{volume}{111}},
  \bibinfo{pages}{185301} (\bibinfo{year}{2013}).

\bibitem[{\citenamefont{Miyake et~al.}(2013)\citenamefont{Miyake, Siviloglou,
  Kennedy, Burton, and Ketterle}}]{Miyake2013}
\bibinfo{author}{\bibfnamefont{H.}~\bibnamefont{Miyake}},
  \bibinfo{author}{\bibfnamefont{G.~A.} \bibnamefont{Siviloglou}},
  \bibinfo{author}{\bibfnamefont{C.~J.} \bibnamefont{Kennedy}},
  \bibinfo{author}{\bibfnamefont{W.~C.} \bibnamefont{Burton}},
  \bibnamefont{and} \bibinfo{author}{\bibfnamefont{W.}~\bibnamefont{Ketterle}},
  \bibinfo{journal}{Phys. Rev. Lett.} \textbf{\bibinfo{volume}{111}},
  \bibinfo{pages}{185302} (\bibinfo{year}{2013}).

\bibitem[{\citenamefont{Dumke et~al.}(2002)\citenamefont{Dumke, Volk, M\"uther,
  Buchkremer, Birkl, and Ertmer}}]{Dumke2002}
\bibinfo{author}{\bibfnamefont{R.}~\bibnamefont{Dumke}},
  \bibinfo{author}{\bibfnamefont{M.}~\bibnamefont{Volk}},
  \bibinfo{author}{\bibfnamefont{T.}~\bibnamefont{M\"uther}},
  \bibinfo{author}{\bibfnamefont{F.~B.~J.} \bibnamefont{Buchkremer}},
  \bibinfo{author}{\bibfnamefont{G.}~\bibnamefont{Birkl}}, \bibnamefont{and}
  \bibinfo{author}{\bibfnamefont{W.}~\bibnamefont{Ertmer}},
  \bibinfo{journal}{Phys. Rev. Lett.} \textbf{\bibinfo{volume}{89}},
  \bibinfo{pages}{097903} (\bibinfo{year}{2002}).

\bibitem[{\citenamefont{Nogrette et~al.}(2014)\citenamefont{Nogrette, Labuhn,
  Ravets, Barredo, B\'eguin, Vernier, Lahaye, and Browaeys}}]{Nogrette2014}
\bibinfo{author}{\bibfnamefont{F.}~\bibnamefont{Nogrette}},
  \bibinfo{author}{\bibfnamefont{H.}~\bibnamefont{Labuhn}},
  \bibinfo{author}{\bibfnamefont{S.}~\bibnamefont{Ravets}},
  \bibinfo{author}{\bibfnamefont{D.}~\bibnamefont{Barredo}},
  \bibinfo{author}{\bibfnamefont{L.}~\bibnamefont{B\'eguin}},
  \bibinfo{author}{\bibfnamefont{A.}~\bibnamefont{Vernier}},
  \bibinfo{author}{\bibfnamefont{T.}~\bibnamefont{Lahaye}}, \bibnamefont{and}
  \bibinfo{author}{\bibfnamefont{A.}~\bibnamefont{Browaeys}},
  \bibinfo{journal}{Phys. Rev. X} \textbf{\bibinfo{volume}{4}},
  \bibinfo{pages}{021034} (\bibinfo{year}{2014}).

\bibitem[{\citenamefont{Periwal et~al.}(2021)\citenamefont{Periwal, Cooper,
  Kunkel, Wienand, Davis, and Schleier-Smith}}]{periwal2021}
\bibinfo{author}{\bibfnamefont{A.}~\bibnamefont{Periwal}},
  \bibinfo{author}{\bibfnamefont{E.~S.} \bibnamefont{Cooper}},
  \bibinfo{author}{\bibfnamefont{P.}~\bibnamefont{Kunkel}},
  \bibinfo{author}{\bibfnamefont{J.~F.} \bibnamefont{Wienand}},
  \bibinfo{author}{\bibfnamefont{E.~J.} \bibnamefont{Davis}}, \bibnamefont{and}
  \bibinfo{author}{\bibfnamefont{M.}~\bibnamefont{Schleier-Smith}},
  \bibinfo{journal}{Nature} \textbf{\bibinfo{volume}{600}},
  \bibinfo{pages}{630} (\bibinfo{year}{2021}).

\bibitem[{\citenamefont{Endres et~al.}(2016)\citenamefont{Endres, Bernien,
  Keesling, Levine, Anschuetz, Krajenbrink, Senko, Vuletic, Greiner, and
  Lukin}}]{Endres2016}
\bibinfo{author}{\bibfnamefont{M.}~\bibnamefont{Endres}},
  \bibinfo{author}{\bibfnamefont{H.}~\bibnamefont{Bernien}},
  \bibinfo{author}{\bibfnamefont{A.}~\bibnamefont{Keesling}},
  \bibinfo{author}{\bibfnamefont{H.}~\bibnamefont{Levine}},
  \bibinfo{author}{\bibfnamefont{E.~R.} \bibnamefont{Anschuetz}},
  \bibinfo{author}{\bibfnamefont{A.}~\bibnamefont{Krajenbrink}},
  \bibinfo{author}{\bibfnamefont{C.}~\bibnamefont{Senko}},
  \bibinfo{author}{\bibfnamefont{V.}~\bibnamefont{Vuletic}},
  \bibinfo{author}{\bibfnamefont{M.}~\bibnamefont{Greiner}}, \bibnamefont{and}
  \bibinfo{author}{\bibfnamefont{M.~D.} \bibnamefont{Lukin}},
  \bibinfo{journal}{Science} \textbf{\bibinfo{volume}{354}},
  \bibinfo{pages}{1024} (\bibinfo{year}{2016}).

\bibitem[{\citenamefont{Barredo et~al.}(2016)\citenamefont{Barredo,
  de~Léséleuc, Lienhard, Lahaye, and Browaeys}}]{barredo2016}
\bibinfo{author}{\bibfnamefont{D.}~\bibnamefont{Barredo}},
  \bibinfo{author}{\bibfnamefont{S.}~\bibnamefont{de~Léséleuc}},
  \bibinfo{author}{\bibfnamefont{V.}~\bibnamefont{Lienhard}},
  \bibinfo{author}{\bibfnamefont{T.}~\bibnamefont{Lahaye}}, \bibnamefont{and}
  \bibinfo{author}{\bibfnamefont{A.}~\bibnamefont{Browaeys}},
  \bibinfo{journal}{Science} \textbf{\bibinfo{volume}{354}},
  \bibinfo{pages}{1021} (\bibinfo{year}{2016}).

\bibitem[{\citenamefont{Barredo et~al.}(2018)\citenamefont{Barredo, Lienhard,
  De~Leseleuc, Lahaye, and Browaeys}}]{barredo2018synthetic}
\bibinfo{author}{\bibfnamefont{D.}~\bibnamefont{Barredo}},
  \bibinfo{author}{\bibfnamefont{V.}~\bibnamefont{Lienhard}},
  \bibinfo{author}{\bibfnamefont{S.}~\bibnamefont{De~Leseleuc}},
  \bibinfo{author}{\bibfnamefont{T.}~\bibnamefont{Lahaye}}, \bibnamefont{and}
  \bibinfo{author}{\bibfnamefont{A.}~\bibnamefont{Browaeys}},
  \bibinfo{journal}{Nature} \textbf{\bibinfo{volume}{561}}, \bibinfo{pages}{79}
  (\bibinfo{year}{2018}).

\bibitem[{\citenamefont{Saffman et~al.}(2010)\citenamefont{Saffman, Walker, and
  M\o{}lmer}}]{Saffman2010}
\bibinfo{author}{\bibfnamefont{M.}~\bibnamefont{Saffman}},
  \bibinfo{author}{\bibfnamefont{T.~G.} \bibnamefont{Walker}},
  \bibnamefont{and}
  \bibinfo{author}{\bibfnamefont{K.}~\bibnamefont{M\o{}lmer}},
  \bibinfo{journal}{Rev. Mod. Phys.} \textbf{\bibinfo{volume}{82}},
  \bibinfo{pages}{2313} (\bibinfo{year}{2010}).

\bibitem[{\citenamefont{Jaksch et~al.}(2000)\citenamefont{Jaksch, Cirac,
  Zoller, Rolston, C\^ot\'e, and Lukin}}]{Jaksch2001}
\bibinfo{author}{\bibfnamefont{D.}~\bibnamefont{Jaksch}},
  \bibinfo{author}{\bibfnamefont{J.~I.} \bibnamefont{Cirac}},
  \bibinfo{author}{\bibfnamefont{P.}~\bibnamefont{Zoller}},
  \bibinfo{author}{\bibfnamefont{S.~L.} \bibnamefont{Rolston}},
  \bibinfo{author}{\bibfnamefont{R.}~\bibnamefont{C\^ot\'e}}, \bibnamefont{and}
  \bibinfo{author}{\bibfnamefont{M.~D.} \bibnamefont{Lukin}},
  \bibinfo{journal}{Phys. Rev. Lett.} \textbf{\bibinfo{volume}{85}},
  \bibinfo{pages}{2208} (\bibinfo{year}{2000}).

\bibitem[{\citenamefont{Urban et~al.}(2009)\citenamefont{Urban, Johnson,
  Henage, Isenhower, Yavuz, Walker, and Saffman}}]{urban2009observation}
\bibinfo{author}{\bibfnamefont{E.}~\bibnamefont{Urban}},
  \bibinfo{author}{\bibfnamefont{T.~A.} \bibnamefont{Johnson}},
  \bibinfo{author}{\bibfnamefont{T.}~\bibnamefont{Henage}},
  \bibinfo{author}{\bibfnamefont{L.}~\bibnamefont{Isenhower}},
  \bibinfo{author}{\bibfnamefont{D.}~\bibnamefont{Yavuz}},
  \bibinfo{author}{\bibfnamefont{T.}~\bibnamefont{Walker}}, \bibnamefont{and}
  \bibinfo{author}{\bibfnamefont{M.}~\bibnamefont{Saffman}},
  \bibinfo{journal}{Nature Physics} \textbf{\bibinfo{volume}{5}},
  \bibinfo{pages}{110} (\bibinfo{year}{2009}).

\bibitem[{\citenamefont{Ga{\"e}tan et~al.}(2009)\citenamefont{Ga{\"e}tan,
  Miroshnychenko, Wilk, Chotia, Viteau, Comparat, Pillet, Browaeys, and
  Grangier}}]{gaetan2009observation}
\bibinfo{author}{\bibfnamefont{A.}~\bibnamefont{Ga{\"e}tan}},
  \bibinfo{author}{\bibfnamefont{Y.}~\bibnamefont{Miroshnychenko}},
  \bibinfo{author}{\bibfnamefont{T.}~\bibnamefont{Wilk}},
  \bibinfo{author}{\bibfnamefont{A.}~\bibnamefont{Chotia}},
  \bibinfo{author}{\bibfnamefont{M.}~\bibnamefont{Viteau}},
  \bibinfo{author}{\bibfnamefont{D.}~\bibnamefont{Comparat}},
  \bibinfo{author}{\bibfnamefont{P.}~\bibnamefont{Pillet}},
  \bibinfo{author}{\bibfnamefont{A.}~\bibnamefont{Browaeys}}, \bibnamefont{and}
  \bibinfo{author}{\bibfnamefont{P.}~\bibnamefont{Grangier}},
  \bibinfo{journal}{Nature Physics} \textbf{\bibinfo{volume}{5}},
  \bibinfo{pages}{115} (\bibinfo{year}{2009}).

\bibitem[{\citenamefont{Walker and Saffman}(2005)}]{walker2005zeros}
\bibinfo{author}{\bibfnamefont{T.~G.} \bibnamefont{Walker}} \bibnamefont{and}
  \bibinfo{author}{\bibfnamefont{M.}~\bibnamefont{Saffman}},
  \bibinfo{journal}{Journal of Physics B: Atomic, Molecular and Optical
  Physics} \textbf{\bibinfo{volume}{38}}, \bibinfo{pages}{S309}
  (\bibinfo{year}{2005}).

\bibitem[{\citenamefont{Barredo et~al.}(2015)\citenamefont{Barredo, Labuhn,
  Ravets, Lahaye, Browaeys, and Adams}}]{Barredo2015}
\bibinfo{author}{\bibfnamefont{D.}~\bibnamefont{Barredo}},
  \bibinfo{author}{\bibfnamefont{H.}~\bibnamefont{Labuhn}},
  \bibinfo{author}{\bibfnamefont{S.}~\bibnamefont{Ravets}},
  \bibinfo{author}{\bibfnamefont{T.}~\bibnamefont{Lahaye}},
  \bibinfo{author}{\bibfnamefont{A.}~\bibnamefont{Browaeys}}, \bibnamefont{and}
  \bibinfo{author}{\bibfnamefont{C.~S.} \bibnamefont{Adams}},
  \bibinfo{journal}{Phys. Rev. Lett.} \textbf{\bibinfo{volume}{114}},
  \bibinfo{pages}{113002} (\bibinfo{year}{2015}).

\bibitem[{\citenamefont{Ravets et~al.}(2014)\citenamefont{Ravets, Labuhn,
  Barredo, B{\'e}guin, Lahaye, and Browaeys}}]{ravets2014coherent}
\bibinfo{author}{\bibfnamefont{S.}~\bibnamefont{Ravets}},
  \bibinfo{author}{\bibfnamefont{H.}~\bibnamefont{Labuhn}},
  \bibinfo{author}{\bibfnamefont{D.}~\bibnamefont{Barredo}},
  \bibinfo{author}{\bibfnamefont{L.}~\bibnamefont{B{\'e}guin}},
  \bibinfo{author}{\bibfnamefont{T.}~\bibnamefont{Lahaye}}, \bibnamefont{and}
  \bibinfo{author}{\bibfnamefont{A.}~\bibnamefont{Browaeys}},
  \bibinfo{journal}{Nature Physics} \textbf{\bibinfo{volume}{10}},
  \bibinfo{pages}{914} (\bibinfo{year}{2014}).

\bibitem[{\citenamefont{Bluvstein et~al.}(2022)\citenamefont{Bluvstein, Levine,
  Semeghini, Wang, Ebadi, Kalinowski, Keesling, Maskara, Pichler, Greiner
  et~al.}}]{bluvstein2022quantum}
\bibinfo{author}{\bibfnamefont{D.}~\bibnamefont{Bluvstein}},
  \bibinfo{author}{\bibfnamefont{H.}~\bibnamefont{Levine}},
  \bibinfo{author}{\bibfnamefont{G.}~\bibnamefont{Semeghini}},
  \bibinfo{author}{\bibfnamefont{T.~T.} \bibnamefont{Wang}},
  \bibinfo{author}{\bibfnamefont{S.}~\bibnamefont{Ebadi}},
  \bibinfo{author}{\bibfnamefont{M.}~\bibnamefont{Kalinowski}},
  \bibinfo{author}{\bibfnamefont{A.}~\bibnamefont{Keesling}},
  \bibinfo{author}{\bibfnamefont{N.}~\bibnamefont{Maskara}},
  \bibinfo{author}{\bibfnamefont{H.}~\bibnamefont{Pichler}},
  \bibinfo{author}{\bibfnamefont{M.}~\bibnamefont{Greiner}},
  \bibnamefont{et~al.}, \bibinfo{journal}{Nature}
  \textbf{\bibinfo{volume}{604}}, \bibinfo{pages}{451} (\bibinfo{year}{2022}).

\bibitem[{\citenamefont{Graham et~al.}(2022)\citenamefont{Graham, Song, Scott,
  Poole, Phuttitarn, Jooya, Eichler, Jiang, Marra, Grinkemeyer
  et~al.}}]{graham2022multi}
\bibinfo{author}{\bibfnamefont{T.}~\bibnamefont{Graham}},
  \bibinfo{author}{\bibfnamefont{Y.}~\bibnamefont{Song}},
  \bibinfo{author}{\bibfnamefont{J.}~\bibnamefont{Scott}},
  \bibinfo{author}{\bibfnamefont{C.}~\bibnamefont{Poole}},
  \bibinfo{author}{\bibfnamefont{L.}~\bibnamefont{Phuttitarn}},
  \bibinfo{author}{\bibfnamefont{K.}~\bibnamefont{Jooya}},
  \bibinfo{author}{\bibfnamefont{P.}~\bibnamefont{Eichler}},
  \bibinfo{author}{\bibfnamefont{X.}~\bibnamefont{Jiang}},
  \bibinfo{author}{\bibfnamefont{A.}~\bibnamefont{Marra}},
  \bibinfo{author}{\bibfnamefont{B.}~\bibnamefont{Grinkemeyer}},
  \bibnamefont{et~al.}, \bibinfo{journal}{Nature}
  \textbf{\bibinfo{volume}{604}}, \bibinfo{pages}{457} (\bibinfo{year}{2022}).

\bibitem[{\citenamefont{Bernien et~al.}(2017)\citenamefont{Bernien, Schwartz,
  Keesling, Levine, Omran, Pichler, Choi, Zibrov, Endres, Greiner
  et~al.}}]{bernien2017probing}
\bibinfo{author}{\bibfnamefont{H.}~\bibnamefont{Bernien}},
  \bibinfo{author}{\bibfnamefont{S.}~\bibnamefont{Schwartz}},
  \bibinfo{author}{\bibfnamefont{A.}~\bibnamefont{Keesling}},
  \bibinfo{author}{\bibfnamefont{H.}~\bibnamefont{Levine}},
  \bibinfo{author}{\bibfnamefont{A.}~\bibnamefont{Omran}},
  \bibinfo{author}{\bibfnamefont{H.}~\bibnamefont{Pichler}},
  \bibinfo{author}{\bibfnamefont{S.}~\bibnamefont{Choi}},
  \bibinfo{author}{\bibfnamefont{A.~S.} \bibnamefont{Zibrov}},
  \bibinfo{author}{\bibfnamefont{M.}~\bibnamefont{Endres}},
  \bibinfo{author}{\bibfnamefont{M.}~\bibnamefont{Greiner}},
  \bibnamefont{et~al.}, \bibinfo{journal}{Nature}
  \textbf{\bibinfo{volume}{551}}, \bibinfo{pages}{579} (\bibinfo{year}{2017}).

\bibitem[{\citenamefont{Ebadi et~al.}(2021)\citenamefont{Ebadi, Wang, Levine,
  Keesling, Semeghini, Omran, Bluvstein, Samajdar, Pichler, Ho
  et~al.}}]{ebadi2021quantum}
\bibinfo{author}{\bibfnamefont{S.}~\bibnamefont{Ebadi}},
  \bibinfo{author}{\bibfnamefont{T.~T.} \bibnamefont{Wang}},
  \bibinfo{author}{\bibfnamefont{H.}~\bibnamefont{Levine}},
  \bibinfo{author}{\bibfnamefont{A.}~\bibnamefont{Keesling}},
  \bibinfo{author}{\bibfnamefont{G.}~\bibnamefont{Semeghini}},
  \bibinfo{author}{\bibfnamefont{A.}~\bibnamefont{Omran}},
  \bibinfo{author}{\bibfnamefont{D.}~\bibnamefont{Bluvstein}},
  \bibinfo{author}{\bibfnamefont{R.}~\bibnamefont{Samajdar}},
  \bibinfo{author}{\bibfnamefont{H.}~\bibnamefont{Pichler}},
  \bibinfo{author}{\bibfnamefont{W.~W.} \bibnamefont{Ho}},
  \bibnamefont{et~al.}, \bibinfo{journal}{Nature}
  \textbf{\bibinfo{volume}{595}}, \bibinfo{pages}{227} (\bibinfo{year}{2021}).

\bibitem[{\citenamefont{Scholl et~al.}(2021)\citenamefont{Scholl, Schuler,
  Williams, Eberharter, Barredo, Schymik, Lienhard, Henry, Lang, Lahaye
  et~al.}}]{scholl2021quantum}
\bibinfo{author}{\bibfnamefont{P.}~\bibnamefont{Scholl}},
  \bibinfo{author}{\bibfnamefont{M.}~\bibnamefont{Schuler}},
  \bibinfo{author}{\bibfnamefont{H.~J.} \bibnamefont{Williams}},
  \bibinfo{author}{\bibfnamefont{A.~A.} \bibnamefont{Eberharter}},
  \bibinfo{author}{\bibfnamefont{D.}~\bibnamefont{Barredo}},
  \bibinfo{author}{\bibfnamefont{K.-N.} \bibnamefont{Schymik}},
  \bibinfo{author}{\bibfnamefont{V.}~\bibnamefont{Lienhard}},
  \bibinfo{author}{\bibfnamefont{L.-P.} \bibnamefont{Henry}},
  \bibinfo{author}{\bibfnamefont{T.~C.} \bibnamefont{Lang}},
  \bibinfo{author}{\bibfnamefont{T.}~\bibnamefont{Lahaye}},
  \bibnamefont{et~al.}, \bibinfo{journal}{Nature}
  \textbf{\bibinfo{volume}{595}}, \bibinfo{pages}{233} (\bibinfo{year}{2021}).

\bibitem[{\citenamefont{Chen et~al.}(2023)\citenamefont{Chen, Bornet, Bintz,
  Emperauger, Leclerc, Liu, Scholl, Barredo, Hauschild, Chatterjee
  et~al.}}]{chen2023continuous}
\bibinfo{author}{\bibfnamefont{C.}~\bibnamefont{Chen}},
  \bibinfo{author}{\bibfnamefont{G.}~\bibnamefont{Bornet}},
  \bibinfo{author}{\bibfnamefont{M.}~\bibnamefont{Bintz}},
  \bibinfo{author}{\bibfnamefont{G.}~\bibnamefont{Emperauger}},
  \bibinfo{author}{\bibfnamefont{L.}~\bibnamefont{Leclerc}},
  \bibinfo{author}{\bibfnamefont{V.~S.} \bibnamefont{Liu}},
  \bibinfo{author}{\bibfnamefont{P.}~\bibnamefont{Scholl}},
  \bibinfo{author}{\bibfnamefont{D.}~\bibnamefont{Barredo}},
  \bibinfo{author}{\bibfnamefont{J.}~\bibnamefont{Hauschild}},
  \bibinfo{author}{\bibfnamefont{S.}~\bibnamefont{Chatterjee}},
  \bibnamefont{et~al.}, \bibinfo{journal}{Nature}
  \textbf{\bibinfo{volume}{616}}, \bibinfo{pages}{691} (\bibinfo{year}{2023}).

\bibitem[{\citenamefont{de~Léséleuc et~al.}(2019)\citenamefont{de~Léséleuc,
  Lienhard, Scholl, Barredo, Weber, Lang, Büchler, Lahaye, and
  Browaeys}}]{Deleseluc2019}
\bibinfo{author}{\bibfnamefont{S.}~\bibnamefont{de~Léséleuc}},
  \bibinfo{author}{\bibfnamefont{V.}~\bibnamefont{Lienhard}},
  \bibinfo{author}{\bibfnamefont{P.}~\bibnamefont{Scholl}},
  \bibinfo{author}{\bibfnamefont{D.}~\bibnamefont{Barredo}},
  \bibinfo{author}{\bibfnamefont{S.}~\bibnamefont{Weber}},
  \bibinfo{author}{\bibfnamefont{N.}~\bibnamefont{Lang}},
  \bibinfo{author}{\bibfnamefont{H.~P.} \bibnamefont{Büchler}},
  \bibinfo{author}{\bibfnamefont{T.}~\bibnamefont{Lahaye}}, \bibnamefont{and}
  \bibinfo{author}{\bibfnamefont{A.}~\bibnamefont{Browaeys}},
  \bibinfo{journal}{Science} \textbf{\bibinfo{volume}{365}},
  \bibinfo{pages}{775} (\bibinfo{year}{2019}).

\bibitem[{\citenamefont{Semeghini et~al.}(2021)\citenamefont{Semeghini, Levine,
  Keesling, Ebadi, Wang, Bluvstein, Verresen, Pichler, Kalinowski, Samajdar
  et~al.}}]{semeghini2021}
\bibinfo{author}{\bibfnamefont{G.}~\bibnamefont{Semeghini}},
  \bibinfo{author}{\bibfnamefont{H.}~\bibnamefont{Levine}},
  \bibinfo{author}{\bibfnamefont{A.}~\bibnamefont{Keesling}},
  \bibinfo{author}{\bibfnamefont{S.}~\bibnamefont{Ebadi}},
  \bibinfo{author}{\bibfnamefont{T.~T.} \bibnamefont{Wang}},
  \bibinfo{author}{\bibfnamefont{D.}~\bibnamefont{Bluvstein}},
  \bibinfo{author}{\bibfnamefont{R.}~\bibnamefont{Verresen}},
  \bibinfo{author}{\bibfnamefont{H.}~\bibnamefont{Pichler}},
  \bibinfo{author}{\bibfnamefont{M.}~\bibnamefont{Kalinowski}},
  \bibinfo{author}{\bibfnamefont{R.}~\bibnamefont{Samajdar}},
  \bibnamefont{et~al.}, \bibinfo{journal}{Science}
  \textbf{\bibinfo{volume}{374}}, \bibinfo{pages}{1242} (\bibinfo{year}{2021}).

\bibitem[{\citenamefont{Weimer et~al.}(2010)\citenamefont{Weimer, M{\"u}ller,
  Lesanovsky, Zoller, and B{\"u}chler}}]{weimer2010rydberg}
\bibinfo{author}{\bibfnamefont{H.}~\bibnamefont{Weimer}},
  \bibinfo{author}{\bibfnamefont{M.}~\bibnamefont{M{\"u}ller}},
  \bibinfo{author}{\bibfnamefont{I.}~\bibnamefont{Lesanovsky}},
  \bibinfo{author}{\bibfnamefont{P.}~\bibnamefont{Zoller}}, \bibnamefont{and}
  \bibinfo{author}{\bibfnamefont{H.~P.} \bibnamefont{B{\"u}chler}},
  \bibinfo{journal}{Nature Physics} \textbf{\bibinfo{volume}{6}},
  \bibinfo{pages}{382} (\bibinfo{year}{2010}).

\bibitem[{\citenamefont{Sheng et~al.}(2022)\citenamefont{Sheng, Hou, He, Wang,
  Guo, Zhuang, Mamat, Xu, Liu, Wang et~al.}}]{Sheng2022}
\bibinfo{author}{\bibfnamefont{C.}~\bibnamefont{Sheng}},
  \bibinfo{author}{\bibfnamefont{J.}~\bibnamefont{Hou}},
  \bibinfo{author}{\bibfnamefont{X.}~\bibnamefont{He}},
  \bibinfo{author}{\bibfnamefont{K.}~\bibnamefont{Wang}},
  \bibinfo{author}{\bibfnamefont{R.}~\bibnamefont{Guo}},
  \bibinfo{author}{\bibfnamefont{J.}~\bibnamefont{Zhuang}},
  \bibinfo{author}{\bibfnamefont{B.}~\bibnamefont{Mamat}},
  \bibinfo{author}{\bibfnamefont{P.}~\bibnamefont{Xu}},
  \bibinfo{author}{\bibfnamefont{M.}~\bibnamefont{Liu}},
  \bibinfo{author}{\bibfnamefont{J.}~\bibnamefont{Wang}}, \bibnamefont{et~al.},
  \bibinfo{journal}{Phys. Rev. Lett.} \textbf{\bibinfo{volume}{128}},
  \bibinfo{pages}{083202} (\bibinfo{year}{2022}).

\bibitem[{\citenamefont{Singh et~al.}(2022{\natexlab{a}})\citenamefont{Singh,
  Anand, Pocklington, Kemp, and Bernien}}]{Singh2022}
\bibinfo{author}{\bibfnamefont{K.}~\bibnamefont{Singh}},
  \bibinfo{author}{\bibfnamefont{S.}~\bibnamefont{Anand}},
  \bibinfo{author}{\bibfnamefont{A.}~\bibnamefont{Pocklington}},
  \bibinfo{author}{\bibfnamefont{J.~T.} \bibnamefont{Kemp}}, \bibnamefont{and}
  \bibinfo{author}{\bibfnamefont{H.}~\bibnamefont{Bernien}},
  \bibinfo{journal}{Phys. Rev. X} \textbf{\bibinfo{volume}{12}},
  \bibinfo{pages}{011040} (\bibinfo{year}{2022}{\natexlab{a}}).

\bibitem[{\citenamefont{Singh et~al.}(2022{\natexlab{b}})\citenamefont{Singh,
  Bradley, Anand, Ramesh, White, and Bernien}}]{singh2022mid}
\bibinfo{author}{\bibfnamefont{K.}~\bibnamefont{Singh}},
  \bibinfo{author}{\bibfnamefont{C.~E.} \bibnamefont{Bradley}},
  \bibinfo{author}{\bibfnamefont{S.}~\bibnamefont{Anand}},
  \bibinfo{author}{\bibfnamefont{V.}~\bibnamefont{Ramesh}},
  \bibinfo{author}{\bibfnamefont{R.}~\bibnamefont{White}}, \bibnamefont{and}
  \bibinfo{author}{\bibfnamefont{H.}~\bibnamefont{Bernien}},
  \bibinfo{journal}{arXiv preprint arXiv:2208.11716}
  (\bibinfo{year}{2022}{\natexlab{b}}).

\bibitem[{\citenamefont{Altman et~al.}(2021)\citenamefont{Altman, Brown,
  Carleo, Carr, Demler, Chin, DeMarco, Economou, Eriksson, Fu
  et~al.}}]{Altman2021}
\bibinfo{author}{\bibfnamefont{E.}~\bibnamefont{Altman}},
  \bibinfo{author}{\bibfnamefont{K.~R.} \bibnamefont{Brown}},
  \bibinfo{author}{\bibfnamefont{G.}~\bibnamefont{Carleo}},
  \bibinfo{author}{\bibfnamefont{L.~D.} \bibnamefont{Carr}},
  \bibinfo{author}{\bibfnamefont{E.}~\bibnamefont{Demler}},
  \bibinfo{author}{\bibfnamefont{C.}~\bibnamefont{Chin}},
  \bibinfo{author}{\bibfnamefont{B.}~\bibnamefont{DeMarco}},
  \bibinfo{author}{\bibfnamefont{S.~E.} \bibnamefont{Economou}},
  \bibinfo{author}{\bibfnamefont{M.~A.} \bibnamefont{Eriksson}},
  \bibinfo{author}{\bibfnamefont{K.-M.~C.} \bibnamefont{Fu}},
  \bibnamefont{et~al.}, \bibinfo{journal}{PRX Quantum}
  \textbf{\bibinfo{volume}{2}}, \bibinfo{pages}{017003} (\bibinfo{year}{2021}).

\bibitem[{\citenamefont{Roessler et~al.}(2006)\citenamefont{Roessler, Bogdanov,
  and Pfleiderer}}]{roessler2006}
\bibinfo{author}{\bibfnamefont{U.~K.} \bibnamefont{Roessler}},
  \bibinfo{author}{\bibfnamefont{A.}~\bibnamefont{Bogdanov}}, \bibnamefont{and}
  \bibinfo{author}{\bibfnamefont{C.}~\bibnamefont{Pfleiderer}},
  \bibinfo{journal}{Nature} \textbf{\bibinfo{volume}{442}},
  \bibinfo{pages}{797} (\bibinfo{year}{2006}).

\bibitem[{\citenamefont{Lohani et~al.}(2019)\citenamefont{Lohani, Hickey,
  Masell, and Rosch}}]{Lohani2019}
\bibinfo{author}{\bibfnamefont{V.}~\bibnamefont{Lohani}},
  \bibinfo{author}{\bibfnamefont{C.}~\bibnamefont{Hickey}},
  \bibinfo{author}{\bibfnamefont{J.}~\bibnamefont{Masell}}, \bibnamefont{and}
  \bibinfo{author}{\bibfnamefont{A.}~\bibnamefont{Rosch}},
  \bibinfo{journal}{Phys. Rev. X} \textbf{\bibinfo{volume}{9}},
  \bibinfo{pages}{041063} (\bibinfo{year}{2019}),
  \urlprefix\url{https://link.aps.org/doi/10.1103/PhysRevX.9.041063}.

\bibitem[{\citenamefont{Kitaev}(2006)}]{kitaev2006}
\bibinfo{author}{\bibfnamefont{A.}~\bibnamefont{Kitaev}},
  \bibinfo{journal}{Annals of Physics} \textbf{\bibinfo{volume}{321}},
  \bibinfo{pages}{2} (\bibinfo{year}{2006}).

\bibitem[{\citenamefont{Goldman and Dalibard}(2014)}]{Goldman2014}
\bibinfo{author}{\bibfnamefont{N.}~\bibnamefont{Goldman}} \bibnamefont{and}
  \bibinfo{author}{\bibfnamefont{J.}~\bibnamefont{Dalibard}},
  \bibinfo{journal}{Phys. Rev. X} \textbf{\bibinfo{volume}{4}},
  \bibinfo{pages}{031027} (\bibinfo{year}{2014}),
  \urlprefix\url{https://link.aps.org/doi/10.1103/PhysRevX.4.031027}.

\bibitem[{\citenamefont{Bukov et~al.}(2015)\citenamefont{Bukov, D'Alessio, and
  Polkovnikov}}]{bukov2015}
\bibinfo{author}{\bibfnamefont{M.}~\bibnamefont{Bukov}},
  \bibinfo{author}{\bibfnamefont{L.}~\bibnamefont{D'Alessio}},
  \bibnamefont{and}
  \bibinfo{author}{\bibfnamefont{A.}~\bibnamefont{Polkovnikov}},
  \bibinfo{journal}{Advances in Physics} \textbf{\bibinfo{volume}{64}},
  \bibinfo{pages}{139} (\bibinfo{year}{2015}).

\bibitem[{\citenamefont{Tsesses et~al.}(2022)\citenamefont{Tsesses, Keselman,
  Browaeys, and Lahaye}}]{tsesses2022}
\bibinfo{author}{\bibfnamefont{S.}~\bibnamefont{Tsesses}},
  \bibinfo{author}{\bibfnamefont{A.}~\bibnamefont{Keselman}},
  \bibinfo{author}{\bibfnamefont{A.}~\bibnamefont{Browaeys}}, \bibnamefont{and}
  \bibinfo{author}{\bibfnamefont{T.}~\bibnamefont{Lahaye}}, in
  \emph{\bibinfo{booktitle}{Quantum 2.0}} (\bibinfo{organization}{Optica
  Publishing Group}, \bibinfo{year}{2022}), pp. \bibinfo{pages}{QW3A--3}.

\bibitem[{\citenamefont{Scholl et~al.}(2022)\citenamefont{Scholl, Williams,
  Bornet, Wallner, Barredo, Henriet, Signoles, Hainaut, Franz, Geier
  et~al.}}]{Scholl2022}
\bibinfo{author}{\bibfnamefont{P.}~\bibnamefont{Scholl}},
  \bibinfo{author}{\bibfnamefont{H.~J.} \bibnamefont{Williams}},
  \bibinfo{author}{\bibfnamefont{G.}~\bibnamefont{Bornet}},
  \bibinfo{author}{\bibfnamefont{F.}~\bibnamefont{Wallner}},
  \bibinfo{author}{\bibfnamefont{D.}~\bibnamefont{Barredo}},
  \bibinfo{author}{\bibfnamefont{L.}~\bibnamefont{Henriet}},
  \bibinfo{author}{\bibfnamefont{A.}~\bibnamefont{Signoles}},
  \bibinfo{author}{\bibfnamefont{C.}~\bibnamefont{Hainaut}},
  \bibinfo{author}{\bibfnamefont{T.}~\bibnamefont{Franz}},
  \bibinfo{author}{\bibfnamefont{S.}~\bibnamefont{Geier}},
  \bibnamefont{et~al.}, \bibinfo{journal}{PRX Quantum}
  \textbf{\bibinfo{volume}{3}}, \bibinfo{pages}{020303} (\bibinfo{year}{2022}),
  \urlprefix\url{https://link.aps.org/doi/10.1103/PRXQuantum.3.020303}.

\bibitem[{\citenamefont{de~L\'es\'eleuc
  et~al.}(2017)\citenamefont{de~L\'es\'eleuc, Barredo, Lienhard, Browaeys, and
  Lahaye}}]{deLeseluc2017}
\bibinfo{author}{\bibfnamefont{S.}~\bibnamefont{de~L\'es\'eleuc}},
  \bibinfo{author}{\bibfnamefont{D.}~\bibnamefont{Barredo}},
  \bibinfo{author}{\bibfnamefont{V.}~\bibnamefont{Lienhard}},
  \bibinfo{author}{\bibfnamefont{A.}~\bibnamefont{Browaeys}}, \bibnamefont{and}
  \bibinfo{author}{\bibfnamefont{T.}~\bibnamefont{Lahaye}},
  \bibinfo{journal}{Phys. Rev. Lett.} \textbf{\bibinfo{volume}{119}},
  \bibinfo{pages}{053202} (\bibinfo{year}{2017}),
  \urlprefix\url{https://link.aps.org/doi/10.1103/PhysRevLett.119.053202}.

\bibitem[{\citenamefont{Dzyaloshinsky}(1958)}]{dzyaloshinsky1958}
\bibinfo{author}{\bibfnamefont{I.}~\bibnamefont{Dzyaloshinsky}},
  \bibinfo{journal}{Journal of physics and chemistry of solids}
  \textbf{\bibinfo{volume}{4}}, \bibinfo{pages}{241} (\bibinfo{year}{1958}).

\bibitem[{\citenamefont{Moriya}(1960)}]{moriya1960}
\bibinfo{author}{\bibfnamefont{T.}~\bibnamefont{Moriya}},
  \bibinfo{journal}{Phys. Rev.} \textbf{\bibinfo{volume}{120}},
  \bibinfo{pages}{91} (\bibinfo{year}{1960}),
  \urlprefix\url{https://link.aps.org/doi/10.1103/PhysRev.120.91}.

\bibitem[{\citenamefont{Browaeys and Lahaye}(2020)}]{browaeys2020}
\bibinfo{author}{\bibfnamefont{A.}~\bibnamefont{Browaeys}} \bibnamefont{and}
  \bibinfo{author}{\bibfnamefont{T.}~\bibnamefont{Lahaye}},
  \bibinfo{journal}{Nature Physics} \textbf{\bibinfo{volume}{16}},
  \bibinfo{pages}{132} (\bibinfo{year}{2020}).

\bibitem[{\citenamefont{Jotzu et~al.}(2014)\citenamefont{Jotzu, Messer,
  Desbuquois, Lebrat, Uehlinger, Greif, and Esslinger}}]{jotzu2014}
\bibinfo{author}{\bibfnamefont{G.}~\bibnamefont{Jotzu}},
  \bibinfo{author}{\bibfnamefont{M.}~\bibnamefont{Messer}},
  \bibinfo{author}{\bibfnamefont{R.}~\bibnamefont{Desbuquois}},
  \bibinfo{author}{\bibfnamefont{M.}~\bibnamefont{Lebrat}},
  \bibinfo{author}{\bibfnamefont{T.}~\bibnamefont{Uehlinger}},
  \bibinfo{author}{\bibfnamefont{D.}~\bibnamefont{Greif}}, \bibnamefont{and}
  \bibinfo{author}{\bibfnamefont{T.}~\bibnamefont{Esslinger}},
  \bibinfo{journal}{Nature} \textbf{\bibinfo{volume}{515}},
  \bibinfo{pages}{237} (\bibinfo{year}{2014}).

\bibitem[{\citenamefont{Geier et~al.}(2021)\citenamefont{Geier, Thaicharoen,
  Hainaut, Franz, Salzinger, Tebben, Grimshandl, Z{\"u}rn, and
  Weidem{\"u}ller}}]{geier2021}
\bibinfo{author}{\bibfnamefont{S.}~\bibnamefont{Geier}},
  \bibinfo{author}{\bibfnamefont{N.}~\bibnamefont{Thaicharoen}},
  \bibinfo{author}{\bibfnamefont{C.}~\bibnamefont{Hainaut}},
  \bibinfo{author}{\bibfnamefont{T.}~\bibnamefont{Franz}},
  \bibinfo{author}{\bibfnamefont{A.}~\bibnamefont{Salzinger}},
  \bibinfo{author}{\bibfnamefont{A.}~\bibnamefont{Tebben}},
  \bibinfo{author}{\bibfnamefont{D.}~\bibnamefont{Grimshandl}},
  \bibinfo{author}{\bibfnamefont{G.}~\bibnamefont{Z{\"u}rn}}, \bibnamefont{and}
  \bibinfo{author}{\bibfnamefont{M.}~\bibnamefont{Weidem{\"u}ller}},
  \bibinfo{journal}{Science} \textbf{\bibinfo{volume}{374}},
  \bibinfo{pages}{1149} (\bibinfo{year}{2021}).

\bibitem[{\citenamefont{Haeberlen and Waugh}(1968)}]{haeberlen1968coherent}
\bibinfo{author}{\bibfnamefont{U.}~\bibnamefont{Haeberlen}} \bibnamefont{and}
  \bibinfo{author}{\bibfnamefont{J.~S.} \bibnamefont{Waugh}},
  \bibinfo{journal}{Physical Review} \textbf{\bibinfo{volume}{175}},
  \bibinfo{pages}{453} (\bibinfo{year}{1968}).

\bibitem[{\citenamefont{Vandersypen and Chuang}(2005)}]{vandersypen2005nmr}
\bibinfo{author}{\bibfnamefont{L.~M.} \bibnamefont{Vandersypen}}
  \bibnamefont{and} \bibinfo{author}{\bibfnamefont{I.~L.}
  \bibnamefont{Chuang}}, \bibinfo{journal}{Reviews of modern physics}
  \textbf{\bibinfo{volume}{76}}, \bibinfo{pages}{1037} (\bibinfo{year}{2005}).

\bibitem[{\citenamefont{De~Lange et~al.}(2010)\citenamefont{De~Lange, Wang,
  Riste, Dobrovitski, and Hanson}}]{de2010universal}
\bibinfo{author}{\bibfnamefont{G.}~\bibnamefont{De~Lange}},
  \bibinfo{author}{\bibfnamefont{Z.-H.} \bibnamefont{Wang}},
  \bibinfo{author}{\bibfnamefont{D.}~\bibnamefont{Riste}},
  \bibinfo{author}{\bibfnamefont{V.}~\bibnamefont{Dobrovitski}},
  \bibnamefont{and} \bibinfo{author}{\bibfnamefont{R.}~\bibnamefont{Hanson}},
  \bibinfo{journal}{Science} \textbf{\bibinfo{volume}{330}},
  \bibinfo{pages}{60} (\bibinfo{year}{2010}).

\bibitem[{\citenamefont{Ryan et~al.}(2010)\citenamefont{Ryan, Hodges, and
  Cory}}]{ryan2010robust}
\bibinfo{author}{\bibfnamefont{C.~A.} \bibnamefont{Ryan}},
  \bibinfo{author}{\bibfnamefont{J.~S.} \bibnamefont{Hodges}},
  \bibnamefont{and} \bibinfo{author}{\bibfnamefont{D.~G.} \bibnamefont{Cory}},
  \bibinfo{journal}{Physical Review Letters} \textbf{\bibinfo{volume}{105}},
  \bibinfo{pages}{200402} (\bibinfo{year}{2010}).

\bibitem[{\citenamefont{Roushan et~al.}(2017)\citenamefont{Roushan, Neill,
  Megrant, Chen, Babbush, Barends, Campbell, Chen, Chiaro, Dunsworth
  et~al.}}]{roushan2017chiral}
\bibinfo{author}{\bibfnamefont{P.}~\bibnamefont{Roushan}},
  \bibinfo{author}{\bibfnamefont{C.}~\bibnamefont{Neill}},
  \bibinfo{author}{\bibfnamefont{A.}~\bibnamefont{Megrant}},
  \bibinfo{author}{\bibfnamefont{Y.}~\bibnamefont{Chen}},
  \bibinfo{author}{\bibfnamefont{R.}~\bibnamefont{Babbush}},
  \bibinfo{author}{\bibfnamefont{R.}~\bibnamefont{Barends}},
  \bibinfo{author}{\bibfnamefont{B.}~\bibnamefont{Campbell}},
  \bibinfo{author}{\bibfnamefont{Z.}~\bibnamefont{Chen}},
  \bibinfo{author}{\bibfnamefont{B.}~\bibnamefont{Chiaro}},
  \bibinfo{author}{\bibfnamefont{A.}~\bibnamefont{Dunsworth}},
  \bibnamefont{et~al.}, \bibinfo{journal}{Nature Physics}
  \textbf{\bibinfo{volume}{13}}, \bibinfo{pages}{146} (\bibinfo{year}{2017}).

\bibitem[{\citenamefont{Wang et~al.}(2019)\citenamefont{Wang, Song, Feng, Cai,
  Xu, Deng, Li, Zheng, Zhu, Wang et~al.}}]{wang2019synthesis}
\bibinfo{author}{\bibfnamefont{D.-W.} \bibnamefont{Wang}},
  \bibinfo{author}{\bibfnamefont{C.}~\bibnamefont{Song}},
  \bibinfo{author}{\bibfnamefont{W.}~\bibnamefont{Feng}},
  \bibinfo{author}{\bibfnamefont{H.}~\bibnamefont{Cai}},
  \bibinfo{author}{\bibfnamefont{D.}~\bibnamefont{Xu}},
  \bibinfo{author}{\bibfnamefont{H.}~\bibnamefont{Deng}},
  \bibinfo{author}{\bibfnamefont{H.}~\bibnamefont{Li}},
  \bibinfo{author}{\bibfnamefont{D.}~\bibnamefont{Zheng}},
  \bibinfo{author}{\bibfnamefont{X.}~\bibnamefont{Zhu}},
  \bibinfo{author}{\bibfnamefont{H.}~\bibnamefont{Wang}}, \bibnamefont{et~al.},
  \bibinfo{journal}{Nature Physics} \textbf{\bibinfo{volume}{15}},
  \bibinfo{pages}{382} (\bibinfo{year}{2019}).

\bibitem[{\citenamefont{Lienhard et~al.}(2020)\citenamefont{Lienhard, Scholl,
  Weber, Barredo, de~L\'es\'eleuc, Bai, Lang, Fleischhauer, B\"uchler, Lahaye
  et~al.}}]{Lienhard2020Peierls}
\bibinfo{author}{\bibfnamefont{V.}~\bibnamefont{Lienhard}},
  \bibinfo{author}{\bibfnamefont{P.}~\bibnamefont{Scholl}},
  \bibinfo{author}{\bibfnamefont{S.}~\bibnamefont{Weber}},
  \bibinfo{author}{\bibfnamefont{D.}~\bibnamefont{Barredo}},
  \bibinfo{author}{\bibfnamefont{S.}~\bibnamefont{de~L\'es\'eleuc}},
  \bibinfo{author}{\bibfnamefont{R.}~\bibnamefont{Bai}},
  \bibinfo{author}{\bibfnamefont{N.}~\bibnamefont{Lang}},
  \bibinfo{author}{\bibfnamefont{M.}~\bibnamefont{Fleischhauer}},
  \bibinfo{author}{\bibfnamefont{H.~P.} \bibnamefont{B\"uchler}},
  \bibinfo{author}{\bibfnamefont{T.}~\bibnamefont{Lahaye}},
  \bibnamefont{et~al.}, \bibinfo{journal}{Phys. Rev. X}
  \textbf{\bibinfo{volume}{10}}, \bibinfo{pages}{021031}
  (\bibinfo{year}{2020}),
  \urlprefix\url{https://link.aps.org/doi/10.1103/PhysRevX.10.021031}.

\bibitem[{\citenamefont{Ohler et~al.}(2022)\citenamefont{Ohler,
  Kiefer-Emmanouilidis, Browaeys, B{\"u}chler, and
  Fleischhauer}}]{ohler2022self}
\bibinfo{author}{\bibfnamefont{S.}~\bibnamefont{Ohler}},
  \bibinfo{author}{\bibfnamefont{M.}~\bibnamefont{Kiefer-Emmanouilidis}},
  \bibinfo{author}{\bibfnamefont{A.}~\bibnamefont{Browaeys}},
  \bibinfo{author}{\bibfnamefont{H.~P.} \bibnamefont{B{\"u}chler}},
  \bibnamefont{and}
  \bibinfo{author}{\bibfnamefont{M.}~\bibnamefont{Fleischhauer}},
  \bibinfo{journal}{New Journal of Physics} \textbf{\bibinfo{volume}{24}},
  \bibinfo{pages}{023017} (\bibinfo{year}{2022}).

\bibitem[{\citenamefont{Weber et~al.}(2022)\citenamefont{Weber, Bai, Makki,
  M\"ogerle, Lahaye, Browaeys, Daghofer, Lang, and B\"uchler}}]{Weber2022}
\bibinfo{author}{\bibfnamefont{S.}~\bibnamefont{Weber}},
  \bibinfo{author}{\bibfnamefont{R.}~\bibnamefont{Bai}},
  \bibinfo{author}{\bibfnamefont{N.}~\bibnamefont{Makki}},
  \bibinfo{author}{\bibfnamefont{J.}~\bibnamefont{M\"ogerle}},
  \bibinfo{author}{\bibfnamefont{T.}~\bibnamefont{Lahaye}},
  \bibinfo{author}{\bibfnamefont{A.}~\bibnamefont{Browaeys}},
  \bibinfo{author}{\bibfnamefont{M.}~\bibnamefont{Daghofer}},
  \bibinfo{author}{\bibfnamefont{N.}~\bibnamefont{Lang}}, \bibnamefont{and}
  \bibinfo{author}{\bibfnamefont{H.~P.} \bibnamefont{B\"uchler}},
  \bibinfo{journal}{PRX Quantum} \textbf{\bibinfo{volume}{3}},
  \bibinfo{pages}{030302} (\bibinfo{year}{2022}),
  \urlprefix\url{https://link.aps.org/doi/10.1103/PRXQuantum.3.030302}.

\bibitem[{\citenamefont{Haegeman et~al.}(2011)\citenamefont{Haegeman, Cirac,
  Osborne, Pi\ifmmode~\check{z}\else \v{z}\fi{}orn, Verschelde, and
  Verstraete}}]{tdvp1}
\bibinfo{author}{\bibfnamefont{J.}~\bibnamefont{Haegeman}},
  \bibinfo{author}{\bibfnamefont{J.~I.} \bibnamefont{Cirac}},
  \bibinfo{author}{\bibfnamefont{T.~J.} \bibnamefont{Osborne}},
  \bibinfo{author}{\bibfnamefont{I.}~\bibnamefont{Pi\ifmmode~\check{z}\else
  \v{z}\fi{}orn}},
  \bibinfo{author}{\bibfnamefont{H.}~\bibnamefont{Verschelde}},
  \bibnamefont{and}
  \bibinfo{author}{\bibfnamefont{F.}~\bibnamefont{Verstraete}},
  \bibinfo{journal}{Phys. Rev. Lett.} \textbf{\bibinfo{volume}{107}},
  \bibinfo{pages}{070601} (\bibinfo{year}{2011}),
  \urlprefix\url{https://link.aps.org/doi/10.1103/PhysRevLett.107.070601}.

\bibitem[{\citenamefont{Haegeman et~al.}(2013)\citenamefont{Haegeman, Osborne,
  and Verstraete}}]{tdvp2}
\bibinfo{author}{\bibfnamefont{J.}~\bibnamefont{Haegeman}},
  \bibinfo{author}{\bibfnamefont{T.~J.} \bibnamefont{Osborne}},
  \bibnamefont{and}
  \bibinfo{author}{\bibfnamefont{F.}~\bibnamefont{Verstraete}},
  \bibinfo{journal}{Phys. Rev. B} \textbf{\bibinfo{volume}{88}},
  \bibinfo{pages}{075133} (\bibinfo{year}{2013}),
  \urlprefix\url{https://link.aps.org/doi/10.1103/PhysRevB.88.075133}.

\bibitem[{\citenamefont{Schollwöck}(2011)}]{Schollwock2011}
\bibinfo{author}{\bibfnamefont{U.}~\bibnamefont{Schollwöck}},
  \bibinfo{journal}{Annals of Physics} \textbf{\bibinfo{volume}{326}},
  \bibinfo{pages}{96} (\bibinfo{year}{2011}), ISSN \bibinfo{issn}{0003-4916},
  \bibinfo{note}{january 2011 Special Issue},
  \urlprefix\url{https://www.sciencedirect.com/science/article/pii/S0003491610001752}.

\bibitem[{\citenamefont{Fishman et~al.}(2022)\citenamefont{Fishman, White, and
  Stoudenmire}}]{itensor}
\bibinfo{author}{\bibfnamefont{M.}~\bibnamefont{Fishman}},
  \bibinfo{author}{\bibfnamefont{S.~R.} \bibnamefont{White}}, \bibnamefont{and}
  \bibinfo{author}{\bibfnamefont{E.~M.} \bibnamefont{Stoudenmire}},
  \bibinfo{journal}{SciPost Phys. Codebases} p.~\bibinfo{pages}{4}
  (\bibinfo{year}{2022}),
  \urlprefix\url{https://scipost.org/10.21468/SciPostPhysCodeb.4}.

\bibitem[{\citenamefont{Choi et~al.}(2020)\citenamefont{Choi, Zhou, Knowles,
  Landig, Choi, and Lukin}}]{choi2020robust}
\bibinfo{author}{\bibfnamefont{J.}~\bibnamefont{Choi}},
  \bibinfo{author}{\bibfnamefont{H.}~\bibnamefont{Zhou}},
  \bibinfo{author}{\bibfnamefont{H.~S.} \bibnamefont{Knowles}},
  \bibinfo{author}{\bibfnamefont{R.}~\bibnamefont{Landig}},
  \bibinfo{author}{\bibfnamefont{S.}~\bibnamefont{Choi}}, \bibnamefont{and}
  \bibinfo{author}{\bibfnamefont{M.~D.} \bibnamefont{Lukin}},
  \bibinfo{journal}{Physical Review X} \textbf{\bibinfo{volume}{10}},
  \bibinfo{pages}{031002} (\bibinfo{year}{2020}).

\bibitem[{\citenamefont{Bogdanov and
  Hubert}(1994)}]{bogdanov1994thermodynamically}
\bibinfo{author}{\bibfnamefont{A.}~\bibnamefont{Bogdanov}} \bibnamefont{and}
  \bibinfo{author}{\bibfnamefont{A.}~\bibnamefont{Hubert}},
  \bibinfo{journal}{Journal of magnetism and magnetic materials}
  \textbf{\bibinfo{volume}{138}}, \bibinfo{pages}{255} (\bibinfo{year}{1994}).

\bibitem[{\citenamefont{Muhlbauer et~al.}(2009)\citenamefont{Muhlbauer, Binz,
  Jonietz, Pfleiderer, Rosch, Neubauer, Georgii, and
  Boni}}]{muhlbauer2009skyrmion}
\bibinfo{author}{\bibfnamefont{S.}~\bibnamefont{Muhlbauer}},
  \bibinfo{author}{\bibfnamefont{B.}~\bibnamefont{Binz}},
  \bibinfo{author}{\bibfnamefont{F.}~\bibnamefont{Jonietz}},
  \bibinfo{author}{\bibfnamefont{C.}~\bibnamefont{Pfleiderer}},
  \bibinfo{author}{\bibfnamefont{A.}~\bibnamefont{Rosch}},
  \bibinfo{author}{\bibfnamefont{A.}~\bibnamefont{Neubauer}},
  \bibinfo{author}{\bibfnamefont{R.}~\bibnamefont{Georgii}}, \bibnamefont{and}
  \bibinfo{author}{\bibfnamefont{P.}~\bibnamefont{Boni}},
  \bibinfo{journal}{Science} \textbf{\bibinfo{volume}{323}},
  \bibinfo{pages}{915} (\bibinfo{year}{2009}).

\bibitem[{\citenamefont{Yu et~al.}(2010)\citenamefont{Yu, Onose, Kanazawa,
  Park, Han, Matsui, Nagaosa, and Tokura}}]{yu2010real}
\bibinfo{author}{\bibfnamefont{X.}~\bibnamefont{Yu}},
  \bibinfo{author}{\bibfnamefont{Y.}~\bibnamefont{Onose}},
  \bibinfo{author}{\bibfnamefont{N.}~\bibnamefont{Kanazawa}},
  \bibinfo{author}{\bibfnamefont{J.~H.} \bibnamefont{Park}},
  \bibinfo{author}{\bibfnamefont{J.}~\bibnamefont{Han}},
  \bibinfo{author}{\bibfnamefont{Y.}~\bibnamefont{Matsui}},
  \bibinfo{author}{\bibfnamefont{N.}~\bibnamefont{Nagaosa}}, \bibnamefont{and}
  \bibinfo{author}{\bibfnamefont{Y.}~\bibnamefont{Tokura}},
  \bibinfo{journal}{Nature} \textbf{\bibinfo{volume}{465}},
  \bibinfo{pages}{901} (\bibinfo{year}{2010}).

\bibitem[{\citenamefont{Nagaosa and Tokura}(2013)}]{nagaosa2013topological}
\bibinfo{author}{\bibfnamefont{N.}~\bibnamefont{Nagaosa}} \bibnamefont{and}
  \bibinfo{author}{\bibfnamefont{Y.}~\bibnamefont{Tokura}},
  \bibinfo{journal}{Nature nanotechnology} \textbf{\bibinfo{volume}{8}},
  \bibinfo{pages}{899} (\bibinfo{year}{2013}).

\bibitem[{\citenamefont{Fert et~al.}(2013)\citenamefont{Fert, Cros, and
  Sampaio}}]{fert2013skyrmions}
\bibinfo{author}{\bibfnamefont{A.}~\bibnamefont{Fert}},
  \bibinfo{author}{\bibfnamefont{V.}~\bibnamefont{Cros}}, \bibnamefont{and}
  \bibinfo{author}{\bibfnamefont{J.}~\bibnamefont{Sampaio}},
  \bibinfo{journal}{Nature nanotechnology} \textbf{\bibinfo{volume}{8}},
  \bibinfo{pages}{152} (\bibinfo{year}{2013}).

\bibitem[{\citenamefont{Back et~al.}(2020)\citenamefont{Back, Cros, Ebert,
  Everschor-Sitte, Fert, Garst, Ma, Mankovsky, Monchesky, Mostovoy
  et~al.}}]{back20202020}
\bibinfo{author}{\bibfnamefont{C.}~\bibnamefont{Back}},
  \bibinfo{author}{\bibfnamefont{V.}~\bibnamefont{Cros}},
  \bibinfo{author}{\bibfnamefont{H.}~\bibnamefont{Ebert}},
  \bibinfo{author}{\bibfnamefont{K.}~\bibnamefont{Everschor-Sitte}},
  \bibinfo{author}{\bibfnamefont{A.}~\bibnamefont{Fert}},
  \bibinfo{author}{\bibfnamefont{M.}~\bibnamefont{Garst}},
  \bibinfo{author}{\bibfnamefont{T.}~\bibnamefont{Ma}},
  \bibinfo{author}{\bibfnamefont{S.}~\bibnamefont{Mankovsky}},
  \bibinfo{author}{\bibfnamefont{T.}~\bibnamefont{Monchesky}},
  \bibinfo{author}{\bibfnamefont{M.}~\bibnamefont{Mostovoy}},
  \bibnamefont{et~al.}, \bibinfo{journal}{Journal of Physics D: Applied
  Physics} \textbf{\bibinfo{volume}{53}}, \bibinfo{pages}{363001}
  (\bibinfo{year}{2020}).

\bibitem[{\citenamefont{Sotnikov et~al.}(2021)\citenamefont{Sotnikov,
  Mazurenko, Colbois, Mila, Katsnelson, and Stepanov}}]{Sotnikov2021}
\bibinfo{author}{\bibfnamefont{O.~M.} \bibnamefont{Sotnikov}},
  \bibinfo{author}{\bibfnamefont{V.~V.} \bibnamefont{Mazurenko}},
  \bibinfo{author}{\bibfnamefont{J.}~\bibnamefont{Colbois}},
  \bibinfo{author}{\bibfnamefont{F.}~\bibnamefont{Mila}},
  \bibinfo{author}{\bibfnamefont{M.~I.} \bibnamefont{Katsnelson}},
  \bibnamefont{and} \bibinfo{author}{\bibfnamefont{E.~A.}
  \bibnamefont{Stepanov}}, \bibinfo{journal}{Phys. Rev. B}
  \textbf{\bibinfo{volume}{103}}, \bibinfo{pages}{L060404}
  (\bibinfo{year}{2021}),
  \urlprefix\url{https://link.aps.org/doi/10.1103/PhysRevB.103.L060404}.

\bibitem[{\citenamefont{Psaroudaki and Panagopoulos}(2021)}]{Psaroudaki2021}
\bibinfo{author}{\bibfnamefont{C.}~\bibnamefont{Psaroudaki}} \bibnamefont{and}
  \bibinfo{author}{\bibfnamefont{C.}~\bibnamefont{Panagopoulos}},
  \bibinfo{journal}{Phys. Rev. Lett.} \textbf{\bibinfo{volume}{127}},
  \bibinfo{pages}{067201} (\bibinfo{year}{2021}),
  \urlprefix\url{https://link.aps.org/doi/10.1103/PhysRevLett.127.067201}.

\bibitem[{\citenamefont{Siegl et~al.}(2022)\citenamefont{Siegl, Vedmedenko,
  Stier, Thorwart, and Posske}}]{Siegl2022}
\bibinfo{author}{\bibfnamefont{P.}~\bibnamefont{Siegl}},
  \bibinfo{author}{\bibfnamefont{E.~Y.} \bibnamefont{Vedmedenko}},
  \bibinfo{author}{\bibfnamefont{M.}~\bibnamefont{Stier}},
  \bibinfo{author}{\bibfnamefont{M.}~\bibnamefont{Thorwart}}, \bibnamefont{and}
  \bibinfo{author}{\bibfnamefont{T.}~\bibnamefont{Posske}},
  \bibinfo{journal}{Phys. Rev. Res.} \textbf{\bibinfo{volume}{4}},
  \bibinfo{pages}{023111} (\bibinfo{year}{2022}),
  \urlprefix\url{https://link.aps.org/doi/10.1103/PhysRevResearch.4.023111}.

\bibitem[{\citenamefont{Haller et~al.}(2022)\citenamefont{Haller, Groenendijk,
  Habibi, Michels, and Schmidt}}]{Haller2022}
\bibinfo{author}{\bibfnamefont{A.}~\bibnamefont{Haller}},
  \bibinfo{author}{\bibfnamefont{S.}~\bibnamefont{Groenendijk}},
  \bibinfo{author}{\bibfnamefont{A.}~\bibnamefont{Habibi}},
  \bibinfo{author}{\bibfnamefont{A.}~\bibnamefont{Michels}}, \bibnamefont{and}
  \bibinfo{author}{\bibfnamefont{T.~L.} \bibnamefont{Schmidt}},
  \bibinfo{journal}{Phys. Rev. Res.} \textbf{\bibinfo{volume}{4}},
  \bibinfo{pages}{043113} (\bibinfo{year}{2022}),
  \urlprefix\url{https://link.aps.org/doi/10.1103/PhysRevResearch.4.043113}.

\bibitem[{\citenamefont{Barredo et~al.}(2020)\citenamefont{Barredo, Lienhard,
  Scholl, de~L\'es\'eleuc, Boulier, Browaeys, and Lahaye}}]{Barredo2020}
\bibinfo{author}{\bibfnamefont{D.}~\bibnamefont{Barredo}},
  \bibinfo{author}{\bibfnamefont{V.}~\bibnamefont{Lienhard}},
  \bibinfo{author}{\bibfnamefont{P.}~\bibnamefont{Scholl}},
  \bibinfo{author}{\bibfnamefont{S.}~\bibnamefont{de~L\'es\'eleuc}},
  \bibinfo{author}{\bibfnamefont{T.}~\bibnamefont{Boulier}},
  \bibinfo{author}{\bibfnamefont{A.}~\bibnamefont{Browaeys}}, \bibnamefont{and}
  \bibinfo{author}{\bibfnamefont{T.}~\bibnamefont{Lahaye}},
  \bibinfo{journal}{Phys. Rev. Lett.} \textbf{\bibinfo{volume}{124}},
  \bibinfo{pages}{023201} (\bibinfo{year}{2020}),
  \urlprefix\url{https://link.aps.org/doi/10.1103/PhysRevLett.124.023201}.

\bibitem[{\citenamefont{Xu et~al.}(2021)\citenamefont{Xu, Venkatramani,
  Cant\'u, \ifmmode~\check{S}\else \v{S}\fi{}umarac, Kl\"usener, Lukin, and
  Vuleti\ifmmode~\acute{c}\else \'{c}\fi{}}}]{Xu2021}
\bibinfo{author}{\bibfnamefont{W.}~\bibnamefont{Xu}},
  \bibinfo{author}{\bibfnamefont{A.~V.} \bibnamefont{Venkatramani}},
  \bibinfo{author}{\bibfnamefont{S.~H.} \bibnamefont{Cant\'u}},
  \bibinfo{author}{\bibfnamefont{T.}~\bibnamefont{\ifmmode~\check{S}\else
  \v{S}\fi{}umarac}},
  \bibinfo{author}{\bibfnamefont{V.}~\bibnamefont{Kl\"usener}},
  \bibinfo{author}{\bibfnamefont{M.~D.} \bibnamefont{Lukin}}, \bibnamefont{and}
  \bibinfo{author}{\bibfnamefont{V.}~\bibnamefont{Vuleti\ifmmode~\acute{c}\else
  \'{c}\fi{}}}, \bibinfo{journal}{Phys. Rev. Lett.}
  \textbf{\bibinfo{volume}{127}}, \bibinfo{pages}{050501}
  (\bibinfo{year}{2021}),
  \urlprefix\url{https://link.aps.org/doi/10.1103/PhysRevLett.127.050501}.

\bibitem[{\citenamefont{Balents}(2010)}]{balents2010spin}
\bibinfo{author}{\bibfnamefont{L.}~\bibnamefont{Balents}},
  \bibinfo{journal}{Nature} \textbf{\bibinfo{volume}{464}},
  \bibinfo{pages}{199} (\bibinfo{year}{2010}).

\bibitem[{\citenamefont{Jackeli and Khaliullin}(2009)}]{Jackeli2009}
\bibinfo{author}{\bibfnamefont{G.}~\bibnamefont{Jackeli}} \bibnamefont{and}
  \bibinfo{author}{\bibfnamefont{G.}~\bibnamefont{Khaliullin}},
  \bibinfo{journal}{Phys. Rev. Lett.} \textbf{\bibinfo{volume}{102}},
  \bibinfo{pages}{017205} (\bibinfo{year}{2009}),
  \urlprefix\url{https://link.aps.org/doi/10.1103/PhysRevLett.102.017205}.

\bibitem[{\citenamefont{Kim et~al.}(2009)\citenamefont{Kim, Ohsumi, Komesu,
  Sakai, Morita, Takagi, and Arima}}]{kim2009phase}
\bibinfo{author}{\bibfnamefont{B.}~\bibnamefont{Kim}},
  \bibinfo{author}{\bibfnamefont{H.}~\bibnamefont{Ohsumi}},
  \bibinfo{author}{\bibfnamefont{T.}~\bibnamefont{Komesu}},
  \bibinfo{author}{\bibfnamefont{S.}~\bibnamefont{Sakai}},
  \bibinfo{author}{\bibfnamefont{T.}~\bibnamefont{Morita}},
  \bibinfo{author}{\bibfnamefont{H.}~\bibnamefont{Takagi}}, \bibnamefont{and}
  \bibinfo{author}{\bibfnamefont{T.-h.} \bibnamefont{Arima}},
  \bibinfo{journal}{Science} \textbf{\bibinfo{volume}{323}},
  \bibinfo{pages}{1329} (\bibinfo{year}{2009}).

\bibitem[{\citenamefont{Singh et~al.}(2012)\citenamefont{Singh, Manni, Reuther,
  Berlijn, Thomale, Ku, Trebst, and Gegenwart}}]{Singh2012}
\bibinfo{author}{\bibfnamefont{Y.}~\bibnamefont{Singh}},
  \bibinfo{author}{\bibfnamefont{S.}~\bibnamefont{Manni}},
  \bibinfo{author}{\bibfnamefont{J.}~\bibnamefont{Reuther}},
  \bibinfo{author}{\bibfnamefont{T.}~\bibnamefont{Berlijn}},
  \bibinfo{author}{\bibfnamefont{R.}~\bibnamefont{Thomale}},
  \bibinfo{author}{\bibfnamefont{W.}~\bibnamefont{Ku}},
  \bibinfo{author}{\bibfnamefont{S.}~\bibnamefont{Trebst}}, \bibnamefont{and}
  \bibinfo{author}{\bibfnamefont{P.}~\bibnamefont{Gegenwart}},
  \bibinfo{journal}{Phys. Rev. Lett.} \textbf{\bibinfo{volume}{108}},
  \bibinfo{pages}{127203} (\bibinfo{year}{2012}),
  \urlprefix\url{https://link.aps.org/doi/10.1103/PhysRevLett.108.127203}.

\bibitem[{\citenamefont{Choi et~al.}(2012)\citenamefont{Choi, Coldea,
  Kolmogorov, Lancaster, Mazin, Blundell, Radaelli, Singh, Gegenwart, Choi
  et~al.}}]{Choi2012}
\bibinfo{author}{\bibfnamefont{S.~K.} \bibnamefont{Choi}},
  \bibinfo{author}{\bibfnamefont{R.}~\bibnamefont{Coldea}},
  \bibinfo{author}{\bibfnamefont{A.~N.} \bibnamefont{Kolmogorov}},
  \bibinfo{author}{\bibfnamefont{T.}~\bibnamefont{Lancaster}},
  \bibinfo{author}{\bibfnamefont{I.~I.} \bibnamefont{Mazin}},
  \bibinfo{author}{\bibfnamefont{S.~J.} \bibnamefont{Blundell}},
  \bibinfo{author}{\bibfnamefont{P.~G.} \bibnamefont{Radaelli}},
  \bibinfo{author}{\bibfnamefont{Y.}~\bibnamefont{Singh}},
  \bibinfo{author}{\bibfnamefont{P.}~\bibnamefont{Gegenwart}},
  \bibinfo{author}{\bibfnamefont{K.~R.} \bibnamefont{Choi}},
  \bibnamefont{et~al.}, \bibinfo{journal}{Phys. Rev. Lett.}
  \textbf{\bibinfo{volume}{108}}, \bibinfo{pages}{127204}
  (\bibinfo{year}{2012}),
  \urlprefix\url{https://link.aps.org/doi/10.1103/PhysRevLett.108.127204}.

\bibitem[{\citenamefont{Plumb et~al.}(2014)\citenamefont{Plumb, Clancy,
  Sandilands, Shankar, Hu, Burch, Kee, and Kim}}]{Plumb2014}
\bibinfo{author}{\bibfnamefont{K.~W.} \bibnamefont{Plumb}},
  \bibinfo{author}{\bibfnamefont{J.~P.} \bibnamefont{Clancy}},
  \bibinfo{author}{\bibfnamefont{L.~J.} \bibnamefont{Sandilands}},
  \bibinfo{author}{\bibfnamefont{V.~V.} \bibnamefont{Shankar}},
  \bibinfo{author}{\bibfnamefont{Y.~F.} \bibnamefont{Hu}},
  \bibinfo{author}{\bibfnamefont{K.~S.} \bibnamefont{Burch}},
  \bibinfo{author}{\bibfnamefont{H.-Y.} \bibnamefont{Kee}}, \bibnamefont{and}
  \bibinfo{author}{\bibfnamefont{Y.-J.} \bibnamefont{Kim}},
  \bibinfo{journal}{Phys. Rev. B} \textbf{\bibinfo{volume}{90}},
  \bibinfo{pages}{041112} (\bibinfo{year}{2014}),
  \urlprefix\url{https://link.aps.org/doi/10.1103/PhysRevB.90.041112}.

\bibitem[{\citenamefont{Hwan~Chun et~al.}(2015)\citenamefont{Hwan~Chun, Kim,
  Kim, Zheng, Stoumpos, Malliakas, Mitchell, Mehlawat, Singh, Choi
  et~al.}}]{hwan2015direct}
\bibinfo{author}{\bibfnamefont{S.}~\bibnamefont{Hwan~Chun}},
  \bibinfo{author}{\bibfnamefont{J.-W.} \bibnamefont{Kim}},
  \bibinfo{author}{\bibfnamefont{J.}~\bibnamefont{Kim}},
  \bibinfo{author}{\bibfnamefont{H.}~\bibnamefont{Zheng}},
  \bibinfo{author}{\bibfnamefont{C.~C.} \bibnamefont{Stoumpos}},
  \bibinfo{author}{\bibfnamefont{C.}~\bibnamefont{Malliakas}},
  \bibinfo{author}{\bibfnamefont{J.}~\bibnamefont{Mitchell}},
  \bibinfo{author}{\bibfnamefont{K.}~\bibnamefont{Mehlawat}},
  \bibinfo{author}{\bibfnamefont{Y.}~\bibnamefont{Singh}},
  \bibinfo{author}{\bibfnamefont{Y.}~\bibnamefont{Choi}}, \bibnamefont{et~al.},
  \bibinfo{journal}{Nature Physics} \textbf{\bibinfo{volume}{11}},
  \bibinfo{pages}{462} (\bibinfo{year}{2015}).

\bibitem[{\citenamefont{Banerjee et~al.}(2016)\citenamefont{Banerjee, Bridges,
  Yan, Aczel, Li, Stone, Granroth, Lumsden, Yiu, Knolle
  et~al.}}]{banerjee2016proximate}
\bibinfo{author}{\bibfnamefont{A.}~\bibnamefont{Banerjee}},
  \bibinfo{author}{\bibfnamefont{C.}~\bibnamefont{Bridges}},
  \bibinfo{author}{\bibfnamefont{J.-Q.} \bibnamefont{Yan}},
  \bibinfo{author}{\bibfnamefont{A.}~\bibnamefont{Aczel}},
  \bibinfo{author}{\bibfnamefont{L.}~\bibnamefont{Li}},
  \bibinfo{author}{\bibfnamefont{M.}~\bibnamefont{Stone}},
  \bibinfo{author}{\bibfnamefont{G.}~\bibnamefont{Granroth}},
  \bibinfo{author}{\bibfnamefont{M.}~\bibnamefont{Lumsden}},
  \bibinfo{author}{\bibfnamefont{Y.}~\bibnamefont{Yiu}},
  \bibinfo{author}{\bibfnamefont{J.}~\bibnamefont{Knolle}},
  \bibnamefont{et~al.}, \bibinfo{journal}{Nature materials}
  \textbf{\bibinfo{volume}{15}}, \bibinfo{pages}{733} (\bibinfo{year}{2016}).

\bibitem[{\citenamefont{Kasahara et~al.}(2018)\citenamefont{Kasahara, Ohnishi,
  Mizukami, Tanaka, Ma, Sugii, Kurita, Tanaka, Nasu, Motome
  et~al.}}]{kasahara2018majorana}
\bibinfo{author}{\bibfnamefont{Y.}~\bibnamefont{Kasahara}},
  \bibinfo{author}{\bibfnamefont{T.}~\bibnamefont{Ohnishi}},
  \bibinfo{author}{\bibfnamefont{Y.}~\bibnamefont{Mizukami}},
  \bibinfo{author}{\bibfnamefont{O.}~\bibnamefont{Tanaka}},
  \bibinfo{author}{\bibfnamefont{S.}~\bibnamefont{Ma}},
  \bibinfo{author}{\bibfnamefont{K.}~\bibnamefont{Sugii}},
  \bibinfo{author}{\bibfnamefont{N.}~\bibnamefont{Kurita}},
  \bibinfo{author}{\bibfnamefont{H.}~\bibnamefont{Tanaka}},
  \bibinfo{author}{\bibfnamefont{J.}~\bibnamefont{Nasu}},
  \bibinfo{author}{\bibfnamefont{Y.}~\bibnamefont{Motome}},
  \bibnamefont{et~al.}, \bibinfo{journal}{Nature}
  \textbf{\bibinfo{volume}{559}}, \bibinfo{pages}{227} (\bibinfo{year}{2018}).

\bibitem[{\citenamefont{Sears et~al.}(2020)\citenamefont{Sears, Chern, Kim,
  Bereciartua, Francoual, Kim, and Kim}}]{sears2020ferromagnetic}
\bibinfo{author}{\bibfnamefont{J.~A.} \bibnamefont{Sears}},
  \bibinfo{author}{\bibfnamefont{L.~E.} \bibnamefont{Chern}},
  \bibinfo{author}{\bibfnamefont{S.}~\bibnamefont{Kim}},
  \bibinfo{author}{\bibfnamefont{P.~J.} \bibnamefont{Bereciartua}},
  \bibinfo{author}{\bibfnamefont{S.}~\bibnamefont{Francoual}},
  \bibinfo{author}{\bibfnamefont{Y.~B.} \bibnamefont{Kim}}, \bibnamefont{and}
  \bibinfo{author}{\bibfnamefont{Y.-J.} \bibnamefont{Kim}},
  \bibinfo{journal}{Nature physics} \textbf{\bibinfo{volume}{16}},
  \bibinfo{pages}{837} (\bibinfo{year}{2020}).

\bibitem[{\citenamefont{Yokoi et~al.}(2021)\citenamefont{Yokoi, Ma, Kasahara,
  Kasahara, Shibauchi, Kurita, Tanaka, Nasu, Motome, Hickey
  et~al.}}]{yokoi2021half}
\bibinfo{author}{\bibfnamefont{T.}~\bibnamefont{Yokoi}},
  \bibinfo{author}{\bibfnamefont{S.}~\bibnamefont{Ma}},
  \bibinfo{author}{\bibfnamefont{Y.}~\bibnamefont{Kasahara}},
  \bibinfo{author}{\bibfnamefont{S.}~\bibnamefont{Kasahara}},
  \bibinfo{author}{\bibfnamefont{T.}~\bibnamefont{Shibauchi}},
  \bibinfo{author}{\bibfnamefont{N.}~\bibnamefont{Kurita}},
  \bibinfo{author}{\bibfnamefont{H.}~\bibnamefont{Tanaka}},
  \bibinfo{author}{\bibfnamefont{J.}~\bibnamefont{Nasu}},
  \bibinfo{author}{\bibfnamefont{Y.}~\bibnamefont{Motome}},
  \bibinfo{author}{\bibfnamefont{C.}~\bibnamefont{Hickey}},
  \bibnamefont{et~al.}, \bibinfo{journal}{Science}
  \textbf{\bibinfo{volume}{373}}, \bibinfo{pages}{568} (\bibinfo{year}{2021}).

\bibitem[{\citenamefont{Kalinowski et~al.}(2022)\citenamefont{Kalinowski,
  Maskara, and Lukin}}]{kalinowski2022non}
\bibinfo{author}{\bibfnamefont{M.}~\bibnamefont{Kalinowski}},
  \bibinfo{author}{\bibfnamefont{N.}~\bibnamefont{Maskara}}, \bibnamefont{and}
  \bibinfo{author}{\bibfnamefont{M.~D.} \bibnamefont{Lukin}},
  \bibinfo{journal}{arXiv preprint arXiv:2211.00017}  (\bibinfo{year}{2022}).

\bibitem[{\citenamefont{Sun et~al.}(2022)\citenamefont{Sun, Goldman,
  Aidelsburger, and Bukov}}]{sun2022engineering}
\bibinfo{author}{\bibfnamefont{B.-Y.} \bibnamefont{Sun}},
  \bibinfo{author}{\bibfnamefont{N.}~\bibnamefont{Goldman}},
  \bibinfo{author}{\bibfnamefont{M.}~\bibnamefont{Aidelsburger}},
  \bibnamefont{and} \bibinfo{author}{\bibfnamefont{M.}~\bibnamefont{Bukov}},
  \bibinfo{journal}{arXiv preprint arXiv:2211.09777}  (\bibinfo{year}{2022}).

\bibitem[{\citenamefont{Young et~al.}(2020)\citenamefont{Young, Eckner, Milner,
  Kedar, Norcia, Oelker, Schine, Ye, and Kaufman}}]{young2020half}
\bibinfo{author}{\bibfnamefont{A.~W.} \bibnamefont{Young}},
  \bibinfo{author}{\bibfnamefont{W.~J.} \bibnamefont{Eckner}},
  \bibinfo{author}{\bibfnamefont{W.~R.} \bibnamefont{Milner}},
  \bibinfo{author}{\bibfnamefont{D.}~\bibnamefont{Kedar}},
  \bibinfo{author}{\bibfnamefont{M.~A.} \bibnamefont{Norcia}},
  \bibinfo{author}{\bibfnamefont{E.}~\bibnamefont{Oelker}},
  \bibinfo{author}{\bibfnamefont{N.}~\bibnamefont{Schine}},
  \bibinfo{author}{\bibfnamefont{J.}~\bibnamefont{Ye}}, \bibnamefont{and}
  \bibinfo{author}{\bibfnamefont{A.~M.} \bibnamefont{Kaufman}},
  \bibinfo{journal}{Nature} \textbf{\bibinfo{volume}{588}},
  \bibinfo{pages}{408} (\bibinfo{year}{2020}).

\bibitem[{\citenamefont{Evered et~al.}(2023)\citenamefont{Evered, Bluvstein,
  Kalinowski, Ebadi, Manovitz, Zhou, Li, Geim, Wang, Maskara
  et~al.}}]{evered2023highfidelity}
\bibinfo{author}{\bibfnamefont{S.~J.} \bibnamefont{Evered}},
  \bibinfo{author}{\bibfnamefont{D.}~\bibnamefont{Bluvstein}},
  \bibinfo{author}{\bibfnamefont{M.}~\bibnamefont{Kalinowski}},
  \bibinfo{author}{\bibfnamefont{S.}~\bibnamefont{Ebadi}},
  \bibinfo{author}{\bibfnamefont{T.}~\bibnamefont{Manovitz}},
  \bibinfo{author}{\bibfnamefont{H.}~\bibnamefont{Zhou}},
  \bibinfo{author}{\bibfnamefont{S.~H.} \bibnamefont{Li}},
  \bibinfo{author}{\bibfnamefont{A.~A.} \bibnamefont{Geim}},
  \bibinfo{author}{\bibfnamefont{T.~T.} \bibnamefont{Wang}},
  \bibinfo{author}{\bibfnamefont{N.}~\bibnamefont{Maskara}},
  \bibnamefont{et~al.}, \emph{\bibinfo{title}{High-fidelity parallel entangling
  gates on a neutral atom quantum computer}} (\bibinfo{year}{2023}),
  \eprint{2304.05420}.

\end{thebibliography}

\clearpage


\onecolumngrid
\setcounter{equation}{0}
\setcounter{figure}{0}
\renewcommand{\theequation}{S\arabic{equation}}
\renewcommand{\thefigure}{S\arabic{figure}}

\renewcommand{\thesection}{S\arabic{section}}
\renewcommand{\thesubsection}{\thesection.\arabic{subsection}}
\renewcommand{\thesubsubsection}{\thesubsection.\arabic{subsubsection}}

\begin{center}
{\Large\bfseries Supplementary Material}
\end{center}

\section{Effective Hamiltonian derivation for time-modulated Rydberg atom arrays}

As explained in the main text, we assume the atoms in the array interact via resonant dipole-dipole interactions, which in the nearest-neighbors approximation results in the interaction Hamiltonian: 
\begin{equation}
    H_{\rm{XY}}=\frac{1}{2}\sum_{\langle ij \rangle}J_{0}(\sigma_i^x\sigma_{j}^x+\sigma_i^y\sigma_{j}^y),
\end{equation}
where $J_{0}$ is the bare interaction strength between the Rydberg atoms and $\sigma_i^x,\sigma_i^y$ are the Pauli-X and Pauli-Y operators of atom $i$, respectively. Therefore, when adding any modulation to the system, the time-dependent Hamiltonian for the array of Rydberg atoms can be expressed as 
\begin{equation}
    H(t) = H_{\rm{XY}}+H_{\rm{drive}}(t).
\end{equation}
Here, $H_{\rm{drive}}(t)$ is a time-dependent, periodic modulation ($i.e\;H_{\rm{drive}}(t+T)=H_{\rm{drive}}(t)$). In the interaction picture, we can define $U_{\rm{drive}}(t)=\mathcal{T}\exp [-i\int_0^t H_{\rm{drive}}(t')dt']$, enabling a description of the system in a rotating frame with the modulation frequency. In the rotating frame, a state of the system can be described by $\tilde{\rho}(t)=U_{\rm{drive}}^\dagger (t)\rho U_{\rm{drive}}(t)$, where $\rho$ is a state of the unperturbed system; and the state $\tilde{\rho}(t)$ evolves under the rotating frame Hamiltonian $\tilde{H}(t)=U_{\rm{drive}}^\dagger (t) H_{XY}U_{\rm{drive}}(t)$. 

Taking the stroboscopic approach, a unitary operator $U$ over one modulation period for the time-dependent Hamiltonian is equivalent to a unitary operator of some time-independent Hamiltonian $H_F$ over time T
\begin{equation}
    U(T)=\exp [-iH_FT].
\end{equation}
$H_F$ is known as the Floquet Hamiltonian, and it produces the same dynamics at integer intervals of $T$ as the time-dependent Hamiltonian $\tilde{H}(t)$. Using the Floquet-Magnus expansion \cite{bukov2015}, $H_F$ can be written as 
\begin{equation}
    H_F=\sum_{n=0}^\infty H^{(n)},
    \label{eq:H_F_sum}
\end{equation}
where the three lowest-order terms of \ref{eq:H_F_sum} being
\begin{align}
    \label{eq:S5}
    H^{(0)} &=\frac{1}{T}\int_0^T\tilde{H}(t)dt\\
    \label{eq:S6}
    H^{(1)} &=\frac{1}{i2T}\int_0^Tdt\int_0^tdt'[\tilde{H}(t),\tilde{H}(t')]\\
    \label{eq:S7}
    H^{(2)} &=-\frac{1}{6T}\int_0^Tdt\int_0^{t}dt'\int_0^{t'}dt''\Big\{[\tilde{H}(t),[\tilde{H}(t'),\tilde{H}(t'')]]+[[\tilde{H}(t),\tilde{H}(t')],\tilde{H}(t'')]\Big\}
\end{align}
Notably, the $n-\rm th$ order in \ref{eq:H_F_sum} scales as $(J_{0}T)^n$. Thus, in the high-frequency regime ($J_{0}T \ll 1$), the zeroth order term $H^{(0)}$ is a good approximation for $H_F$, with first-order corrections scaling as $T$, as given in Eq. (2) of the main text. 

\section{Instantaneous Hamiltonian in the Rotated frame}

We derive the expression for the Hamiltonian in the rotated frame after applying a general combination of global and local modulation.  The influence of     the global microwave modulation on the Rydberg atoms, as verified in \cite{geier2021,Scholl2022}, acts as a unitary of the form $\exp (i\Theta\sum_j\hat{n}\cdot\Vec{\sigma_j}/2)$ where $\hat{n}$ is the unit vector along the axis of rotation and $\Theta$ is the rotation angle. The Hamiltonian in the rotated frame when considering the global pulse is
\begin{align}
    H(t) &= e^{i\Theta\sum_j\hat{n}\cdot\Vec{\sigma_j}}H_{\rm{XY}} e^{-i\Theta\sum_j\hat{n}\cdot\Vec{\sigma_j}}\\
    &=\frac{1}{2}\sum_{\langle ij \rangle}J_{0}(\tilde{\sigma}_i^x\tilde{\sigma}_{j}^x+\tilde{\sigma}_i^y\tilde{\sigma}_{j}^y)
\end{align}
where $\tilde{\sigma}^\alpha=\exp (i\Theta\hat{n}\cdot\Vec{\sigma})\sigma^\alpha\exp (-i\Theta\hat{n}\cdot\Vec{\sigma})$. On the other hand, the local modulation due to AC Stark shifts results in rotation about the axis perpendicular to the plane of interaction \cite{deLeseluc2017}, which has the form $\exp (i\sum_j\phi_j\tilde{\sigma}_j^z)$. Transformed $\tilde{\sigma}_j^x$ and $\tilde{\sigma}_j^y$ operators under this unitary are 
\begin{align}
    \tilde{\sigma}_j^x & \xrightarrow{} \cos \phi_j\tilde{\sigma}_j^x + \sin \phi_j\tilde{\sigma}_j^y
    \label{eq:S1}\\
    \tilde{\sigma}_j^y & \xrightarrow{} -\sin \phi_j\tilde{\sigma}_j^x + \cos \phi_j\tilde{\sigma}_j^y
    \label{eq:S2}
\end{align}
and the Hamiltonian in the doubly-rotated frame, including the local modulation, is exactly Eq. (3) in the main text, where $\Delta\phi_{ij}=\phi_{j}-\phi_{i}$, with $i$ and $j$ denoting two different atoms.

\section{Mitigating errors in effective Hamiltonian engineering}

\begin{figure*}
  \includegraphics[width=0.5\textwidth]{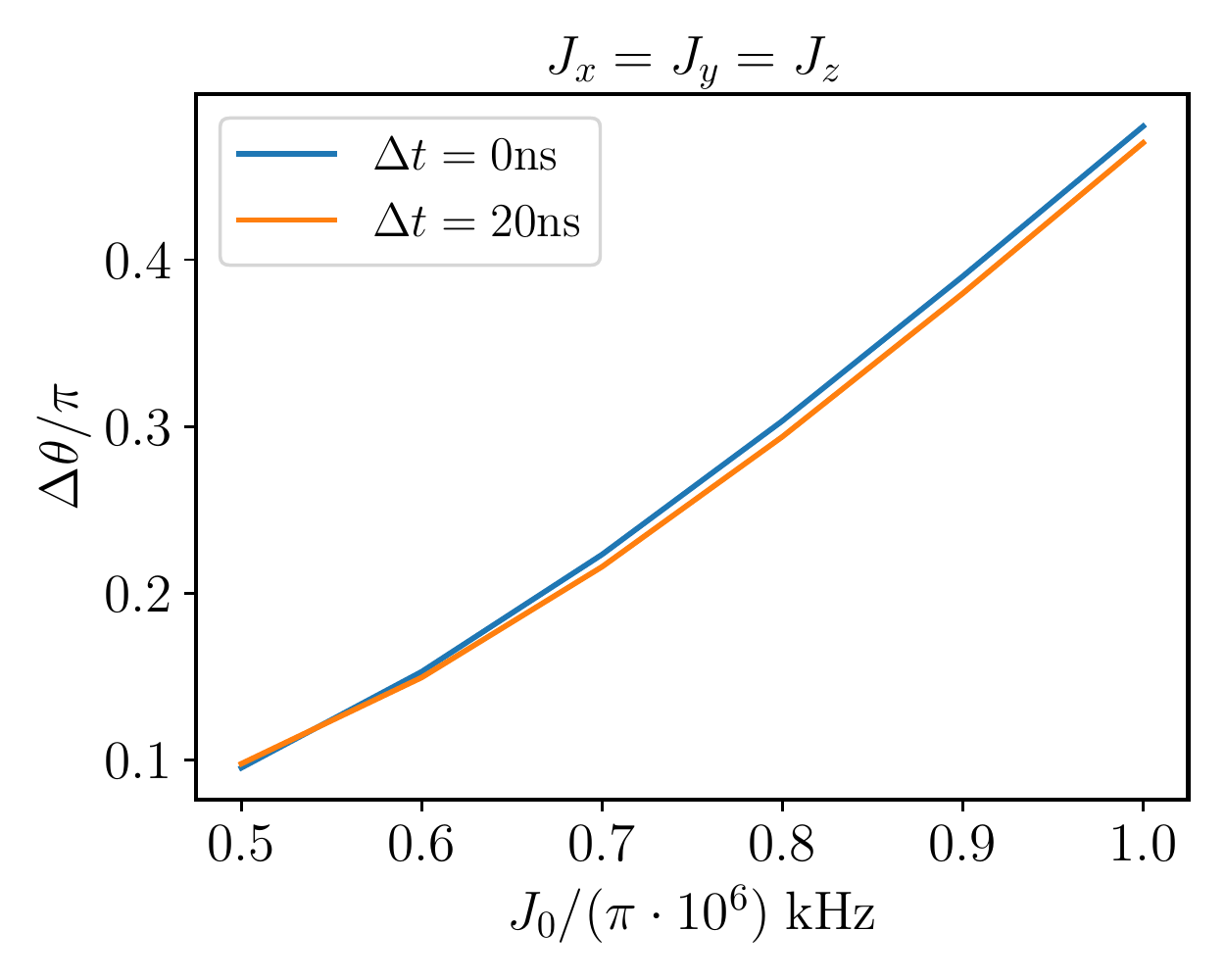}
  \caption{Dependence of the engineering accuracy of XYZ interaction on the bare Rydberg atom interaction strength. The curves show the $\Delta \theta$ parameter after a single modulation period for both ideal modulation ($\Delta t =0 \rm ns$) and practical modulation ($\Delta t =20 \rm ns$) with the symmetric pulse sequence described in Fig. 3 of the main text, and agree very well with a second-order polynomial function. The contribution of the quadratic term to the error is by far the most dominant in both cases. The assumed modulation period was $T=600 \rm ns$.}
  \label{fig:DeltaTheta_scaling_with_J0}
\end{figure*}


First and foremost, should the engineered and initial Hamiltonians commute with each other (i.e., $[H_{XY},H_F]=0$), then $\it{all}$ higher order corrections are nullified for ideal modulation (i.e., with infinitely short pulses). Such is the case in our scheme when using only local modulation (see Fig. 2 in the main text), since the symmetric XY Heisenberg and antisymmetric Z DM interactions fully commute. In contrast, when attempting to use the global modulation to produce an XYZ Heisenberg or Kitaev interaction (as in Figs. 3 and 4 of the main text), the Hamiltonians do not inherently commute, and higher-order corrections arise. Thus, using local modulation in our scheme is $\it{fundamentally}$ more robust than using global modulation.

For an ideal modulation and in the high-frequency regime, constructing a symmetric pulse sequence cancels the first order correction $H^{(1)}$ to the effective Hamiltonian \cite{choi2020robust}. For this reason, all pulse sequences used in this work are symmetric (see additional considerations below). We verified that such is indeed the case by simulating the $\Delta \theta$ parameter for a single modulation period generating the XYZ Hamiltonian, as a  function of $J_0$ (Fig. \ref{fig:DeltaTheta_scaling_with_J0}). $\Delta \theta$, as defined in the main text, gives a quantitative measure for the high-order corrections to the effective Hamiltonian. As expected from Eqs.~\eqref{eq:S5}-\eqref{eq:S7}, a quadratic dependence on $J_0$ is observed, illustrating that second-order corrections dominate. Intriguingly, this scaling occurs even for practical modulation and when the high-frequency regime is not strictly applicable. We note that second-order, and even higher-order corrections to the effective Hamiltonian can be nullified by an appropriate modulation scheme \cite{choi2020robust}, at the cost of a more complex pulse sequence. 

\begin{figure*}
  \includegraphics[width=0.5\textwidth]{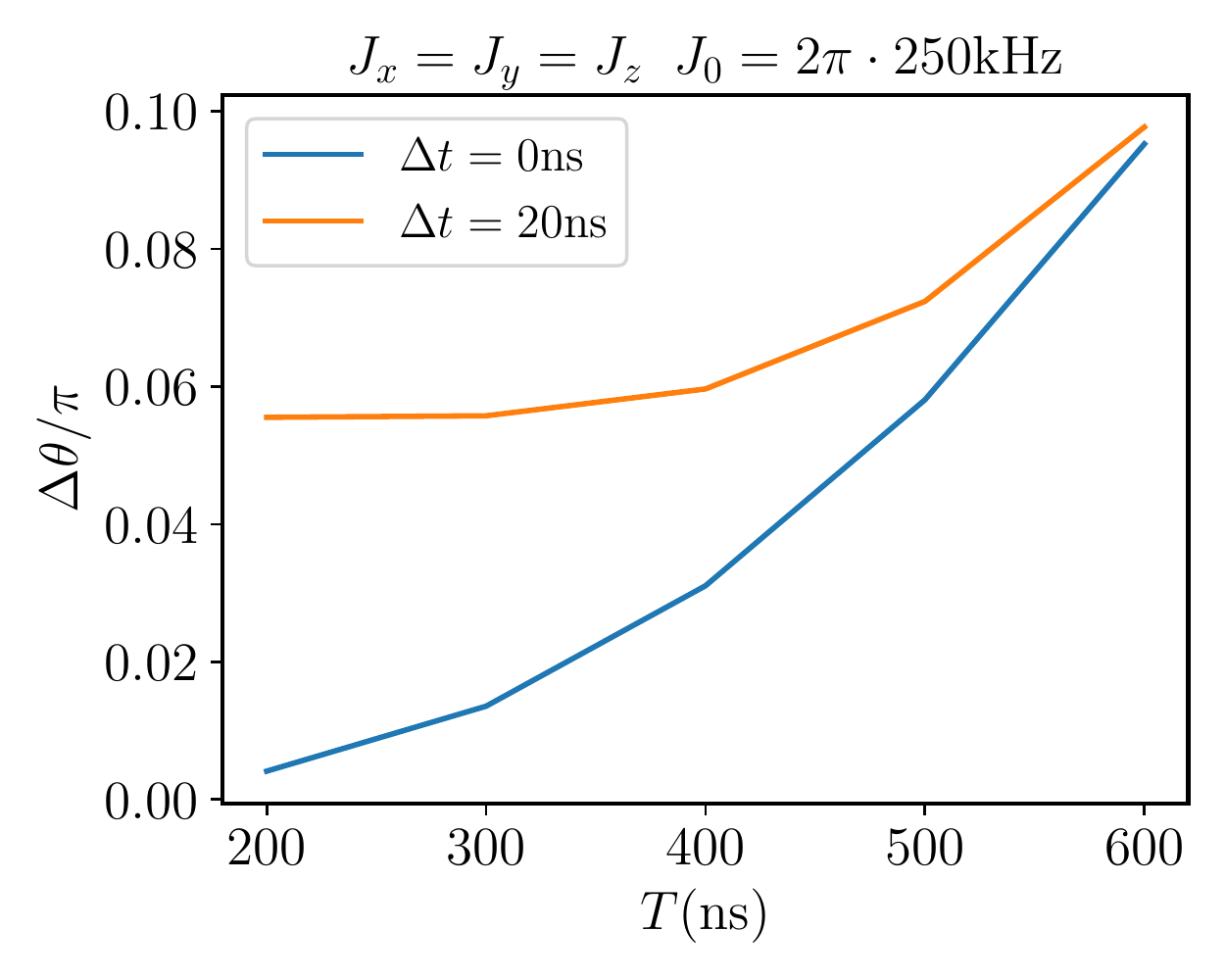}
  \caption{Dependence of the engineering accuracy of XYZ interaction on the modulation period. The curves show the $\Delta \theta$ parameter for the ideal modulation ($\Delta t =0 \rm ns$) and practical modulation ($\Delta t =20 \rm ns$), considering the symmetric pulse sequence described in Fig. 3 of the main text. In the ideal case, the error scales as a second-order polynomial in $T$, as expected from Eqs.~\eqref{eq:S5}-\eqref{eq:S7} for symmetric pulse sequences. In the practical case, however, a convergence to a constant, modulation-period-independent error is observed, owing to the non-vanishing pulse application time. The assumed bare Rydberg atom interaction strength was $J_0=2\pi \cdot 250 \rm kHz$.}
  \label{fig:DeltaTheta_scaling_with_T}
\end{figure*}

For a more practical modulation scheme, where applied pulses have finite widths, additional errors can accumulate and change the dependence of Hamiltonian engineering accuracy on the modulation period $T$. This is a direct implication of the non-vanishing pulse width $\Delta t$, as reducing $T$ decreases the relative time the system freely evolves. A demonstration of this behavior is given in Fig. \ref{fig:DeltaTheta_scaling_with_T}, by plotting the $\Delta \theta$ parameter against the modulation period for a single modulation cycle generating the XYZ Hamiltonian. As suggested above, the dependence is quadratic in the ideal case, but in the more practical case a pulse-width-dependent threshold is apparent, below which the accuracy of the engineered Hamiltonian does not improve. In our simulations, we opted to work with the smallest modulation period in which the practical and ideal modulation converged in their dependence ($T=0.5-0.6 \rm \mu s$ for our chosen parameters). In addition, further optimization of the pulse sequence itself can assist in generating the Hamiltonian more accurately, as discussed below.

\section{Pulse sequence optimization process}

\begin{figure*}
  \includegraphics[width=\textwidth]{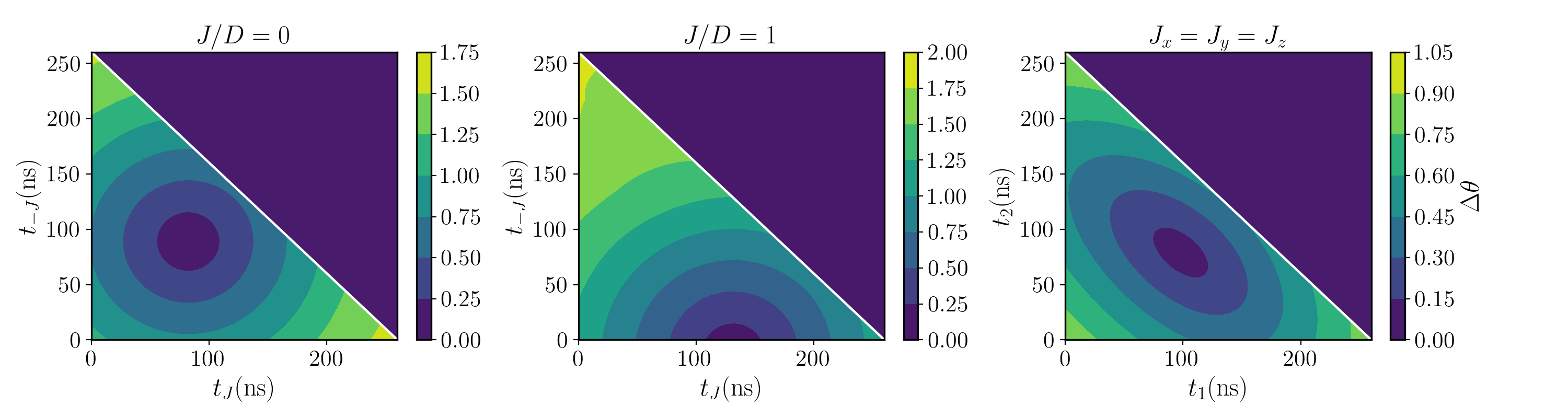}
  \caption{Visualization of the optimization process for practical Floquet engineering. Heat maps depicting optimization of free evolution times to engineer only DM interaction (left panel), equal strength of XY and DM interaction (middle panel) and an isotropic XYZ interaction (right panel) are shown above. The optimal point (dark blue) is the position of minimal $\Delta\theta$ for a single modulation period. The visualization is two-dimensional since both engineering schemes have only two independent free evolution times. Left and middle panel optimize over $t_J,\;t_{-J}$, while the right panel optimizes over $t_1,\;t_2$ (these parameters are defined in the main text). In all the cases presented here $T=600 \rm ns$, $\Delta t = 20 \rm ns$ and $J_0=2\pi\cdot 250\rm{kHz}$. It is clear that the nonzero value of $\Delta t$ alters the optimal free evolution times from their ideal values. For example, one would expect $t_J=t_{-J}$ to fully cancel out the XY interaction or that $t_1=t_{2}$ will lead to isotropic XYZ interaction; However, that is not the case when the system evolution during pulse application is considered as well.}
  \label{fig:Optimization_Delta_theta}
\end{figure*}

We assume that the modulation period in the practical case is made up of $n$ pulses of width $\Delta t$ and free evolution times $t_i$ ,  $i=1,2,\dots, n+1$ . We treat the pulse widths as constant (as they are usually an experimental constraint) and optimize the free evolution times $t_i$ to minimize the quantity $\| U_F^{\dagger} U_S - \mathbbm{1}\|$, as defined in the main text. Through the relation between $\| U_F^{\dagger} U_S - \mathbbm{1}\|$ and the $\Delta \theta$ parameter, it is clear that minimizing one minimizes the other, and we thus present the visualization of our optimization in terms of $\Delta \theta$ after one modulation period. Notably, $0\leq t_i\leq T-n\Delta t$, constraining the maximal free evolution time, so our optimization spans each $t_i$ in values ranging from $0$ to $T-n\Delta t$ while satisfying the condition for the total modulation period stated in the main text. Figure \ref{fig:Optimization_Delta_theta} illustrates the variation of $\Delta\theta$ for three examples given in Figs. 2 and 3 in the main text, which conveniently have only two free parameters, allowing a 2D heatmap to visualize the optimal parameters. Noticeably, the optimized parameters are different from their value in the ideal pulse sequence with $\Delta t=0$, due to the effects of system evolution during the pulse application time $\Delta t$.   

\begin{figure*}
  \includegraphics[width=0.45\textwidth]{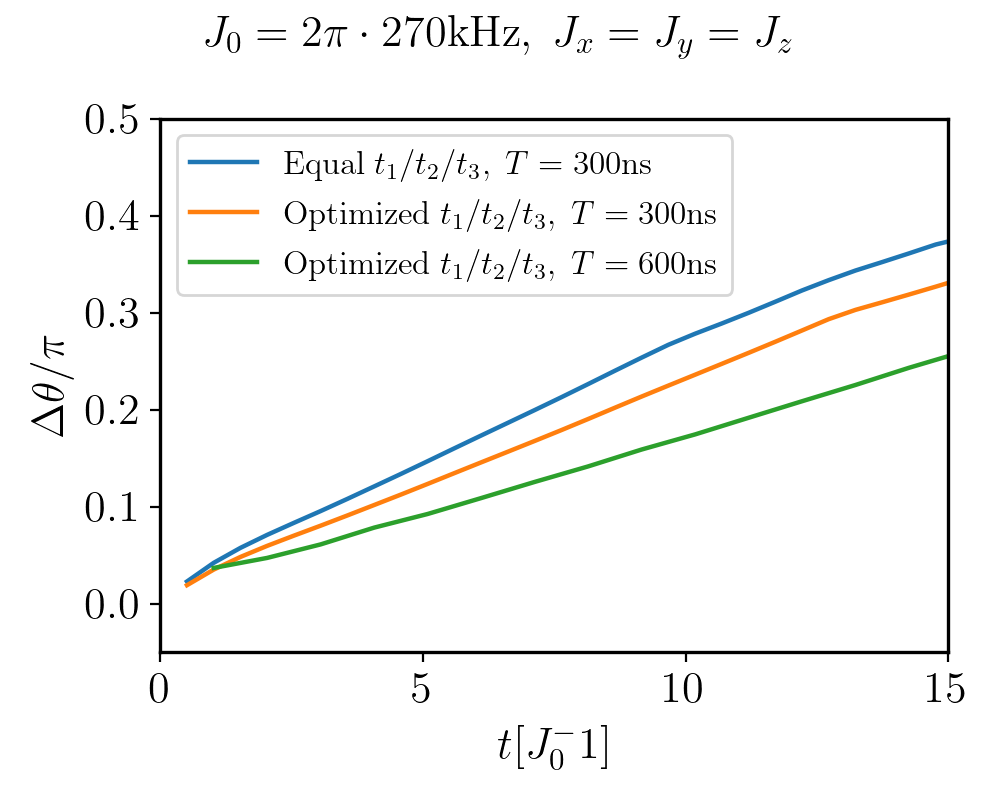}
  \caption{Comparison of optimized and non-optimized pulse sequences. The figure shows the time-dependent $\Delta \theta$ parameter as a function of time, for 3 different pulse sequences. First, a short, non-optimized pulse sequence (blue), whose parameters are the same as in the experimental demonstration \cite{Scholl2022}, producing XYZ interaction on a ring of Rydberg atoms. Second, a short, optimized pulse seuqence (orange), demonstrates $\approx J_0^{-1}$ increase in the coherence time of the engineered Hamiltonian (an increase of ~$20\%$). Lastly, a longer, optimized pulse sequence virtually doubles the engineered Hamiltonian coherence time, which is the same factor by which the modulation period $T$ was extended (a ~$100\%$ increase). This figure illustrates that taking into account the errors of practical Floquet engineering parameters can greatly extend quantum simulation times.}
  \label{fig:Optimized_vs_unoptimized}
\end{figure*} 
The importance of optimizing both the total modulation period and the pulse sequence itself is illustrated in Fig. \ref{fig:Optimized_vs_unoptimized}, where we compare the optimal parameters we reached to engineer an XYZ interaction between Rydberg atoms and the parameters used in the recent experimental demonstration \cite{Scholl2022}. For the same modulation period, it is clear that the optimized parameters slightly increase the coherence time of the engineered Hamiltonian, extending it by $\approx J_{0}^{-1}$. A greater effect, as could also be predicted by Fig. \ref{fig:DeltaTheta_scaling_with_T}, is achieved by doubling the modulation period, which extends the Hamiltonian engineering coherence time by a factor of 2. Thus, the time frame for quantum simulation using our scheme can be greatly extended by appropriately optimizing the modulation. We also note that no optimization was performed for the pulse shapes themselves, which can vastly improve the current results (as stated in \cite{Scholl2022} and also explored in many other works, e.g. \cite{evered2023highfidelity}) and eliminate many errors in the practical case, bringing the result much closer to the ideal Hamiltonian coherence times.

\section{Effect of Next-Nearest Neighbor Interaction}

As mentioned in the main text, our simulation model considers only nearest-neighbor (NN) interactions, for simplicity of the calculations. That said, the resonant dipole-dipole interaction between Rydberg atoms is highly nonlocal (decays as $1/R^3$), and it is essential to check that including the higher order interactions does not significantly change our results. We therefore perform a simulation including the next-nearest neighbor (NNN) interactions, as it applies to the generation of Dzyaloshinskii-Moriya interactions in a ring of atoms (similarly to Fig. 2 in the main text). 

\begin{figure*}
  \includegraphics[width=\textwidth]{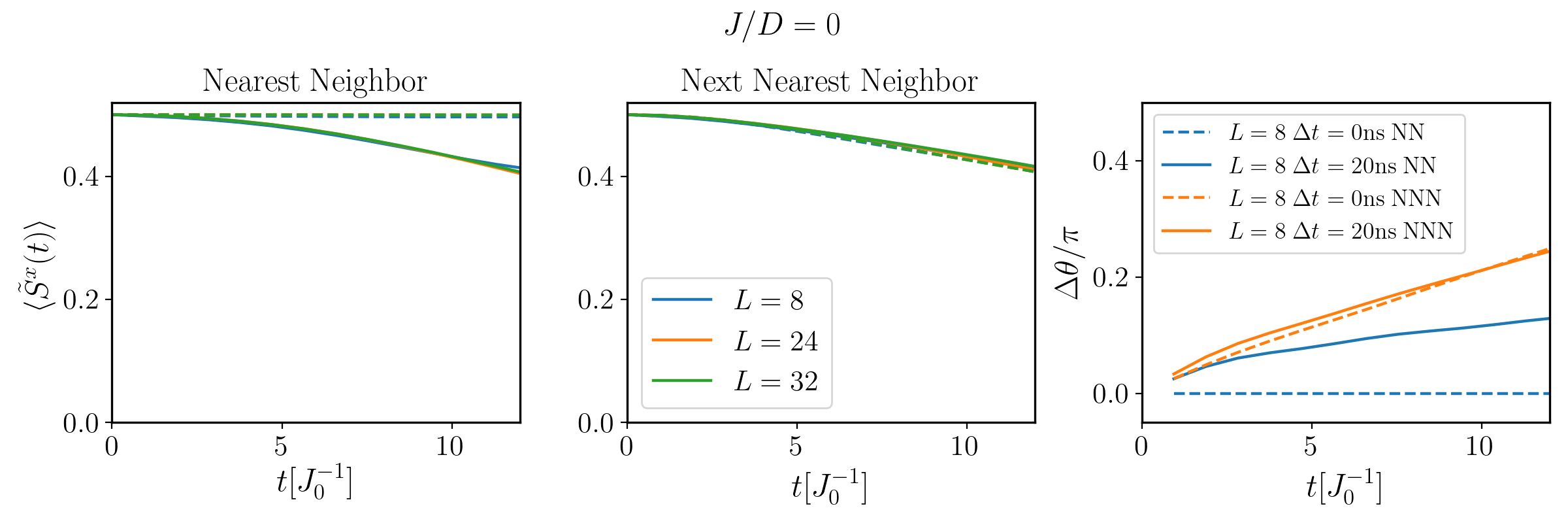}
  \caption{Effect of next-nearest neighbor interaction on the engineering of Dzyaloshinskii-Moriya interaction between Rydberg atoms. The simulation is similar to the one performed in Fig. 2(c) of the main text, calculating the average local magnetization change while including the next order of the long range interaction between Rydberg atoms (middle panel). The result in Fig. 2(c) is given in the left panel for comparison. The right panel compares the time dependent $\Delta \theta$ in both cases, for either an ideal or a practical modulation scheme. The parameters for these simulations were $T=600 \rm ns$ and $J_0=2\pi\cdot 250\rm{kHz}$.}
  \label{fig:Effect of NNN}
\end{figure*}

The simulation results are summarized in Fig. \ref{fig:Effect of NNN}, where it is directly visible that the experimental observable - the average magnetization - remains unchanged. The NNN interaction does, however, affect the coherence of the engineered Hamiltonian, as can be seen in the time-dependence of $\Delta \theta$. Not only is it a very dominant effect, overtaking the effect of the finite pulse length $\Delta t$, but it also degrades the Hamiltonian coherence. Therefore, one can surmise that the magnetization dynamics for a general initial state may not be as similar. That said, the Hamiltonian still remains coherent for times far longer than the Rydberg coherence time, given the parameters considered, such that the effect of NNN interaction on experiment is currently limited. 

Since interactions over an even longer range are at least 3 times weaker than the NNN, we believe this simulation validates the accuracy of the simulations performed throughout the main text and their relevance in supporting experimental endeavors. Even so, should higher-order interaction terms ever become an issue, it is essentially possible to include both control of NN and cancellation of NNN interaction in the Floquet engineering pulse sequence, at the cost of increased complexity (a longer modulation period, as well as additional local modulation pulses).

\begin{figure*}
  \includegraphics[width=0.45\textwidth]{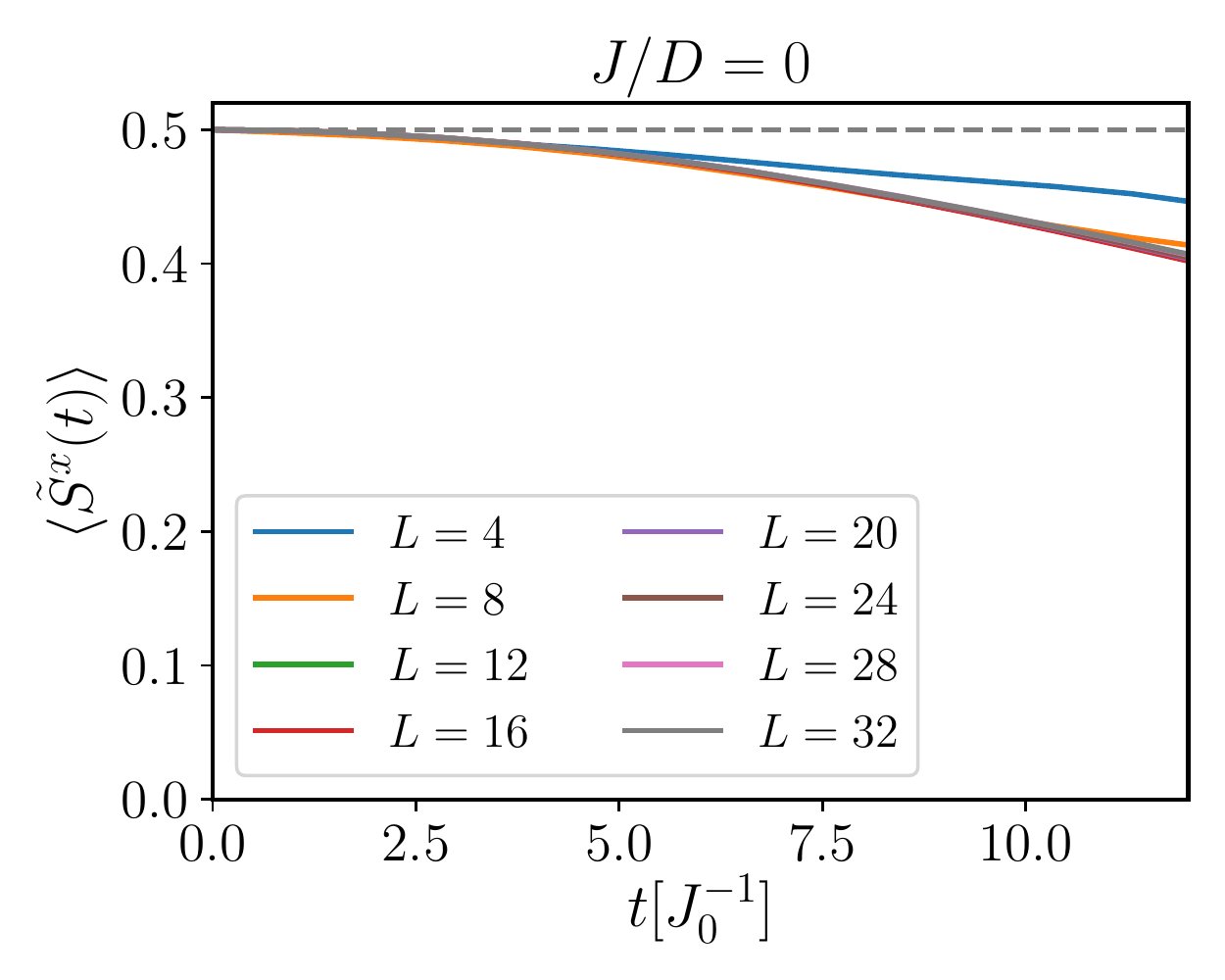}
  \caption{Scaling up quantum simulation with Floquet engineering. The figure shows the local magnetization dynamics of a zero-energy eigenstate in a ring of $L$ atoms, undergoing the modulation shown in Fig. 2 of the main text. The atom number ranges from 4 to 32, in multiples of 4. Aside from boundary effects for low $L$, the dynamics of the system remain virtually unchanged. Thus, it is proven that our method for Floquet engineering is indeed scalable to many atoms. The parameters for these simulations were $T=600 \rm ns$, $\Delta t = 20 \rm ns$ and $J_0=2\pi\cdot 250\rm{kHz}$.}
  \label{fig:Dependence on N}
\end{figure*}

\section{General Considerations for pulse sequence structuring}

As mentioned above, all pulse sequences appearing in this manuscript are inherently symmetric, greatly reducing the first-order corrections to the effective Hamiltonian picture. Otherwise, two other considerations were employed when constructing pulse sequences for the various engineered interactions: the symmetry of the Hamiltonian, as derived from the system geometry; and the boundary conditions of the problem. 

The geometry of the system, when assuming a periodic array, constrains the number of interacting atoms in each of the array's unit cells. Hence, the need to engineer the interaction between all of the atoms in the unit cell constrains the number of pulses required. A good example for scaling with the number of atoms in a unit cell, which is a geometry-dependent property, is given by the sequence used to engineer a DM interaction in a 1D geometry (as in Fig. 2 of the main text), as opposed to a triangular 2D geometry (as shown in \cite{tsesses2022}). At any rate, it is important to note that even if the required resources (i.e., number of addressing fields and their power) scale with the number of atoms, the length of the pulse sequence does not, such that our scheme is indeed scalable. Several examples were already given in the manuscript, and we append a more detailed investigation of one of them here (Fig. \ref{fig:Dependence on N}).  

\begin{figure*}
  \includegraphics[width=0.75\textwidth]{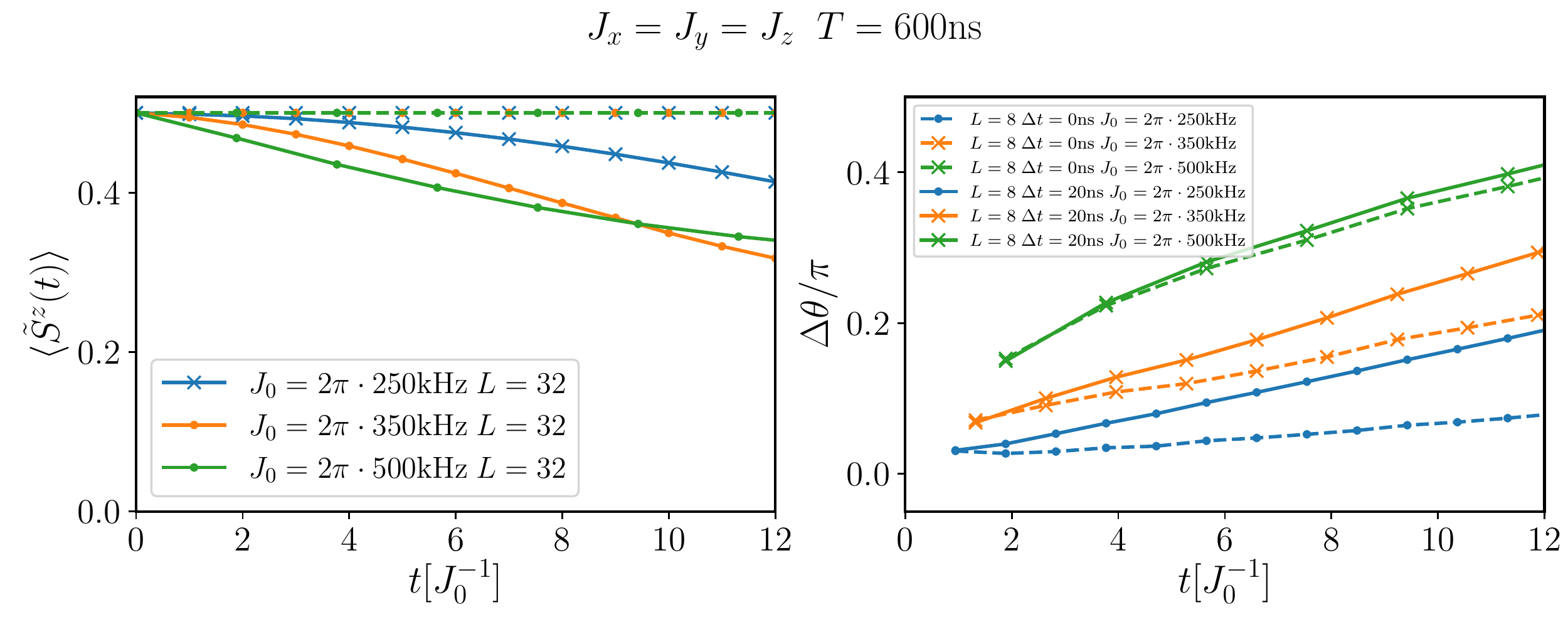}
  \caption{Effect of changing the bare Rydberg interaction strength $J_0$ on the engineered Hamiltonian coherence time and magnetization dynamics. As $J_0$ increases, the magnetization decays faster and the decoherence time, fulfilling the condition $\Delta \theta = \pi/7$, decreases. This is a direct result of exiting the high-frequency regime of Floquet engineering. For $J_0=250 \rm kHz$ (blue curves), the results are as in the main text (Fig. 3). For $J_0=350 \rm kHz$ (orange curves), the Hamiltonian remains coherent up to $t \approx 5J_0^{-1}$, in accordance with the quadratic dependence of the coherence on $J_0$, established in Fig. \ref{fig:DeltaTheta_scaling_with_J0}. Additionally, there is still a visible difference between ideal and practical modulation. For $J_0=500 \rm kHz$ (blue curves), we are completely out of the high-frequency regime, and the practical and ideal cases appear to have the same dependence on $\Delta \theta$ and a coherence time less than $J_0^{-1}$. This shows that controllable quantum simulation in our scheme is only possible in the high-frequency regime. The pulse width in the simulations was taken as $\Delta t = 20 \rm ns$.}
  \label{fig:Effect_of_J0}
\end{figure*}

The boundary conditions, on the other hand, do not change the number of applied pulses but can require additional resources. For example, DM interaction in a 1D geometry with $\it open$ boundary conditions would require only 2-atom segments, instead of the 4-atom segments used for $closed$ boundary conditions in Fig. 2 of the main text, shortening the pulse sequence by half. Specifically in this example, this change is necessary in order to correctly engineer the interaction between the first and last atoms of the 1D chain. We stress that, while the pulse sequences presented in this work are sufficient for meaningful quantum simulation, we do not claim that they are necessarily the most efficient in the number of applied pulses, and there may yet be more possible optimization in this regard.

\section{Hamiltonian-specific Considerations for pulse sequence structuring}

\begin{figure*}
  \includegraphics[width=0.4\textwidth]{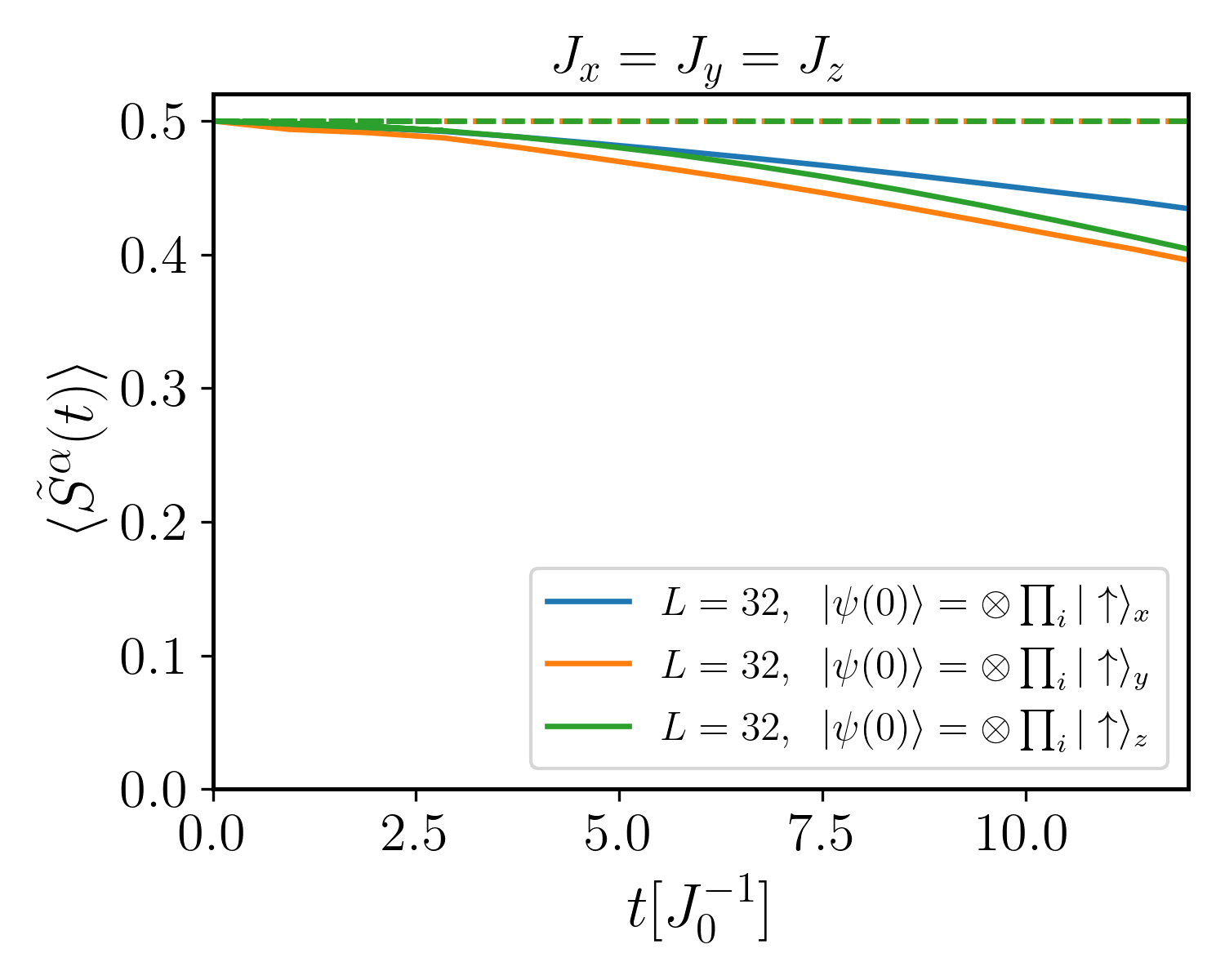}
\caption{Influence of choosing different initial states on demonstrating zero-energy eigenstate dynamics in the engineered XYZ interaction. Each curve shows the magnetization dynamics of a ferromagnetic state initialized along a different axis when engineering an isotropic XYZ interaction Hamiltonian. Ideally, the curves should all coincide, yet the pulse widths in a practical modulation scheme cause some variance in their behavior. That said, the magnetization decays in a fairly similar fashion in all curves, which is a result that could only be achieved when optimizing the engineered Hamiltonian (e.g., through the $\Delta \theta$ parameter) instead of the dynamics of a single state. The parameters for these simulations were $T=600 \rm ns$, $\Delta t = 20 \rm ns$ and $J_0=2\pi\cdot 250\rm{kHz}$, the same as in Fig. 3 of the main text.}
  \label{fig:Global_rotation_diff_state}
\end{figure*}

Aside from the general considerations laid out above, the pulse sequences we used had task-specific considerations. In the case of engineering an XYZ interaction, the considerations are the same as those in \cite{geier2021,Scholl2022}, and the same sequence was used both for comparison and for simplicity, even though it does not allow to engineer a general anisotropy between the coupling strength in different axes. We chose a feasible value of the bare Rydberg atom interaction strength $J_0$ in the simulation, though it should be noted that higher $J_0$ values are possible, at the cost of reducing the engineered Hamiltonian coherence time (see Fig. \ref{fig:Effect_of_J0}). 

\begin{figure*}
  \includegraphics[width=0.4\textwidth]{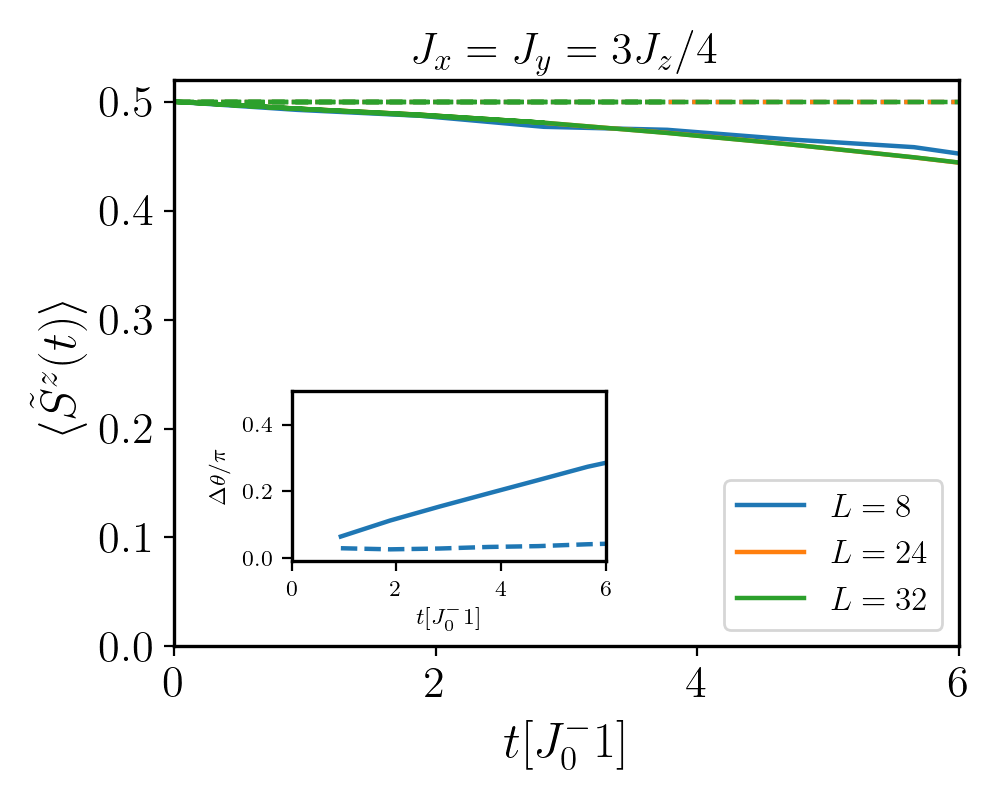}
  \caption{Engineering anisotorpic XYZ interaction between Rydberg atoms. The figure shows an anisotropic XYZ interaction in a ring of atoms, with the system initialized to a ferromagnetic state along the z axis. The initial state is a zero-energy eigenstate of the Hamiltonian, and should thus, ideally, exhibit no dynamics. In practice, the magnetization in the system decays slowly, reaching half of its initial value at $t \approx 14J_0^{-1}$. However, the Hamiltonian coherence time is much shorter ($t \approx 3J_0^{-1}$). We attribute this to a secondary effect of the pulse width $\Delta t$, as larger anisotropy require some of the free evolution times to become very short, thus accruing more errors (as was also exemplified in Fig. \ref{fig:DeltaTheta_scaling_with_T}. The parameters for these simulations were $T=600 \rm ns$, $\Delta t = 20 \rm ns$ and $J_0=2\pi\cdot 250\rm{kHz}$, the same as in Fig. 3 of the main text (though the optimized pulse sequence is different).}
  \label{fig:Global_rotation_XXZ}
\end{figure*}

It is also important to note that optimizing the pulse sequence according to the $\Delta \theta$ parameter instead of the dynamics of a single initial state, was crucial to achieve correct Hamiltonian engineering, as is exhibited by the similar dynamics of the three degenerate zero-energy eigenstates in the case of $J_x=J_y=J_z$ (Fig. \ref{fig:Global_rotation_diff_state}). This is further exemplified in Fig. \ref{fig:Global_rotation_XXZ}, where the dynamics of the zero-energy state for an anisotropic XYZ interaction is simulated. At first glance, the dynamics appear very close to the ideal case, suggesting a long coherence time, yet the $\Delta \theta$ parameter proves that this behavior is only state-specific, and the engineered Hamiltonian is only coherent for a time $t \approx 3J_0^{-1}$. 

In the case of engineering DM interaction, the free evolution time $t_{-J}$ was necessary to allow cancellation of the Heisenberg coupling strength $J$ and span the $J/D$ ratio between 0 and $\infty$. Furthermore, the constraint for 4-atom segments, arising from the required boundary conditions, also limits the number of atoms that may exhibit a zero-energy eigenstate for a certain $J/D$ ratio. Thus, the two cases presented in Fig. 2 of the main text ($J/D=0$ and $J/D=1$) are the only two $J/D$ ratios that have a zero-energy eigenstate in an 8-atom ring.S

\begin{figure*}
  \includegraphics[width=\textwidth]{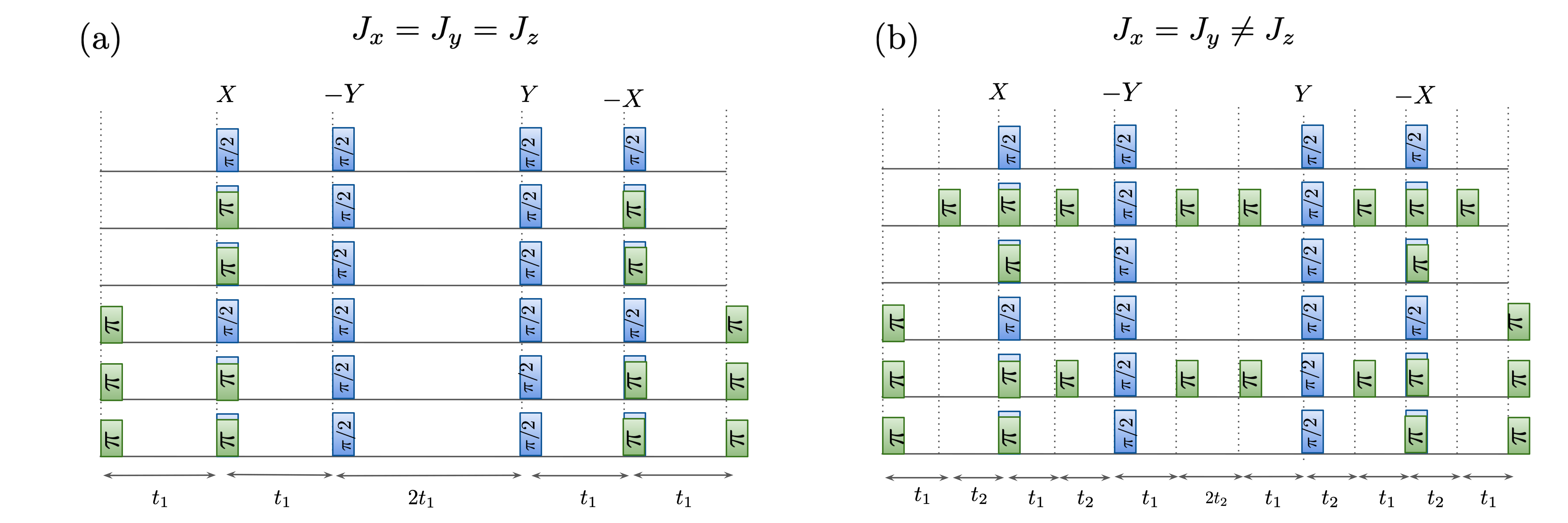}
  \caption{Pulse sequence for engineering the Kitaev interaction (defined in the main text). (a) Global modulation is applied similarly to generating an XYZ interaction, with each pulse rotating the interaction frame of reference for all atoms around either the x or y axes, such that during each free evolution time period the instantaneous interaction is along a different set of two axes. at any given free evolution, the relative phase between the ladder operators of two adjacent atoms is shifted $\it simultaneously$ with the global modulation through local modulation pulses, which rotate along the z axis. Thus, the interaction along unwanted axes cancels out, and each set of two atoms interact only along a single axis, as defined by the Kitaev Hamiltonian. This pulse sequence carries the same limitation on the interaction strength anisotorpy as the sequence creating the XYZ interaction (i.e., $|J_k| \leq 2(|J_l|+|J_m|)$,  $k,l,m \in {x,y,z}$), which is exactly the condition for the B phase of the Kitaev model \cite{kitaev2006}. (b) In order to exceed the anisotropy afforded only by the global modulation, we design another pulse sequence that deliberately reduces the interaction strength along two predefined axes, via additional local modulation pulses. Thus, any required anisotropy can be designed, enabling the simulation of the A phase of the Kitaev model as well. Blue pulses represent global modulation, acting simultaneously on all of the atoms, while green pulses represent local modulation, applied to each atom separately.}
  \label{fig:kitaev_sequence}
\end{figure*}

In the case of Kitaev interaction, the constraints are many and varied, resulting in two possible pulse sequences, both of which are illustrated in Fig. \ref{fig:kitaev_sequence}. These sequences are based on the global modulation generating an XYZ interaction (see Fig. 3 of the main text), and make use of local modulation to produce position-dependent interactions along specific axes. Thus, in the nearest-neighbours approximation, one only needs to control the XYZ interaction strengths via the global modulation, while forcing the operators of two adjacent atoms to be in-phase when interacting along the preferred axis or $\pi$ out-of-phase at any other time. This is illustrated in Fig. \ref{fig:kitaev_sequence}(a), where local modulation is applied only during global modulation. This scheme, used to generate the results in Fig. 4 of the main text, also has the unexpected added value of nullifying next nearest-neighbours interaction. We note that by adding a detuned microwave drive in the Rydberg manifold while the pulse sequence in Fig. \ref{fig:kitaev_sequence}(a) is applied, it is possible to create the effective magnetic field required to generate anyonic excitations in the Kitaev model \cite{kitaev2006}. 

\begin{figure*}
  \includegraphics[width=\textwidth]{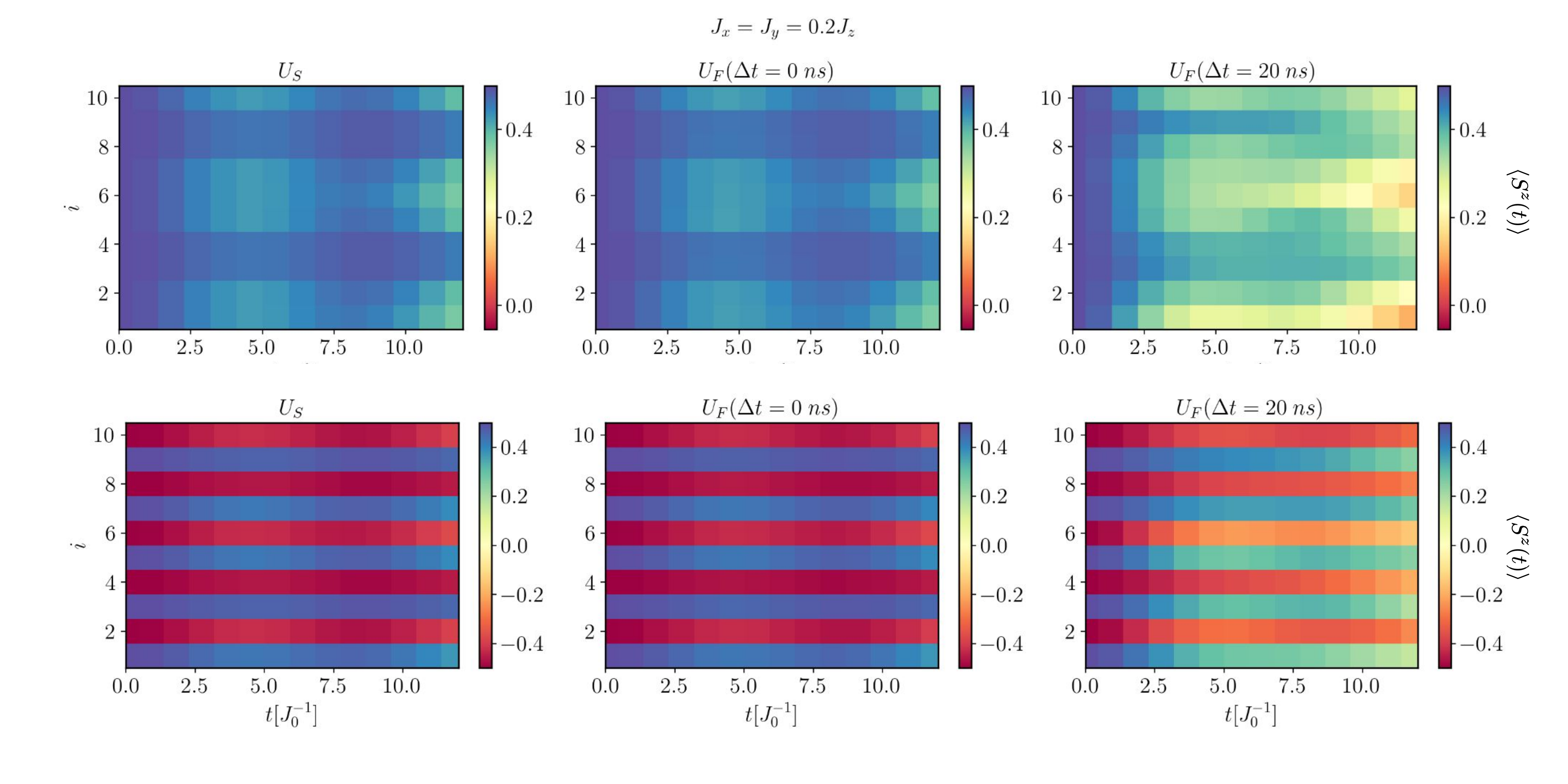}
\caption{Engineering the A (gapped) phase of the Kitaev model. The simulation shows magnetization dynamics along the z axis in color coding, where the vertical axis is the atom number, as defined in Fig. 4 of the main text. We utilize the pulse sequence in Fig. \ref{fig:kitaev_sequence} (b), to produce interaction along the z axis that is 5 times larger than along x or y (strictly speaking, the A phase requires only a ratio larger than 2). The initial states considered here are the same as in Fig. 4 (top: a ferromagnetic state along z; bottom: an antiferromagnetic state along z). In the ideal case the dynamics is virtually the same as in the actual Kitaev model. However, for a practical modulation with pulse widths $\Delta t = 20 \rm ns$, the dynamics follow that which is prescribed by the Kitaev Hamiltonian only until $t \approx 3J_0^{-1}$. The parameters for these simulations were $T=580 \rm ns$,  and $J_0=2\pi\cdot 250\rm{kHz}$, the same as in Fig. 4.}
  \label{fig:Kitaev_A}
\end{figure*}

That said, the pulse sequence in Fig. \ref{fig:kitaev_sequence}(b) suffers from the same limitation described above for the XYZ interaction - it is limited in the possible anisotropy of the coupling strengths. In fact, the limit is exactly the transition point between the B and A phases of the Kitaev model \cite{kitaev2006}. To allow full access to the Kitaev model, we also include a more complex pulse sequence, illustrated in Fig. \ref{fig:kitaev_sequence}(b), whereby local modulation pulses in between global modulation pulses assist in achieving an arbitrary anisotropy of the interaction strengths (at least in theory). Since the pulse sequence in Fig. \ref{fig:kitaev_sequence}(b) requires the application of more pulses, and we constrain the modulation period $T$ to remain the same, its results are degraded compared to those appearing in the main text, as can be viewed in Fig. \ref{fig:Kitaev_A}. Nevertheless, the engineered Hamiltonian is still able to follow the required dynamics for $t \approx 3J_0^{-1}$, which is only slightly below the current Rydberg coherence time ($4J_0^{-1}$ at the most in our chosen parameters).

\end{document}